%% file: main.tex
\documentclass[10pt]{article}

\usepackage{newtxtext,newtxmath}

\usepackage{graphicx}

\usepackage[letterpaper,margin=2cm]{geometry}
\renewenvironment{abstract}
	{\quotation}
	{\endquotation}

\date{}

\makeatletter
\renewcommand{\fnum@figure}{\textbf{Figure \thefigure}}
\renewcommand{\fnum@table}{\textbf{Table \thetable}}
\makeatother

\usepackage{scicite}

\usepackage[hyphens]{url}
\usepackage{hyperref}
\hypersetup{
    colorlinks=true,
    linkcolor=blue,
    citecolor=blue,
    urlcolor=blue
}
\usepackage{xstring}

\usepackage{etoc}
\usepackage{caption}
\usepackage{subcaption}
\usepackage[section]{placeins}

\usepackage{fancyhdr}
\usepackage{booktabs}

\def\scititle{Community Notes undermoderate polarizing content by design creating risks in electoral processes}

\title{\bfseries \boldmath \scititle}

\author{
    Paul Bouchaud$^{1,2,3,\dag}$,
	Pedro Ramaciotti$^{1,3,4,\ddag}$\and
	\small$^{1}$Complex Systems Institute of Paris Ile-de-France CNRS, Paris, France.\and
	\small$^{2}$CAMS EHESS, Paris, France.\and
    \small$^{3}$médialab, Sciences Po, Paris, France.\and
    \small$^{4}$Learning Planet Institute, CY Cergy Paris University, Paris, France.\and
    \small \dag\, paul.bouchaud@cnrs.fr\and
    \small \ddag\, pedro.ramaciotti-morales@cnrs.fr
}

\begin{document} 

\maketitle

\begin{abstract} 
\begin{center}
\vspace{1cm}
\textbf{Abstract}
\vspace{1cm}
\end{center}

\input{Contents/0_Abstract}
\end{abstract}

\vspace{0.5cm}

\twocolumn
\pagestyle{fancy}

\input{Contents/1_Introduction}

\input{Contents/2_Data}

\input{Contents/3_Results}

\input{Contents/4_Conclusions}

\input{Contents/7_Methods}


\bibliography{references} 
\bibliographystyle{sciencemag}


\input{Contents/5_Additional_Information}

\clearpage
\newpage



\renewcommand{\thefigure}{S\arabic{figure}}
\renewcommand{\thetable}{S\arabic{table}}
\renewcommand{\theequation}{S\arabic{equation}}
\renewcommand{\thepage}{S\arabic{page}}
\setcounter{figure}{0}
\setcounter{table}{0}
\setcounter{equation}{0}
\setcounter{page}{1} 

\appendix


\onecolumn

\begin{center}
\section*{Supplementary Materials for\\ \scititle}

Paul Bouchaud and Pedro Ramaciotti
\end{center}

\etocsettocstyle{}{}

\localtableofcontents

\newpage


\input{Contents/6_SI}

\end{document}

%% file: Contents/0_Abstract.tex
Community Notes (CNs) of X enables users to collaboratively moderate misleading content. To resolve conflicting moderation, CNs infers a latent ideological dimension and selects notes garnering cross-partisan support. As this system is now deployed worldwide, we evaluate its operation across diverse polarization contexts. We analyze all 1.9 million moderation notes receiving 135 million ratings by March 2025, cross-referencing ideological scaling data on 13 countries. Our results show that the CNs algorithm effectively captures the main polarizing dimensions across countries, surfacing notes that garner cross-partisan support. This also means that, by design, CNs systematically under-moderate polarizing content. We analyze notes relating to four recent elections in the US (2024), the UK (2024), France (2024) and Germany (2025) and demonstrate that they are systematically under-moderated when compared to other notes, posing potential risks to civic discourse and electoral processes.

%% file: Contents/1_Introduction.tex
\noindent{}Social media platforms have become ubiquitous sources for the consumption of news content \cite{Walker_2021}, where users may also participate in content creation and dissemination.
A persistent challenge in these platforms is the dissemination of misinformation \cite{vanderLinden2024,Ceylan2023,Allen2024}, prompting platforms to deploy means for content moderation \cite{Kozyreva2023, Jhaver2023}. While expert fact-checking has been shown to successfully decrease the extent to which people agree with misleading claims \cite{Nyhan2019,Pennycook2020, Porter2022}, this approach faces significant limitations in practice: i) professional fact-checking operations cannot scale to meet the volume of content requiring verification \cite{Rodrigo_2020, Stray_Sneider_2025}, ii) their efforts may be disconnected from online attention \cite{ribeirozannettougogabenevenutowest2022}, iii) moderation may be perceived as having partisan bias by the public \cite{Walker_2021}, and iv) they may suffer from low visibility \cite{Robertson2020}. Additionally, professional moderation may backfire by further ``entrenching'' polarized individuals \cite{Reinero2023}. 
On the other hand, research has shown that crowd-sourcing fact-checking assessments from regular users can effectively identify misinformation, with politically balanced laypeople ratings correlating strongly with those of professional fact-checkers \cite{Pennycook2019, Bhuiyan2020, Allen2021,Martel2023}.

Motivated by these findings, X (formerly Twitter) piloted a crowd-sourced fact-checking system in the United States in early 2021 \cite{birdwatch}, later deployed worldwide in 2023 under the name Community Notes \cite{CommunityNotes}. This system enables self-selected contributors to attach contextual notes to posts they consider ``misinformed or potentially misleading''. Other contributors can then rate the helpfulness of these notes based on their subjective assessment of the content and its context. A core challenge for the platform is moderating polarizing content, for which consensus on usefulness may not emerge between raters across the ideological spectrum (e.g., when raters preferentially seek to moderate posts by political adversaries).

To address this challenge, X first learns to predict how contributors rate notes fitting latent ideology parameters for users and notes, then algorithmically identifies notes that appeal to users across ideological divides rather than only to like-minded individuals \cite{birdwatch, ovadya2023}. Instead of merely reflecting the idiosyncratic ``helpfulness'' of a note, contributor ratings are subjective assessments that may incorporate ideological alignment or misalignment between the rater and the claim being discussed. To account for this, X's Community Notes system infers latent ideological positions for both the notes ($\theta_n$) and the raters ($\theta_r$) under the key hypothesis that greater ideological alignment increases the probability of a rater rating the note as \textit{Helpful}.

Computationally, X first converts ratings given to notes to numerical values: 1 for \textit{Helpful}, 0.5 for \textit{Somewhat Helpful}, and 0 for \textit{Not Helpful} \cite{TwitterGithubMain}. Then, to predict these empirical ratings, the system infers several parameters: note latent ideologies ($\theta_n$), rater latent ideologies ($\theta_r$), individual rater leniency ($\beta_r$), the tendency of notes to be rated helpful regardless of raters' ideologies ($\beta_n$), and a global helpfulness baseline ($\beta_0$). These parameters are estimated, from the ratings alone, to best predict the rating $\eta$ that rater $r$ will assign to note $n$ through the predictive formula $\hat{\eta}_{rn}=\beta_0+\beta_n+\beta_r+\theta_n\cdot\theta_r \label{eq:predictive_formula}$.

Consequently, a note rated \textit{Helpful} by diverse raters (i.e., with $\theta_r$ values spread across positive and negative values) will have a higher $\beta_n$ value and lower absolute value $|\theta_n|$ when compared to notes rated \textit{Helpful} by only one side of the spectrum. X's Community Notes service uses these parameters to algorithmically resolve conflicting ratings by selecting notes with the highest $\beta_n$ values \cite{birdwatch} and displaying them under the post, visible to all users on the platform \cite{X_2024}.

Results from X's pilot phase in the United States revealed that notes selected for display garnered broad appeal across ideological positions \cite{birdwatch} and that users exposed to such notes were less likely to share the posts and less inclined to agree with the posts' misleading claims. Independent research found high agreement between these algorithmically selected notes and expert fact-checker classifications \cite{Saeed2022}. Additionally, Allen et al. \cite{Allen2022} showed that political partisanship was highly predictive of contributors' ratings in the US, with raters preferentially challenging content from those with whom they disagree politically. Following X's Community Notes global deployment in 2023, subsequent studies corroborated these findings, documenting reduced sharing of flagged content \cite{Renault2024} and accurate sourcing of credible references for notes addressing COVID-19 vaccination misinformation \cite{Allen2024}. Moreover, moderation resulting from the Community Notes service was shown to be less prone to backfire effects that entrench individuals \cite{Kim2025, Reinero2023} and generated greater user trust than warnings issued by expert fact-checkers \cite{Drolsbach2024}.

X's Community Notes system was developed and tested in the United States \cite{birdwatch, Allen2022}, where online polarization is traditionally modeled using a single ideology dimension \cite{barbera2015tweeting, bond2015quantifying}. 
However, the relevance of this single-dimensional representation of note and rater ideology ---central to the algorithmic moderation process in Community Notes--- remains largely unexplored in other national settings that might be structured along several ideology dimensions \cite{Bakker2012,ramaciotti2022embedding,barbera2015understanding}. 
Whether the Community Notes model captures relevant political dimensions in different countries has a direct impact in how the service functions, and it determines what type of polarizing content will not be algorithmically moderated due to lack of sufficient agreement among raters.
In this study, we examine X's Community Notes usage and algorithmic outcomes across all countries in which the service is available.
Leveraging ideology scaling of follower networks \cite{barbera2015birds} calibrated with political survey data \cite{ramaciotti2022inferring}, we compare the latent ideological parameter inferred by Community Note's algorithm with the main dimensions of political competition in 13 countries.
Our results show that in all 13 countries we study, the Community Notes model reliably identifies the main ideological divide that structures political communication networks on X. This remains true even after accounting for a second dimension related to misinformation: trust in elites and institutions \cite{ramaciotti2023geometry,arceneaux2021some}.
This implies that notes are proposed, rated, and surfaced as \textit{Helpful} across the ideological spectrum. However, because this system only surfaces notes that receive broad cross-ideological agreement, content that strongly activates polarization is systematically under-moderated.
We demonstrate the importance of this intrinsic limitation by showing that the share of notes related to elections is consistently lower than for other topics during four major elections in our study period: the 2024 U.S. Presidential election, the 2024 U.K. general election, the 2024 French legislative elections, and the 2025 German federal elections.
While Community Notes offers several important benefits, this limitation poses a significant risk for information integrity during electoral processes.

%% file: Contents/2_Data.tex
\section*{Data and definitions} 

\subsection*{Community Notes data}

To enable third-party scrutiny, X publicly discloses all proposed Community Notes, their ratings, the helpfulness status algorithmically assigned to them, and user requests soliciting Community Notes on posts they believe would benefit from additional context.\newline

As of March 1, 2025, X had disclosed data on 1.88 million Community Notes, which received 135.17 million ratings from 1.15 million contributors. Additionally, 1.49 million X users submitted 4.83 million requests for Community Notes associated with 1.91 million posts.
Following X's preprocessing protocol \cite{birdwatch, TwitterGithubMain}, we retained notes with at least 5 ratings and contributors who rated at least 10 notes. This resulted in 1.45 million notes associated with 886\,370 X posts, rated 133.20 million times by 821\,250 contributors (see descriptive statistics in SM.S\ref{subsec:data_collection_cn}); 59.5\% of ratings classify the proposed notes as \textit{Helpful}, 37.5\% as \textit{Not Helpful}, and 3.0\% as \textit{Somewhat Helpful}. To enable contextual analysis beyond ratings, the posts associated with 83.5\% of proposed notes were retrieved, while the remainder had been deleted by March 2025 (see SM.S\ref{subsec:data_collection_post}).
Hereafter, we use ``notes'' to refer to Community Notes flagging the original post as ``misinformed or potentially misleading'' (73.0\% of all notes; the others are not meant for public display but allow contributors to argue why a note is not needed; see SM.S\ref{sec:presentation_cn}).

\subsection*{Inferring the parameters of the Community Notes model}

We applied the regularized matrix factorization outlined in X's publication \cite{birdwatch} and implemented in their open-source code \cite{TwitterGithubMain}, to infer the global helpfulness bias $\beta_0$, rater biases $\beta_r$, note biases $\beta_n$, and the latent ideologies of notes and raters, $\theta_n$ and $\theta_r$ (see details in SM.S\ref{sec:training_cn}).
The parameters mirror those reported in X's publication \cite{birdwatch} and align closely with note outcomes publicly displayed on X. Quantitatively, the note biases $\beta_n$ we inferred from ratings predict whether a note was algorithmically selected as \textit{Helpful} by X (hereinafter, achieving \textit{Helpful Status}) with an AUC of 0.92, and predict \textit{Not Helpful Status} with an AUC of 0.97 (see the validation of model parameters in SM.S\ref{subsec:training_results}).

\subsection*{Multidimensional ideology scaling of accounts on X}

To examine the ideological scale inferred by X's system, we rely on the method developed by Ramaciotti et al. \cite{ramaciotti2022inferring}, which adapts Barberá's \cite{barbera2015birds} methodology to multidimensional settings.
Specifically, we identify X accounts following Members of Parliament (MPs) from 23 countries across five continents, retaining those that followed at least 3 MPs and were followed by at least 25 accounts, following Barberá's protocol \cite{barbera2015birds,barbera2015understanding,ramaciotti2022inferring}. We then infer, through Correspondence Analysis, a multidimensional latent space arranging MPs and their followers by homophily (see details in SM.S\ref{subsec:epo_dataset}).

For the analysis of Community Notes latent ideology parameter, we consider the position of X accounts along $\delta_1$, the first principal component of the Correspondence Analysis, which best explains which MPs they follow. Additionally, we map the positions of accounts onto interpretable political dimensions of the Global Party Survey \cite{norris2020measuring} ---an expert survey positioning parties along several dimensions endowed with a reference frame making them comparable across countries--- namely i) the Left-Right dimension, a leading dimension of polarization relevant for studies of misinformation in social media \cite{roozenbeek2020susceptibility,mosleh2024differences,grinberg2019fake}, and ii) the Anti-Elite dimension, which measures criticism toward elites and institutions. In contrast to results regarding online misinformation in the US \cite{grinberg2019fake}, the degree of negative attitudes towards and institutions has been shown to play a role in the probability of disseminating misinformation \cite{arceneaux2021some,ramaciotti2023geometry}.
When comparing latent ideologies $\theta_n$ and $\theta_r$ with Left-Right and Anti-Elite dimensions, we focus on 13 countries where at least a thousand X accounts with inferred political positions have authored posts with proposed Community Notes. These countries represent a geographically diverse sample (see SM.S\ref{subsubsec:epo_x_cn}): \textit{United States, United Kingdom, Japan, Spain, France, Brazil, Canada, Germany, Argentina, Israel, Australia, Poland,} and \textit{Mexico}.

%% file: Contents/3_Results.tex
\section*{Results} 
\subsection*{Notes and ratings} 
Community Notes usage reveals highly unequal distributions (see SM.S\ref{subsec:data_collection_cn}). Among the 1.1 million users self-enrolled in the program as of March 2025, 28.9\% have rated fewer than ten notes. Just 1\% of the 405 thousand contributors with note-writing privileges authored 31.2\% of all proposed Community Notes. Similarly, 35.8\% of all proposed Community Notes are associated with posts published by only one thousand X accounts (see most commented accounts per country in SM.S\ref{subsubsec:country_split_methods}). Of the posts for which users requested Community Notes, 79.3\% were flagged by only a single user, and a note was proposed for only 12.5\% of these posts.\newline

\begin{figure*}
\centering
\includegraphics[width=\textwidth]{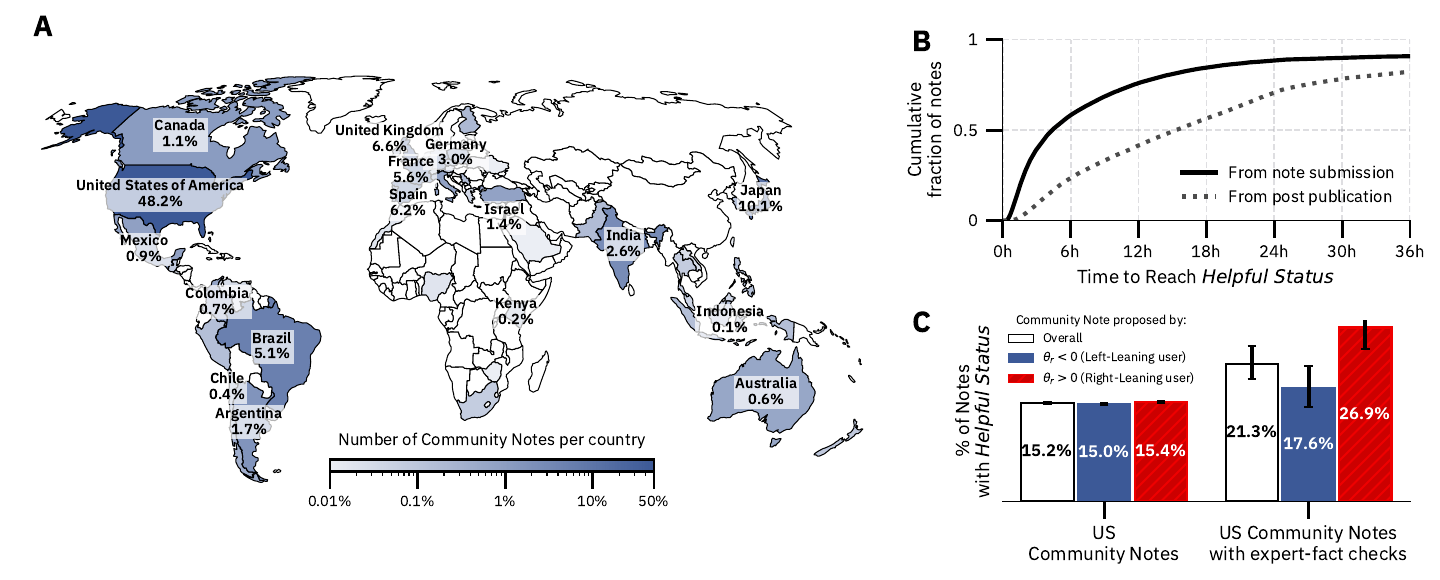}
\caption{\textbf{Community Notes usage and outcomes.} (\textbf{A}) Fraction of Community Notes proposed per country. To be associated with a country, a note must be predominantly evaluated by contributors who typically rate notes in that country's language and that reference national websites (see SM.S\ref{subsubsec:country_split_methods}). (\textbf{B}) Cumulative distribution of time needed for a proposed note to reach \textit{Helpful Status} via X's algorithm, measured from the publication of the original post and from the publication of the note (truncated tails after 36 hours, representing 9.2\% and 17.9\% of notes, respectively). (\textbf{C}) Fraction of notes reaching \textit{Helpful Status} in the US (left) and in the US when referencing expert fact-checks (right), segmented by the ideological leaning of note authors. Error bars show 95\% CI via bootstrapping over notes and fact-checking organizations (see SM.S\ref{paragraph:fact_checks}).}
\label{fig:fig_1_description}
\end{figure*}

Figure~\ref{fig:fig_1_description}.A reveals the worldwide usage of Community Notes. The majority of notes relate to the United States, followed by Japan, the United Kingdom, Brazil, and France (see the country segmentation methodology in SM.S\ref{subsec:country_split}). In the European Union, the number of Community Notes contributors per country correlates strongly with the number of active logged-in users (Pearson $\rho=0.823, ~ p<10^{-4}$), as disclosed under European transparency obligations.\newline

96.2\% of Community Notes, after X's preprocessing \cite{birdwatch, TwitterGithubMain}, contain URLs pointing to supporting sources that aim to contextualize the original post. To examine news media citation practices in Community Notes, we compiled a list of 2\,134 global outlets from BBC Media Guides, supplemented by manual review of the most frequently referenced domains in notes. These established media outlets constitute the predominant sources in notes, appearing in at least 25.9\% of notes, followed by X posts (17.4\%, where contributors cite other posts as evidence) and Wikipedia articles (8.9\% of notes).

Additionally, we curated a list of web domains from professional fact-checking organizations that are International Fact-Checking Network signatories or appear in Duke Reporter's Lab directory, revealing that expert fact-checking articles appeared in only 3.5\% of proposed notes (see the detailed curation of different types of sources in SM.S\ref{subsec:sources}).\newline


Political content accounts for 65.2\% of posts and 76.6\% of Community Notes ratings, with sports as the second most frequent category (6.3\% of posts) (see complete distribution and validation in SM.S\ref{subsec:post_topic}).
To further characterize notes discussing politics, we use a topic model \cite{Burst:2023} trained on political party manifestos from the Manifesto Project \cite{MANIFESTO}. Among the topics included in this scheme, the ones most discussed in notes are \textit{Political Authority} (25.6\% of posts, including praise and criticism of specific parties), \textit{Law and Order} (12.4\%, including mention of crimes, the police, and the judiciary), \textit{Equality} (7.2\%, mentioning underprivileged groups, social classes, economic redistribution and discrimination), \textit{Environmental Protection} (6.6\%, including natural resources, natural preservation, and animal rights) and \textit{Freedom and Human Rights} (6.2\%, mentioning freedom of speech, press and assembly) (see a detailed distribution and validation of political topics in SM.S\ref{subsec:post_manifesto}).\newline

Finally, we observe in Figures \ref{fig:request_submit_helpful}.A \& \ref{fig:request_submit_helpful}.B that, across analyzed countries, Community Notes are requested and proposed under posts authored by accounts throughout the political spectrum. Of the non-deleted posts that received proposed Community Notes and have authors with inferred ideological leanings, 58\% are from right-leaning accounts in the United States, 54\% in Japan, and 53\% in France (see left-right leaning distribution per country in SM.S\ref{subsubsec:epo_x_cn}).

\subsection*{Algorithmic resolution outcomes} 

We now examine the status assigned by X's Community Notes algorithm to proposed notes.
Of all proposed Community Notes arguing the original post is ``misinformed or potentially misleading'', only 11.97\% garner sufficient favorable ratings from ``diverse" contributors to attain \textit{Helpful Status} and be made visible on X. The remaining notes are not publicly displayed, either because consensus determined them to be \textit{Not Helpful} (2.81\%; hereinafter, achieving \textit{Not Helpful Status}), or because they lacked sufficient ratings from diverse contributors to reach a conclusive status (85.22\%). 17.1\% of posts for which at least 5 users requested Community Notes ultimately received a proposed note that reached \textit{Helpful Status}. Between January and March 2025, half of the notes that achieved \textit{Helpful Status} did so within 4h47min from note submission and 15h17min after the original post publication; see distribution in Figure~\ref{fig:fig_1_description}.B.\newline

At the account level, we observe significant variability in the proportion of posts with notes achieving \textit{Helpful Status}. 
Politically divisive accounts show fewer notes achieving Helpful Status, for example, those of Elon Musk or Kamala Harris’s campaign have 3\% or less than 0.1\% of their annotated posts receiving such notes, respectively.
Conversely, parody, entertainment, and conspiracy theory accounts exhibit a high proportion of notes achieving \textit{Helpful Status} (see lists of salient accounts and metrics in SM.S\ref{subsec:outcomes_account}). More generally, 9.7\% of annotated posts authored by accounts that held blue check verification prior to X's 2022 policy change and were deemed of ``public interest'' \cite{X_2016_blue} received notes with \textit{Helpful Status}, compared to 14.0\% for unverified accounts; computations restricted to non-deleted posts.\newline

\begin{figure*}
\centering
\includegraphics[width=0.85\textwidth]{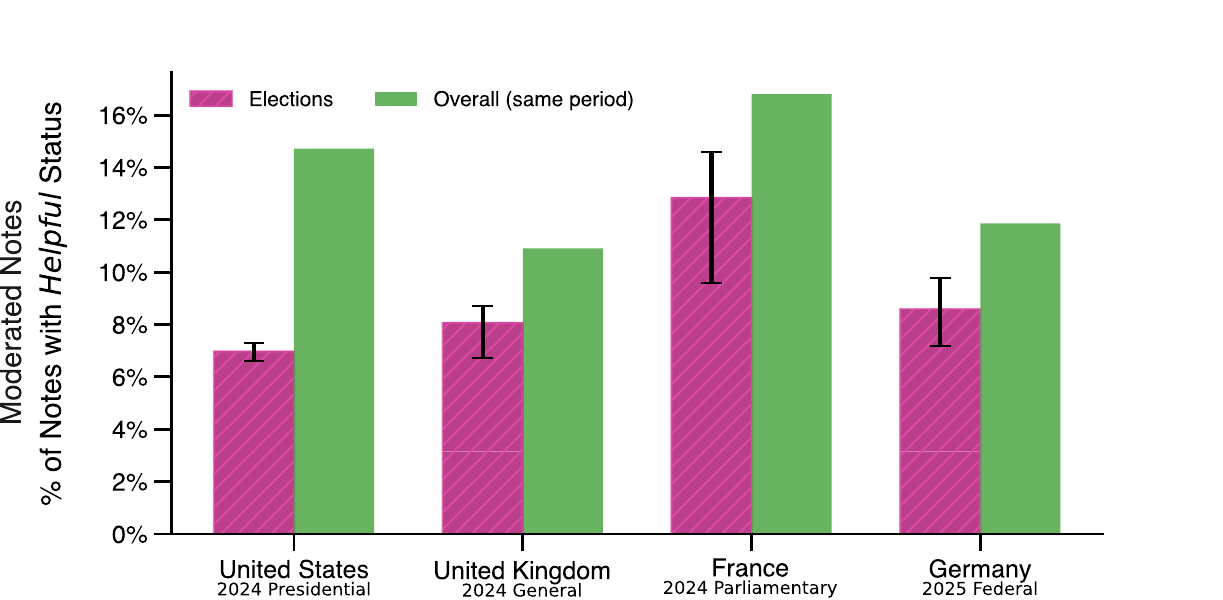}
\caption{Fraction of notes reaching \textit{Helpful Status} per country during elections. The analysis compares election-related notes with the overall corpus (in the same period, ranging from one month before to one week after the election) during the 2024 United States presidential, 2024 United Kingdom general, 2024 France legislative, and 2025 Germany federal elections. Error bars represent 95\% confidence intervals, estimated by bootstrapping over the keywords used to identify election-related notes.}
\label{fig:election_us_uk_fr_de}
\end{figure*}

Similarly, we observe that notes discussing politically divisive issues struggle to garner consensus and achieve \textit{Helpful Status} across each analyzed country, while those reporting scams and other misleading practices commonly do (see SM.S\ref{subsec:outcomes_topic} for country-specific lists of over- and under-represented terms in notes with \textit{Helpful Status}). 
To demonstrate this, we analyzed notes related to four elections taking place during the period of observation of our study.
Figure \ref{fig:election_us_uk_fr_de} shows a comparison between the fraction of notes reaching \textit{Helpful Status} for election-related notes and for overall notes in the same period, for the 2024 United States presidential (6.9\% vs 14.7\%), 2024 United Kingdom general (8.2\% vs 10.9\%), 2024 France legislative (12.8\% vs 16.8\%), and 2025 Germany federal elections (8.6\% vs 11.9\%). For each selected election, consensus is reached less frequently on election-related community notes compared to overall notes published during the same period. For comparison, 39\% of notes reporting scams, frauds, and impersonations reached \textit{Helpful Status}. See SM.S\ref{subsec:outcomes_topic} for a curation of keywords used to identify notes relating to elections and other topics.\newline

\begin{figure*}
\centering
\includegraphics[width=\textwidth]{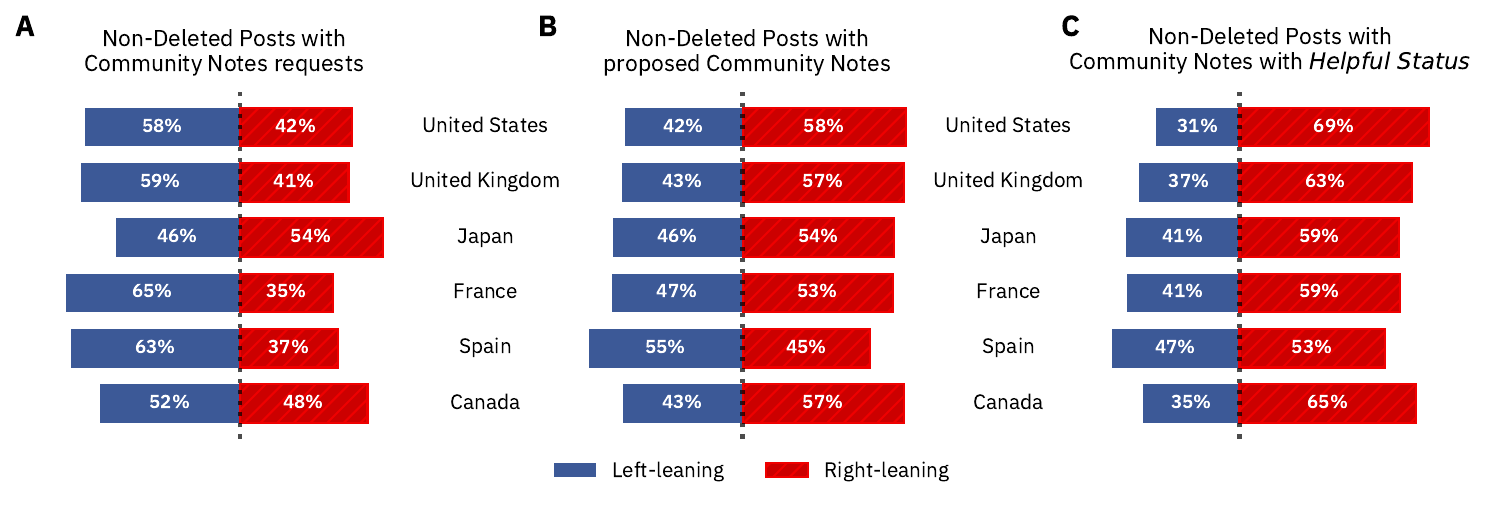}
\caption{Distribution of Left- and Right-leaning X accounts that authored posts for which (\textbf{A}) Community Notes were requested by users; (\textbf{B}) Community Notes were proposed by contributors; and (\textbf{C}) Community Notes reached \textit{Helpful Status}. Statistics computed over non-deleted posts authored by accounts with an inferred ideological leaning (see SM.S\ref{subsec:epo_dataset}).}
\label{fig:request_submit_helpful}
\end{figure*}

Finally, as reported in Figure \ref{fig:request_submit_helpful}.C, posts with notes that reached \textit{Helpful Status} are authored by accounts across the political spectrum in the 13 countries for which we computed ideological positions (see SM.S\ref{subsubsec:epo_x_cn} for Left-Right distributions across all countries). In the United States, 8.3\% of annotated posts authored by Left-leaning users receive notes that achieve \textit{Helpful Status}, compared to 12.9\% for Right-leaning users, as observed in \cite{Renault2025}. However, this disparity should not be interpreted as definitive evidence that content from Right-leaning users is flagged more frequently. Since our analysis is restricted to non-deleted posts, the observed difference could be explained, at least partially, by differential deletion patterns, whereby Left-leaning users may delete their posts with \textit{Helpful Status} notes more frequently than Right-leaning users (see discussion in SM.S\ref{paragraph:deletion_rate}).\newline

\subsection*{Latent ideological dimension in Community Notes} 

\begin{figure*}[!htbp]
\centering
\includegraphics[width=\textwidth]{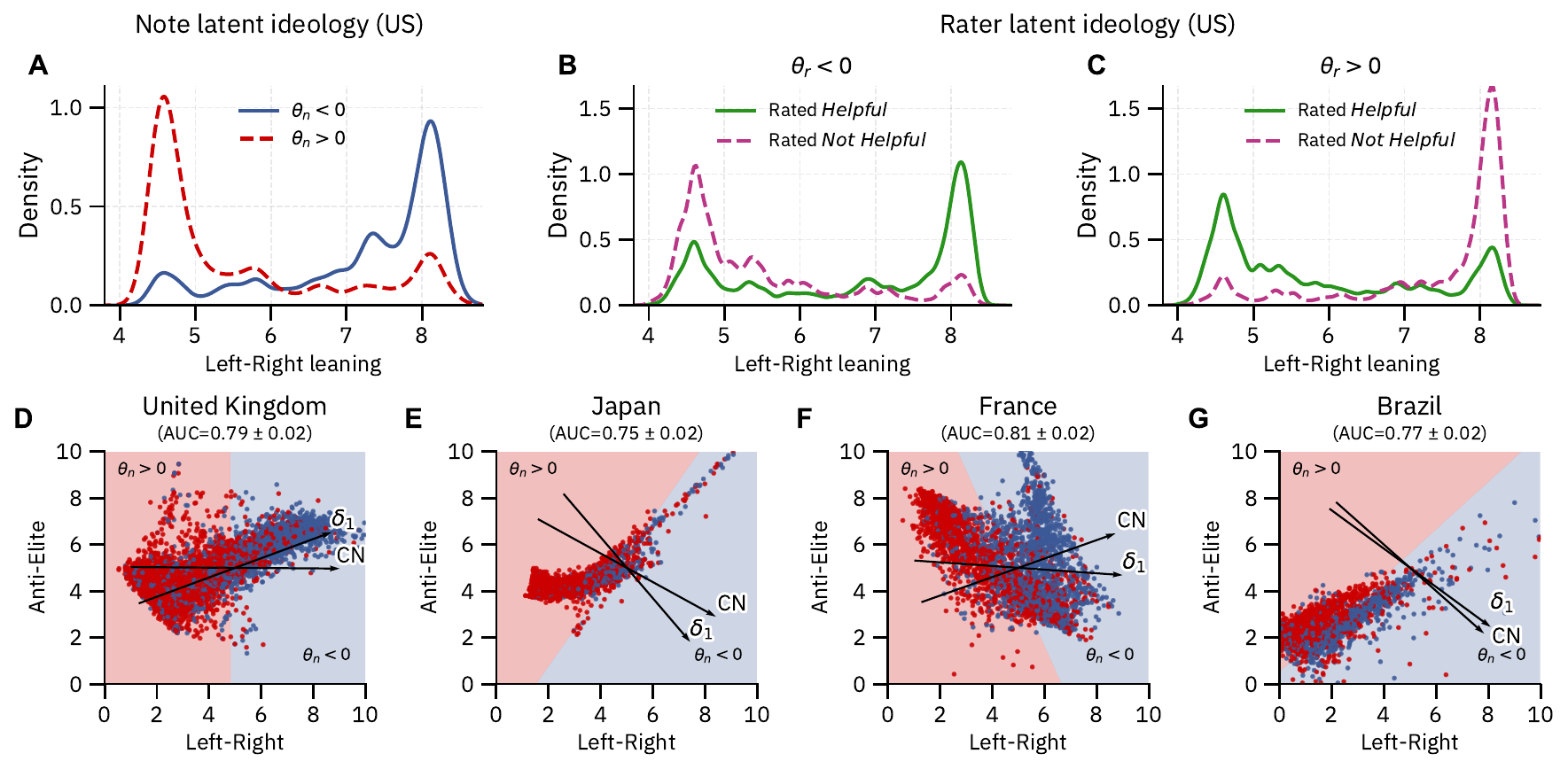}
\caption{(\textbf{A} to \textbf{C}) Distribution of Left-Right leaning of X accounts (\textbf{A}) segmented by the sign of the community note latent ideology $\theta_n$, (\textbf{B}) rated helpful/not-helpful by raters with negative latent ideology $\theta_r<0$, (\textbf{C}) rated helpful/not-helpful by raters with positive latent ideology $\theta_r>0$. (\textbf{D} to \textbf{G}) X accounts positioned in the (Left-Right; Anti-Elite) 2D plane, segmented by the majority sign of the latent ideology $\theta_n$ of their associated Community Notes (red for $\theta_n<0$ and blue for $\theta_n>0$). The direction that best separates notes by majority outcome, learned through logistic regression, is displayed as $CN$ along with the corresponding AUC score for sign prediction based on account positions in the 2D plane (standard deviation from 10-fold cross-validation). The direction structuring each country's political landscape $\delta_1$ is projected onto the plane.}
\label{fig:cn_epo}
\end{figure*}

In the United States, the single-dimensional latent ideological space learned by X's Community Notes system closely aligns with users' Left-Right political leanings, notes with $\theta_n>0$ appear under posts published by Left-leaning accounts, while notes with $\theta_n<0$ are associated with Right-leaning accounts (Fig. \ref{fig:cn_epo}.A). 95.0\% of notes associated with posts authored by Republican members of Congress have $\theta_n<0$, while 94.6\% of those associated with Democratic members of Congress have $\theta_n>0$ (see the average latent ideologies $\theta_n$ of notes associated with key political figures across multiple countries in SM.S\ref{subsec:op_1d}).
Contributors with $\theta_r<0$ tend to rate notes as \textit{Helpful} when they appear under Left-leaning accounts but rate notes as \textit{Not Helpful} when they appear under Right-leaning accounts (Fig. \ref{fig:cn_epo}.B), and the opposite pattern holds for raters with $\theta_r>0$ (Fig. \ref{fig:cn_epo}.C), aligning with the observations made by Allen et al. during the pilot program \cite{Allen2021}.
In the United States, the Left-Right political leaning of users predicts the sign of the latent ideologies $\theta_n$ of the notes associated with their posts, with an AUC of $0.822$. Additionally, news outlets cited in Community Notes align ideologically with contributor leanings; we found a Pearson correlation of $\rho=0.744$ ($p<10^{-4}$) between \textit{Ad Fontes} media bias ratings \cite{Otero2025} and the latent ideology $\theta_r$ of contributors citing these media outlets as sources in their notes (see SM.S\ref{paragraph:media_bias}).\newline

In the United States, expert fact-checking articles are used as sources by contributors across the ideological spectrum, though used predominantly ($60.3\%~[58.0\%,~62.8\%]$) by Left-leaning contributors ($\theta_r<0$). As displayed in Figure~\ref{fig:fig_1_description}.C, Community Notes incorporating expert fact-checking sources attain \textit{Helpful Status} at significantly higher rates ($21.3\%~[18.3\%,~24.1\%]$) than average, with this effect more pronounced for notes authored by contributors deemed Right-leaning according to the Community Notes model ($\theta_r>0$), $26.9\%~[23.3\%,~30.8\%]$ versus $17.6\%~[14.0\%,~20.7\%]$ for Left-leaning contributors (95\% CIs obtained by bootstrapping over notes and fact-checking organization; see SM.S\ref{paragraph:fact_checks}). \newline

The latent ideology computed by the Community Notes model ($\theta_n$ and $\theta_r$) shows varying degrees of alignment with the Left-Right and the Anti-Elite dimensions computed with ideology scaling and political surveys.
Figures \ref{fig:cn_epo}.D to \ref{fig:cn_epo}.G map X users with proposed Community Notes along the Left-Right and Anti-Elite dimensions in the United Kingdom, Japan, France, and Brazil. The latent dimension learned by X's system ($CN$ in Fig.~\ref{fig:cn_epo}) ---along which notes and raters are positioned--- aligns closely with each country's most structuring direction ($\delta_1$ in Fig.~\ref{fig:cn_epo}) in this Left-Right/Anti-Elite plane.
We quantify this alignment by measuring how well the position of a user along $\delta_1$ predicts the sign of the latent ideology  $\theta_n$ of notes associated with their posts. AUC values range from 0.850 in Poland to 0.729 in Israel, averaging $0.808\pm0.037$ across 13 countries. Complementarily, user positions in the Left-Right/Anti-Elite plane predict the sign of the latent ideology $\theta_n$ of notes with an AUC of $0.813 \pm 0.035$ across all 13 countries (see breakdown for all countries in our study in SM.S\ref{subsec:op_multi}).\newline

We find that the direction learned by X’s system to structure content veracity disputes (direction $CN$ in Fig.~\ref{fig:cn_epo}) aligns with the Left-Right dimension but cannot always be reduced to it, as anticipated by recent literature \cite{arceneaux2021some,ramaciotti2023geometry}.
Instead, it may present additional alignments with attitudes towards elites and institutions (and, to a lesser degree, possibly additional issue dimensions not considered in our study), as highlighted by the cases of Japan and Brazil (Fig. \ref{fig:cn_epo}.F \&{} \ref{fig:cn_epo}.G).

%% file: Contents/4_Conclusions.tex
\section*{Discussion} 

We combined X's Community Notes data, the algorithmic representation learned by this service, and ideological scaling for characterizing its operation in polarized settings. We showed, in 13 countries, that notes are requested, proposed, and displayed under posts authored by accounts throughout the political spectrum.
The majority of proposed Community Notes discuss politics, and predominantly rely on news outlet articles as sources, followed by X posts and Wikipedia articles.

X's Community Notes system positions notes and raters along a learned latent ideological axis that we show closely aligns with each country's most structuring political dimension. This latent axis corresponds to the Left-Right dimension and, with differences across countries, to an Anti-Elite dimension. These findings provide the first empirical support for the key hypothesis underlying X's Community Notes' design in a global comparative context.

Our results highlight shortcomings of this crowd-sourced system. While notes displayed on X (i.e., having reached \textit{Helpful status}) have been shown to align with professional fact-checking assessments \cite{birdwatch,Allen2024}, factual but polarizing content may still fail to be resolved as \textit{Helpful}. Only one-tenth of proposed notes ultimately appear on X, with even lower rates for notes associated with politically prominent figures, popular accounts, and divisive issues. More generally, and by design, the most polarizing content relating to the structuring dimension of national politics typically cannots be algorithmically resolved by Community Notes, regardless of their factuality. This limitation is especially relevant to events that activate such structural lines of division, such as national elections, pointing toward the potential complementarity between expert and crowd-sourced fact-checking, as originally intended by X \cite{WiredCommunityNotes}.

Furthermore, the fraction of polarizing content that will remain unresolved (regardless of factuality) is a design choice made by the platform, both explicitly by fixing the ideological diversity threshold for a note to reach \textit{Helpful Status} and implicitly by adjusting the regularization strength of the various latent parameters of the Community Notes model.

We acknowledge limitations in our study. In particular, our analyses rely on a snapshot of the data produced by the service until March 1, 2025, and disregard potential fine-grained temporal dynamics. Additionally, our segmentation of notes and raters per country is based on their involvement in the national political discussions by selecting those that follow MPs, combined with their use of the country's national language. Given the absence of additional metadata disclosed by the platform, we argue that this heuristic is reasonable and necessary for analyzing ideological dynamics in Community Notes.

Importantly, our post analysis was limited to content that remained available at the time of collection in March 2025, introducing a potential bias in the analysis, as posts having notes achieving \textit{Helpful Status} are more likely to be removed by their authors \cite{Renault2024}. Finally, X's Community Notes contributors are self-selected, and the platform provides no information to characterize this population demographically, let alone politically.

As platforms transition from expert fact-checking to crowd-sourced moderation, with YouTube \cite{Youtube2024}, TikTok \cite{TikTokFootnotes}, and Meta \cite{Meta_2025} deploying similar systems ---the latter explicitly using X's open source algorithm as the basis of their rating system--- this study provides a timely and comparative characterization of their operation in polarized settings across diverse countries.
A number of arguments exist for considering alternative to expert fact-checking, as these approaches accumulate challenges with regard to the required scale, reactivity, and efficiency \cite{Stray_Sneider_2025}, but also concerning perceived legitimacy and credibility \cite{Walker_2021}. 
Yet, our results show that X's Community Notes system, by relying on consensus rather than factuality, presents inherent limitations in addressing polarizing content. The use of crowd-sourced moderation as a substitute for expert fact-checking therefore warrants careful consideration and calls for risk assessment comparing the moderation outcomes of these approaches over contested polarizing content widely disseminated during events such as elections.

%% file: Contents/7_Methods.tex
\section*{Materials and Methods}

\subsection*{Data collection}

X publicly and continuously releases all Community Notes contributions enabling third-party scrutiny. As of March 1, 2025, 1.49 million X users have submitted 4.83 million requests for Community Notes across 1.91 million posts, 1.88 millions notes were submitted and rated over 135.17 million times. \newline

To comply with platform policies, user privacy rights, and deletion requests, X's Community Notes dataset only discloses post IDs associated with notes, without further information on post content or author. 
To enable contextual analysis beyond ratings, we collected post content and author information, associated to those notes. As of March 2025, 616\,104 X posts associated with 83.5\% of non-deleted Community Notes were collected. The remaining posts were unavailable at the time of collection, i.e., having been removed. Additionally, out of 966k posts with at least two user requests for a Community Note, we collected 804k posts (83.2\%); the remaining posts were unavailable at the time of collection in May 2025.
Importantly, it should be noted that such collection process (as opposed to live stream collection or access to data reserved to the platform) introduces an inherent observational bias. Indeed, some analyses are limited to posts that remain available at collection time, which then captures the tendency of users to remove their posts after the note reached \textit{Helpful Status}. This observational bias should be considered when interpreting the following results. Analyses that rely solely on the content of Community Notes, however, such as source analysis, are not marred by such limitation.\newline

\subsection*{Topic Analysis}

To classify community note topics, we employed a two-tiered approach. We first translated all posts to English using Facebook's \textit{M2M100} translation model \cite{m2m100}, ensuring consistent classification performance across original languages. For general topic classification, we used Antypas et al.'s topic taxonomy and model \cite{dimosthenis}, selected for its open-weight architectures and top-tier performance. For fine-grained political classification, we employed \textit{Manifestoberta} \cite{Burst:2023}, a language model based on \textit{XLM-RoBERTa} \cite{xlm_roberta}, trained on 1.6 million annotated statements from the Manifesto Corpus \cite{ManifestoProject} following the Manifesto Project coding scheme \cite{MANIFESTO}. We report validation results for both approaches in Supplementary Information. Finally, using a curated set of keywords listed in Supplementary Information, we extracted community notes related to elections held in the United States (2024 presidential election), United Kingdom (2024 general election), France (2024 legislative election), and Germany (2025 federal elections).

\subsection*{Latent Ideology}

We learn biases $\beta_0,~\beta_r,~\beta_n$ and latent ideologies $\theta_r,~\theta_n$ for notes and raters from the publicly released community notes and ratings. We replicate X's approach \cite{birdwatch, TwitterGithubMain} by implementing the matrix factorization described in Equation~\ref{eq:predictive_formula}; computational details and validation are provided in Supplementary Information.

To interpret the learned latent ideology dimension in terms of Left-Right and Anti-Elite positions, we use the ideological scaling of X users developed by Ramaciotti et al. \cite{ramaciotti2022inferring}. This approach extends Barberá's \cite{barbera2015birds} methodology to identify multidimensional ideological positions of users using political survey data across countries. We collected X users following Members of Parliament (MPs) across 23 countries: \textit{France, Germany, United States, United Kingdom, Spain, Brazil, Australia, Italy, Israel, Argentina, Denmark, Sweden, Canada, Poland, the Netherlands, Switzerland, South Africa, Nigeria, Finland, Mexico, New Zealand, Japan, and Belgium.} We retained only users following at least three MPs and followed by at least 25 accounts, following Barberá's protocol \cite{barbera2015birds, Barber2015Posting, barbera2015understanding}. We then estimate ideal positions for users and MPs using Correspondence Analysis \cite{Greenacre2007} on the follower-MP adjacency matrix. Validation and interpretation are reported in Supplementary Information.

%% file: Contents/5_Additional_Information.tex
\section*{Acknowledgments}

The authors thank Jean-Philippe Cointet and Armin Pournaki for their comments, and Hiroki Yamashita, Jimena Royo-Letelier and Armin Pournaki for their assistant in data collection and treatment.

\subsection*{Funding}

This work has been partially funded by the ``European Polarisation Observatory'' (EPO) of CIVICA Research (co-)funded by EU’s Horizon 2020 programme under grant agreement No 101017201, by European Union Horizon program project ``Social Media for Democracy'' under grant agreement No 101094752 (\url{www.some4dem.eu}), by the \textit{Very Large Research Infrastructure} (TGIR) Huma-Num of CNRS, Aix-Marseille Université and Campus Condorcet, and by Project Liberty Institute project ``AI-Political Machines'' (AIPM). P.B. acknowledges support from the Jean-Pierre Aguilar fellowship from the CFM Foundation for Research.

\subsection*{Author contributions}

P.B. and P.R. collected the data, designed the research, performed the analyses, and wrote the paper.

\subsection*{Competing interests}

There are no competing interests to declare.

\subsection*{Ethics and compliance declarations}

Our study did not involve experimentation with human subjects.
All data used used in our study is publicly available through X's API and through the Community Notes program at \url{https://x.com/i/communitynotes/download-data}.

The processing of political attributes data was declared on 19 March 2020 and 15 July 2021 at the registry of data processing at the \textit{Fondation Nationale de Sciences Politiques} (Sciences Po) in accordance with General Data Protection Regulation 2016/679 (GDPR) and X policy. For further details and the respective legal notice, please visit the web page of project EPO: \url{medialab.sciencespo.fr/en/activities/epo/}.
The processing of the data has been approved by the Research Ethics Committee of the Paris Institute of Political Studies in its decision nº2023-038.

%% file: Contents/6_SI.tex

\section{The Community Notes service}
\label{sec:presentation_cn}

The declared aim of X's Community Notes is to improve information quality by ``empowering people on X to collaboratively add helpful notes to posts that might be misleading" \cite{CommunityNotes}. The system combines collaborative annotation with an algorithmic selection process that determines which notes are displayed publicly alongside the posts they moderate.\newline

\begin{figure}[h]
\centering
\includegraphics[width=\textwidth]{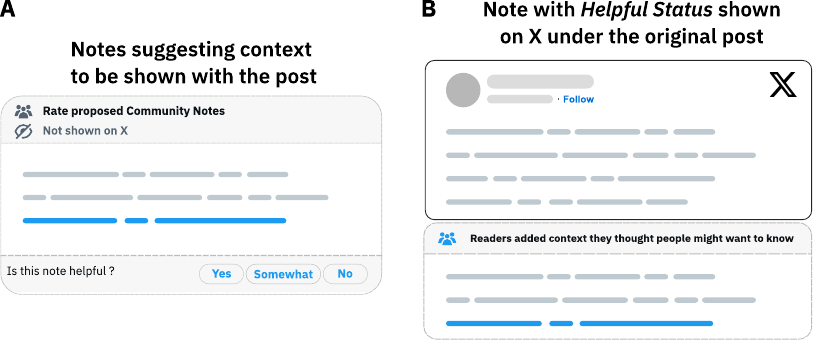}
\caption{\textbf{(A)} An example of a Community Note flagging a post as ``misinformed or potentially misleading", prompting contributors to rate its ``helpfulness". \textbf{(B)} An example of a Community Note that has reached \textit{Helpful Status} and is displayed beneath the original post on X.}
\label{fig:misleading_post_notes}
\end{figure}

X users, having signed up as contributors, can write notes explaining why a post may be ``misinformed or potentially misleading". Other contributors then rate these proposed notes as \textit{Helpful}, \textit{Somewhat Helpful}, \textit{Not Helpful}; as displayed in Figure \ref{fig:misleading_post_notes}.A. If sufficient contributors from ``diverse viewpoints" rate a note as \textit{Helpful}, X assigns it the status of being \textit{Helpful} (denoted as \textit{Helpful Status} through this article to differentiate from a rating given by a user to a note) and it becomes visible to all users on X; as displayed in Figure \ref{fig:misleading_post_notes}.B. Conversely, if enough contributors from different perspectives rate a note as \textit{Not Helpful}, the note remains hidden from public view. Contributors who repeatedly author notes assigned \textit{Not Helpful Status} may lose their ability to contribute further notes. Notes that have not yet received sufficient ratings from contributors with diverse perspectives remain invisible to users outside the Community Notes program. This ensures that only well-vetted content reaches the broader X audience.\newline

\begin{figure}[h]
\centering
\includegraphics[width=\textwidth]{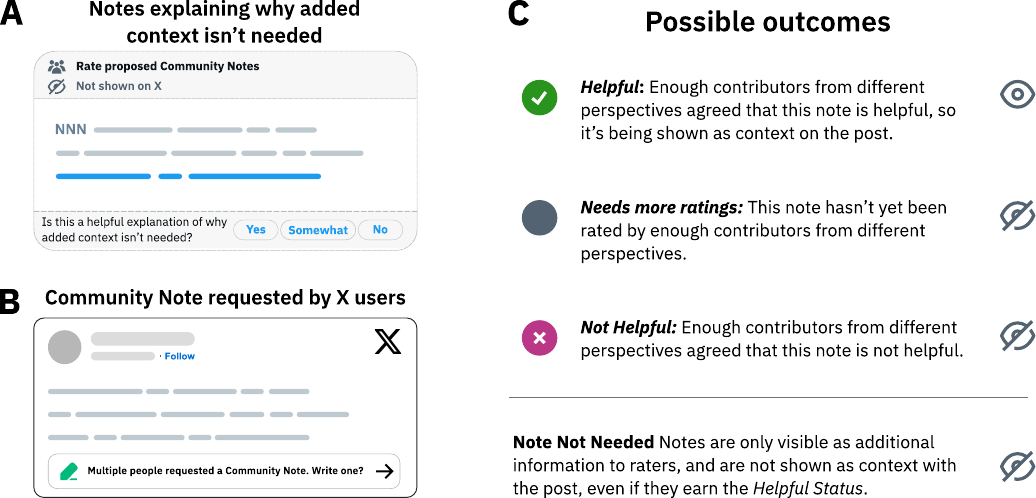}
\caption{\textbf{(A)} An example of a Community Note arguing that the original post is not misinformed or potentially misleading" and that no notes are needed. \textbf{(B)} Signal displayed under a post for which regular X users requested a community note to be proposed. \textbf{(C)} Taxonomy of outcomes for Community Notes; only notes arguing that the original post is misinformed or potentially misleading" and that reach \textit{Helpful Status} are displayed on X.}
\label{fig:nnn_requests_outcomes}
\end{figure}

The system also includes a type of note called ``Note Not Needed", which allow contributors to argue against the necessity of a proposed note; as displayed in Figure \ref{fig:nnn_requests_outcomes}.A. These notes are never displayed publicly on X regardless of their status, serving solely as internal discussions among contributors about whether a note addressing alleged misinformation is warranted.\newline

Since June 2024, X has expanded participation by allowing everyone on the platform with a verified phone number (and not only users enrolled in the Community Notes program) to request Community Notes on posts they believe would benefit from additional context \cite{CommunityNotesRequests}. When sufficient requests accumulate, ``Top Writers" (contributors recognized for writing a significant number of notes that are found \textit{Helpful} by others) receive alerts and can choose to propose notes; as displayed in Figure \ref{fig:nnn_requests_outcomes}.B. This mechanism enables broader user participation while letting contributors know where notes might be most helpful.


\section{Data collection}
\label{sec:data_collection}


\subsection{Community Notes}
\label{subsec:data_collection_cn}

\subsubsection{Notes and ratings}
\begin{figure}[h]
\centering
\includegraphics[width=\textwidth]{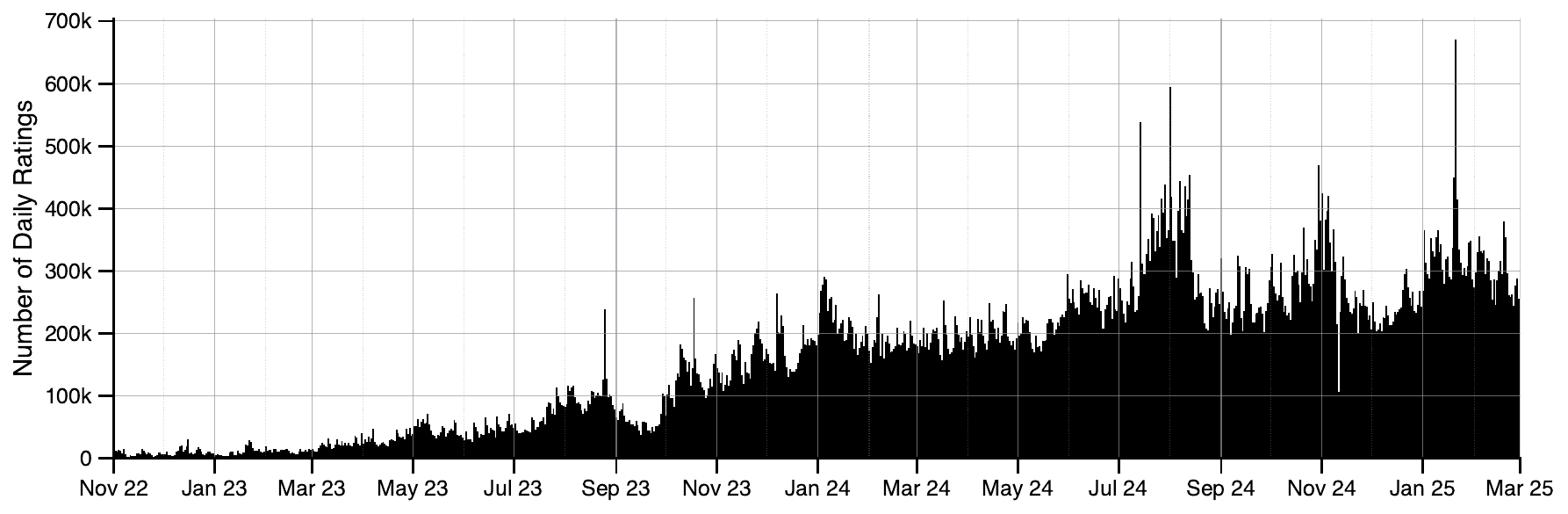}
\caption{Number of ratings submitted daily on X's Community Notes between November 2022 and March 2025.}
\label{fig:nb_rating_chronology}
\end{figure}

X publicly and continuously releases all Community Notes contributions enabling third-party scrutiny \cite{CN_Downloading}.
We consider in this study 1.88 millions notes submitted prior to March 1 2025, and their associated 135.17 million ratings. We display in Figure~\ref{fig:nb_rating_chronology}, the number of ratings submitted daily between November 2022 ---when the program was renamed Community Notes--- to March 1, 2025. The number of ratings is observed to increase within this period, as the program was deployed in more countries \cite{X_europe}.\newline

\begin{figure}[h]
\centering
\includegraphics[width=\textwidth]{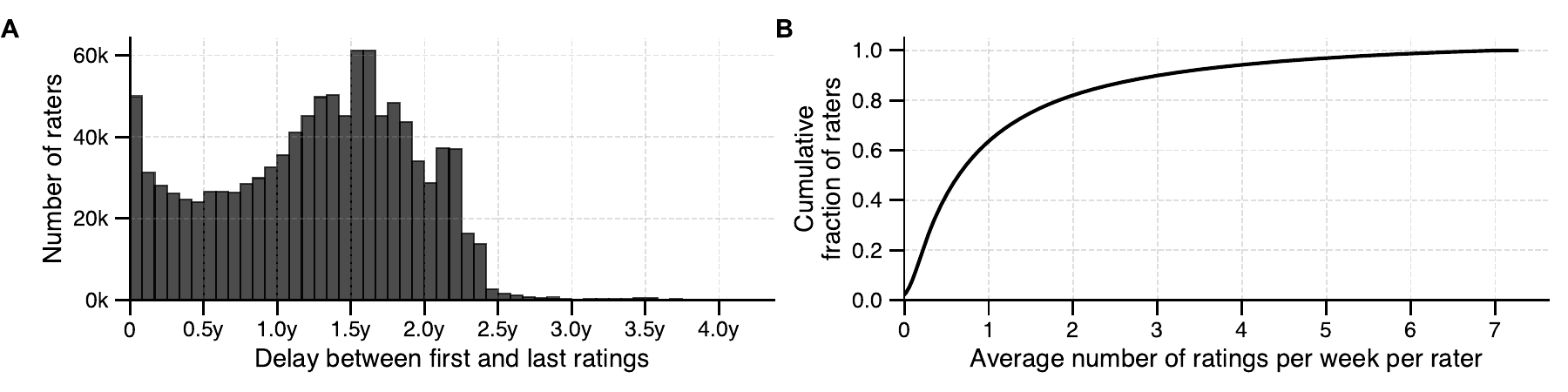}
\caption{Distribution of time spans between first and last ratings per rater (left), and of average weekly rating frequency (right); restricted to raters with multiple submissions}
\label{fig:distrib_delay_actitvity}
\end{figure}

Figure \ref{fig:distrib_delay_actitvity} shows the distribution of time spans between first and last ratings per rater, along with average weekly rating frequency; restricted to raters with multiple submissions. Results demonstrate sustained engagement: 82.6\% of raters maintained activity across at least six months, while 40.2\% averaged more than one rating per week.

\begin{figure}[h]
\centering
\includegraphics[width=\textwidth]{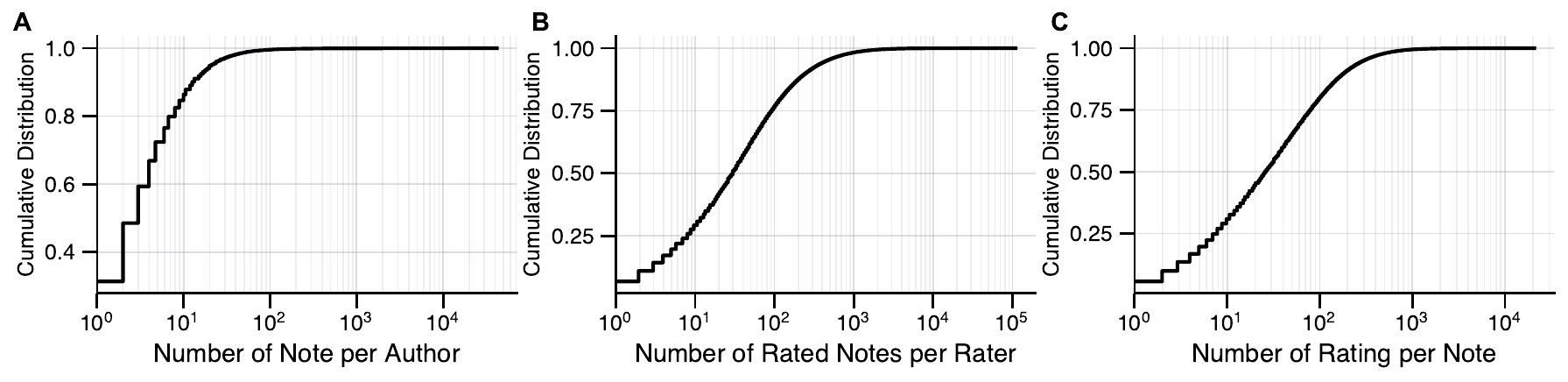}
\caption{Cumulative distribution of notes per author (\textbf{A}), rated notes per rater (\textbf{B}) and ratings per note (\textbf{C}).}
\label{fig:distrib_raters}
\end{figure}

Up to March 2025, X's Community Notes were rated by 1,150,257 raters, with a median of 29 (mean: 117.5) notes rated per rater (where a rater is defined as a contributor who rated at least one Community Note). Overall, 28.6\% of raters evaluated fewer than 10 notes. The cumulative distribution of ratings per rater is displayed in Figure~\ref{fig:distrib_raters}.B.\newline

At the note level, Community Notes received a median of 26 ratings (mean: 77.2), with 16.9\% receiving fewer than 5 ratings, we display the cumulative distribution of rating per note on Figure~\ref{fig:distrib_raters}.C.\newline

To unlock note writing privileges, a contributor must achieve a ``rating impact" of at least 5, meaning they must rate 5 notes as \textit{Helpful} (or \textit{Not Helpful}) that eventually reach \textit{Helpful Status} (or \textit{Not Helpful Status}). As of March 2025, 36.0\% of users enrolled in the Community Notes program had note writing privileges; corresponding to 405k contributors. Among those, 253,002 authored at least one Community Notes, with a median of 3 (mean: 7.4) notes per author (where an author is defined as a contributor who submitted at least one note). Figure~\ref{fig:distrib_raters}.A shows the cumulative distribution of notes per author. The distribution exhibits high inequality, with the top 10\% of authors writing 57.6\% of all Community Notes.\newline

Following X's pre-processing procedure \cite{birdwatch, TwitterGithubMain}, we retained raters who rated at least 10 notes and notes with at least 5 ratings; resulting in a final dataset of 1.45 million notes rated 133.2 million times by 821,251 contributors. We will consider this set of notes and ratings when training the Community Notes algorithm in Section \ref{sec:training_cn}.

\subsubsection{Note requests}

\begin{figure}[h]
\centering
\includegraphics[width=\textwidth]{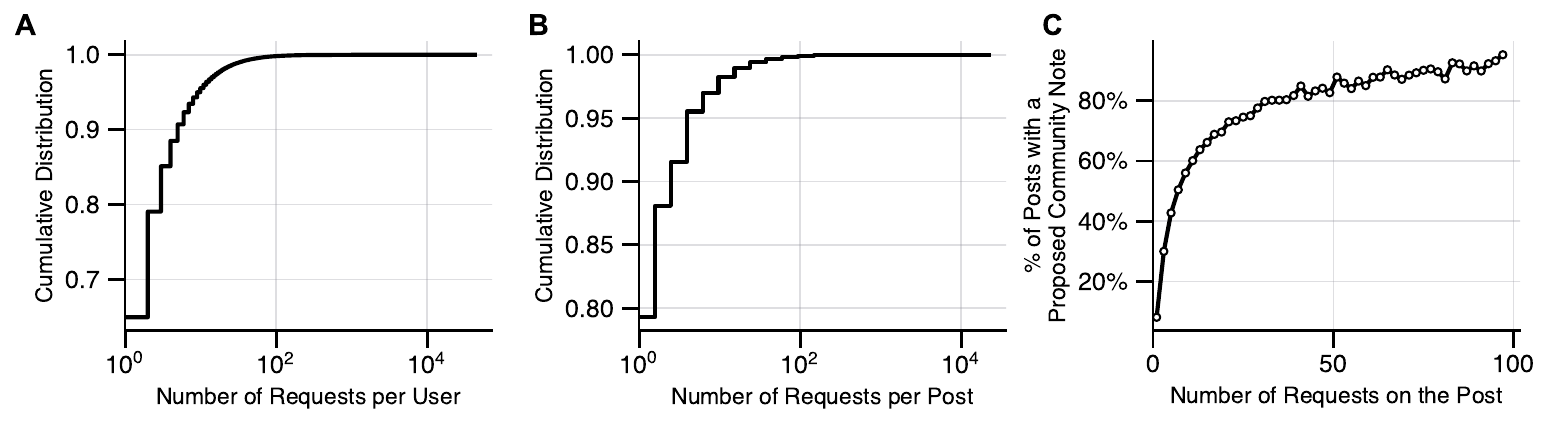}
\caption{Cumulative distribution of notes requested per author (\textbf{A}), and number of requests per post (\textbf{B}). Fraction of posts receiving a Community Note as a function of the number of requests they received; computation restricted to posts having received requests; truncated tail (above 50 requests) represents 0.4\% of requested posts (\textbf{C}).}
\label{fig:distrib_requests}
\end{figure}

As of March 1, 2025, 1.49 million X users have submitted 4.83 million requests for Community Notes across 1.91 million posts. Among users who have requested notes at least once, 65.0\% submit only a single request; see cumulative distribution in Figure \ref{fig:distrib_requests}.A. Similarly, 79.3\% of flagged posts receive only one request for a note, while 93.5\% receive fewer than 5 requests; see cumulative distribution in Figure \ref{fig:distrib_requests}.B. We display in Figure \ref{fig:distrib_requests}.C the fraction of posts receiving a proposed Community Notes in function of the number of requests they received; computation restricted to posts having receive request(s).

\subsection{Posts associated with notes}
\label{subsec:data_collection_post}

To comply with platform policies, user privacy rights, and deletion requests, X's Community Notes dataset only discloses post IDs associated with notes, without further information on post content or author. To enable contextual analysis beyond ratings, we collected post content and author information, associated to those notes. As of March 2025, 616,104 X posts associated with 83.5\% of non-deleted Community Notes were collected. The remaining posts were unavailable at the time of collection, i.e., having been removed.\newline

Additionally, out of 966k posts with at least two user requests for a Community Note, we collected 804k posts (83.2\%); the remaining posts were unavailable at the time of collection in May 2025.\newline

Importantly, it should be noted that such collection process (as opposed to live stream collection or access to data reserved to the platform) introduces an inherent observational bias. Indeed, some analyses are limited to posts that remain available at collection time, which then captures the tendency of users to remove their posts after the note reached \textit{Helpful Status}. Specifically, we observe that 33.4\% of posts with notes having reached \textit{Helpful Status} were deleted by the time of collection, compared to only 15.1\% for posts associated with note with \textit{Needs More Ratings Status} or \textit{Not Helpful Status}. This observational bias should be considered when interpreting the following results. Analyses that rely solely on the content of Community Notes, however, such as source analysis, are not marred by such limitation. \newline

\begin{figure}[h]
\centering
\includegraphics[width=\textwidth]{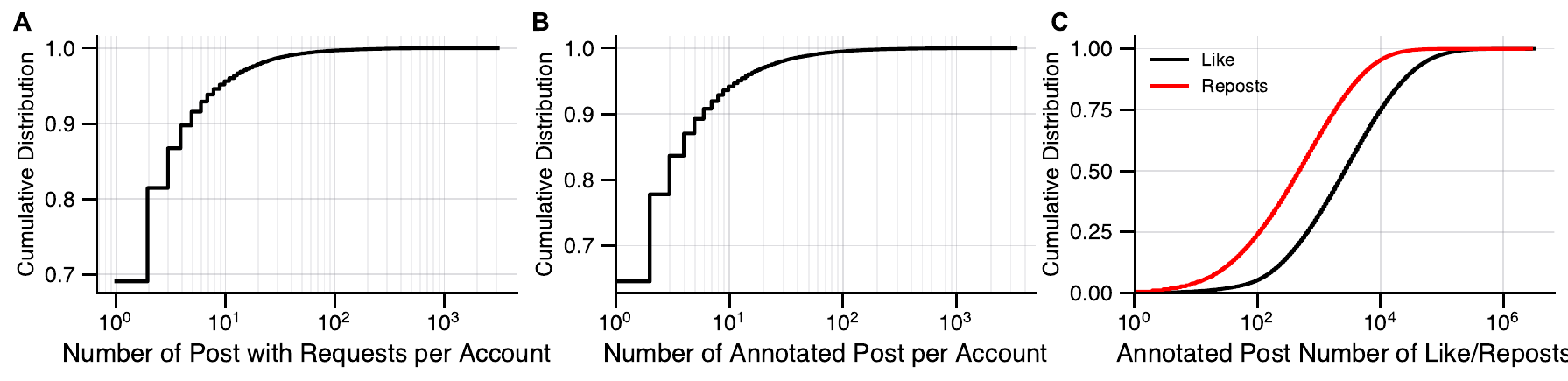}
\caption{Cumulative distribution of number, per X account, of posts with requests for Community Notes (\textbf{A}) and with a submitted Community Notes  (\textbf{B}); conditional on having at least one request/annotated post. (\textbf{C}) Cumulative distribution of annotated posts' number of likes and reposts.}
\label{fig:distrib_account_engagement}
\end{figure}

\begin{figure}[h]
     \centering
     \begin{subfigure}[b]{0.3\textwidth}
         \centering
         \vfill
         \includegraphics[width=\textwidth]{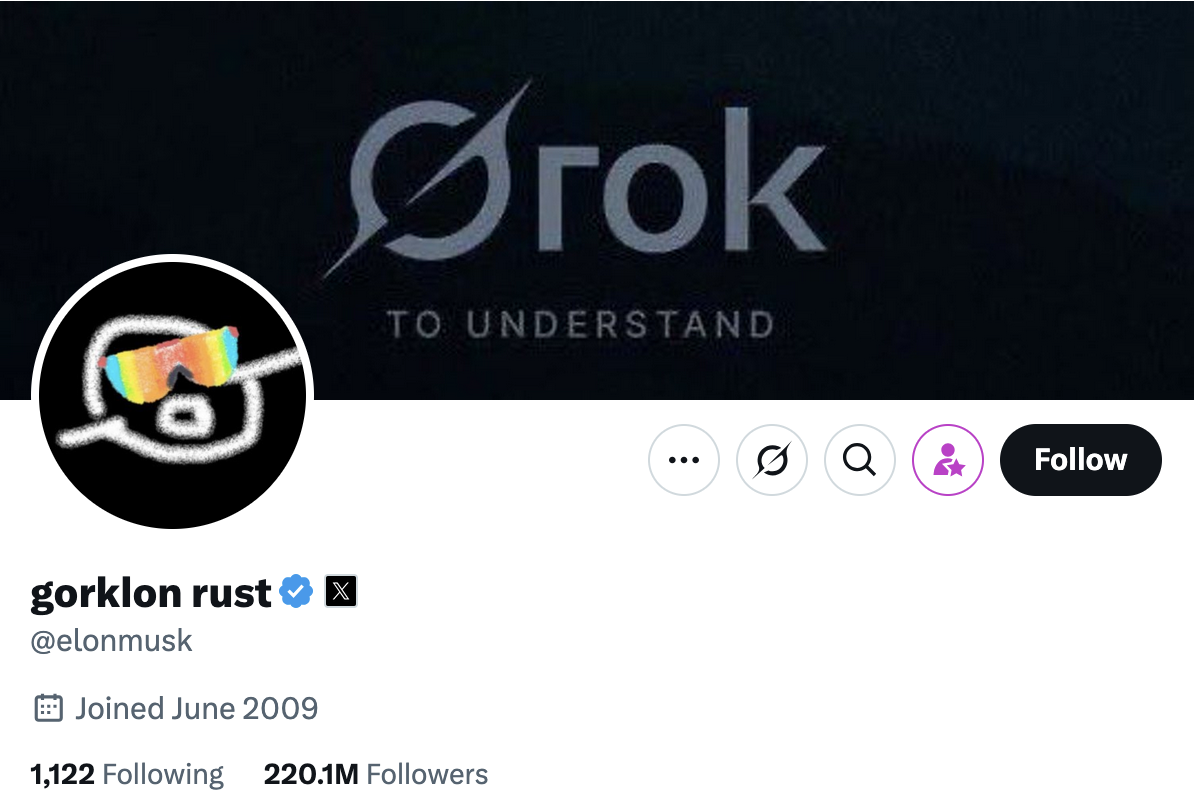}
         \caption{Profile of X account: \texttt{@elonmusk}}
     \end{subfigure}
     \hfill
     \begin{subfigure}[b]{0.3\textwidth}
         \centering
         \vfill
         \includegraphics[width=\textwidth]{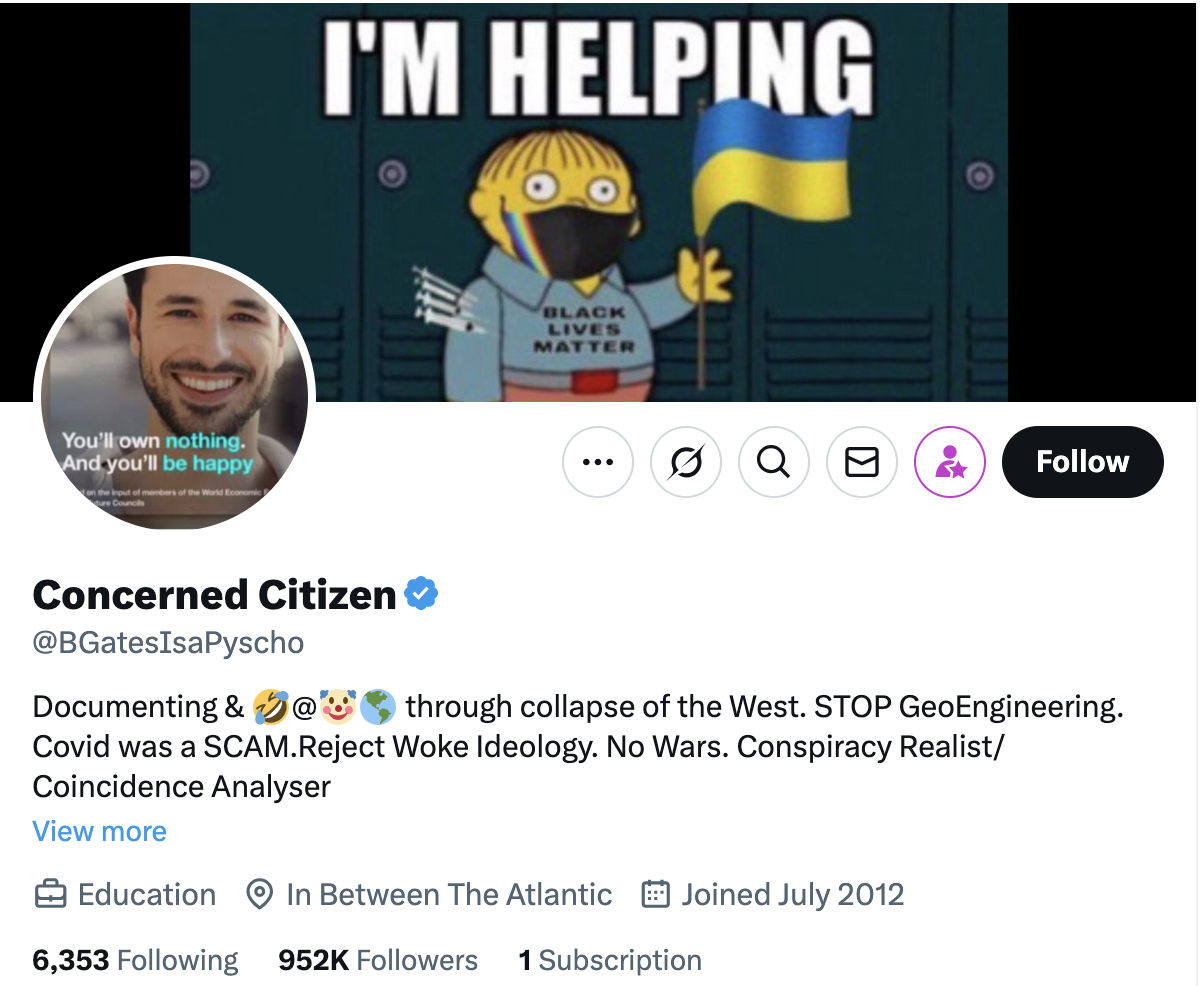}
         \caption{Profile of X account: \texttt{@BGatesIsaPyscho}}
     \end{subfigure}
     \hfill
      \begin{subfigure}[b]{0.3\textwidth}
         \centering
         \vfill
         \includegraphics[width=\textwidth]{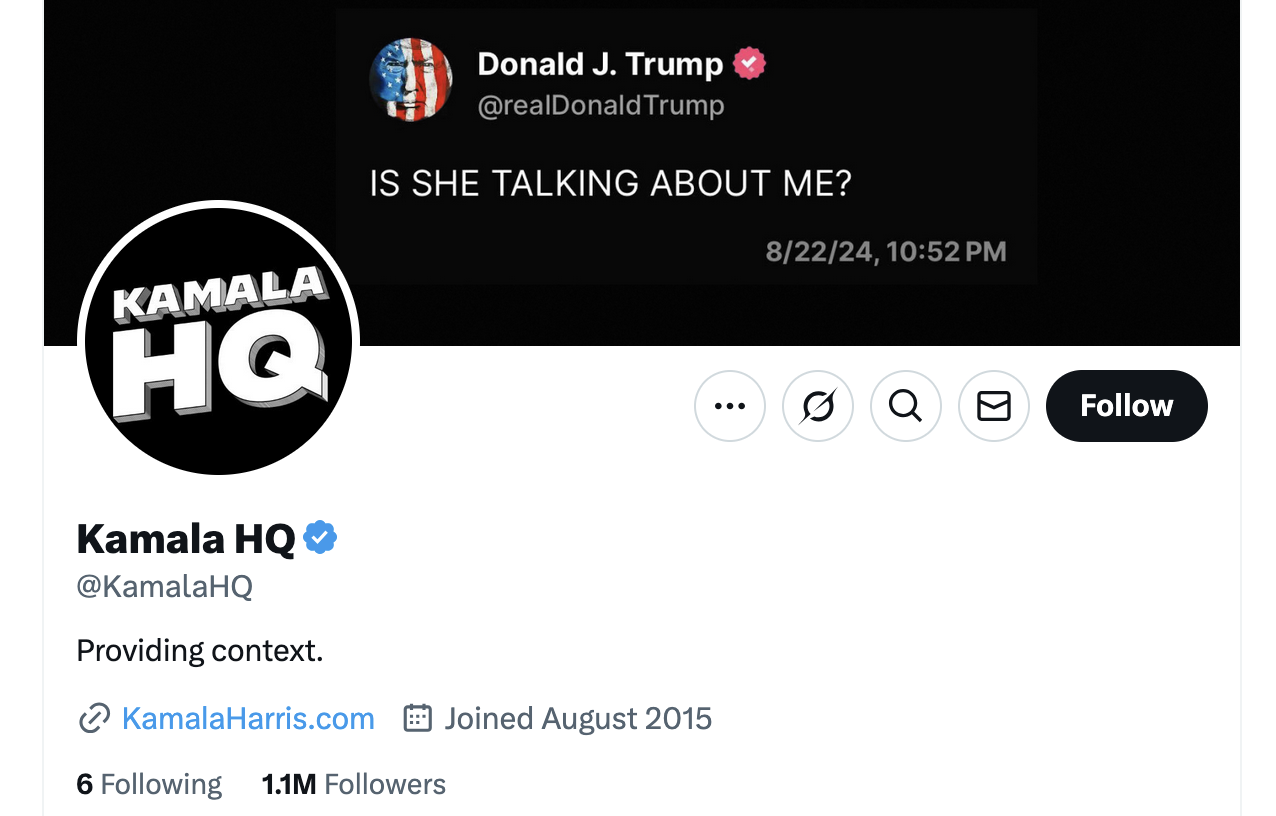}
         \caption{Profile of X account: \texttt{@KamalaHQ}}
     \end{subfigure}
    \caption{Profiles of the X accounts having received the highest number of Community Notes, and number of requested Community Notes as of March 2025.}
    \label{fig:elon_BGatesIsaPyscho}
\end{figure}

The 616,943 collected posts associated with a proposed Community Notes (hereinafter refer to as ``annotated posts") were authored by 140,913 X accounts, averaging 7.8 notes per author (with median equal to 2); among accounts with at least one annotated post. Elon Musk's account (\texttt{@elonmusk}) had the highest number of notes, with over nineteen thousand notes proposed by the community under more than 3,300 posts, followed by conspiracy theorist account \texttt{@BGatesIsaPyscho} with over four thousand notes proposed by the community under more than 2,700 posts; we display these account profiles in Figure \ref{fig:elon_BGatesIsaPyscho}. Similarly, Elon Musk's account (\texttt{@elonmusk}) had the highest number of posts (at least 3,087) with Community Notes being requested, followed by Kamala Harris's campaign account (\texttt{@KamalaHQ}) with at least 1,203 posts with requests. We display in Figure \ref{fig:distrib_account_engagement}.A the distribution per account of posts with requests for Community Notes. \newline

Figure~\ref{fig:distrib_account_engagement} shows the distribution of number of annotated post per X account, along with the distributions of number of likes and reposts for annotated posts. The collected posts averaged 13,693 likes and 2,224 reposts (medians of 2,705 and 480 respectively). We found a positive Spearman correlation, $\rho = 0.444, p < 10^{-4}$, between a post's number of likes and the number of ratings of notes associated to it; and similarly with the number of reposts $\rho = 0.437, p < 10^{-4}$.


\section{Characterization of notes}


\subsection{Misleading nature of posts}
\label{subsec:misleading_tag}

Upon submitting Community Notes, contributors respond to ``Why do you believe this post may be misleading?". For each (non-exclusive) proposed reason, we report the corresponding fraction of notes and the proportion of them reaching \textit{Helpful Status}:

\begin{itemize}
    \item ``It is a misrepresentation or missing important context": 61.7\% of notes; 11.5\% reached \textit{Helpful Status}
    \item ``It contains a factual error": 56.4\% of notes; 12.1\% reached \textit{Helpful Status}
    \item ``It presents an unverified claim as a fact": 35.5\% of notes; 10.7\% reached \textit{Helpful Status}
    \item ``It contains outdated information that may be misleading": 20.8\% of notes; 12.0\% reached \textit{Helpful Status}
    \item ``Other": 13.4\% of notes; 12.7\% reached \textit{Helpful Status}
    \item ``It contains a digitally altered photo or video": 9.2\% of notes; 17.4\% reached \textit{Helpful Status}
    \item ``It is a joke or satire that might be misinterpreted as a fact": 7.2\% of notes; 13.4\% reached \textit{Helpful Status}
\end{itemize}

Statistics were computed across the whole set of Community Notes arguing the original post is ``misinformed or potentially misleading".


\subsection{Web sources used in notes}
\label{subsec:sources}

\begin{figure}[h]
\centering
\includegraphics[width=\textwidth]{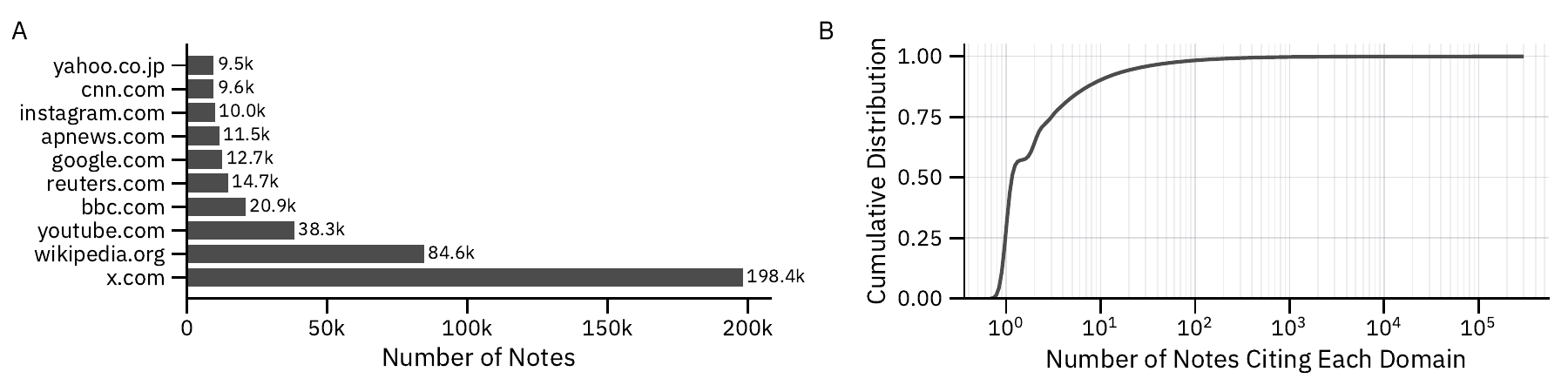}
\caption{(\textbf{A}) Cumulative distribution of number of notes referencing each domains. (\textbf{A}) Number of notes referencing the 10 most frequently cited domains. Statistics computed on the set of Community Notes processed by X's Community Notes algorithms, following their pre-processing procedure \cite{birdwatch, TwitterGithubMain}.}
\label{fig:distrib_domain}
\end{figure}

Community Notes contributors are encouraged to use sources to justify their proposed notes \cite{X_2024_tips}. To characterize their usage, we extract all URLs used in Community Notes qualifying the original post as ``misinformed or potentially misleading"; resolving X's link shortener (\texttt{t.co}) and Google ``amp" redirections. Overall, 96.2\% of Community Notes contain URLs pointing to supporting sources that aim to contextualize the original post. We identified 76.0 thousand unique domains; Figure~\ref{fig:distrib_domain}.B presents the distribution of Community Notes referencing them.\newline

As displayed in Figure~\ref{fig:distrib_domain}.A, it emerges that \texttt{x.com} (aggregated with \texttt{twitter.com}) is the domain most frequently referred to, followed by \texttt{wikipedia.org}, \texttt{youtube.com} (aggregated with \texttt{youtu.be}) and \texttt{bbc.com} (aggregated with \texttt{bbc.co.uk}) and \texttt{reuters.com}.\newline

References to \texttt{x.com} can be further disaggregated into user-generated content (posts, profiles, search results) and platform documentation. Specifically, 18.3\% of Community Notes referencing to \texttt{x.com} or \texttt{twitter.com} refer to one of the following ressources: \texttt{communitynotes.x.com}, \texttt{help.x.com}, \texttt{blog.x.com}, \texttt{business.x.com}, and \texttt{x.com/en/tos} (combined with \texttt{twitter.com} variations).

\subsubsection{News outlets}

We manually inspected the first 500 web domains most cited in notes to identify those pointing towards media outlets, excluding website publishing exclusively fact-checks, and report on the first 139 most cited among them:\newline

\noindent{}\textit{reuters.com, apnews.com, news.yahoo.com, news.yahoo.co.jp, bbc.co.uk, nhk.or.jp, bbc.com, theguardian.com, cnn.com, nytimes.com, impress.co.jp, forbes.com, globo.com, vice.com, usatoday.com, washingtonpost.com, nbcnews.com, ft.com, dailymail.co.uk, npr.org, news.yahoo.com, timesofisrael.com, newsweek.com, uol.com.br, cbsnews.com, afp.com, asahi.com, aljazeera.com, independent.co.uk, abcnews.go.com, infobae.com, lemonde.fr, politico.com, telegraph.co.uk, foxnews.com, wsj.com, cnbc.com, pbs.org, cbc.ca, france24.com, jpost.com, time.com, francetvinfo.fr, thehill.com, businessinsider.com, elpais.com, cnnbrasil.com.br, liberation.fr, indiatimes.com, euronews.com, elmundo.es, haaretz.com, politico.eu, estadao.com.br, leparisien.fr, bloomberg.com, lefigaro.fr, bfmtv.com, ctvnews.ca, dw.com, eldiario.es, gazetadopovo.com.br, hindustantimes.com, indiatoday.in, jiji.com, lanacion.com.ar, latimes.com, mainichi.jp, metropoles.com, ndtv.com, nikkei.com, nypost.com, spiegel.de, poder360.com.br, thehindu.com, yomiuri.co.jp, ynet.co.il, tagesschau.de, sankei.com, 20minutes.fr, 20minutos.es, aa.com.tr, abc.es, bild.de, br.de, cnn.co.jp, economist.com, elconfidencial.com, haaretz.co.il, i24news.tv, irishtimes.com, lepoint.fr, kyivindependent.com, lavanguardia.com, nationalpost.com, ndr.de, theatlantic.com, theverge.com, zeit.de, wired.com, express.co.uk, vox.com, rtve.es, rte.ie, rp.pl, publico.es, news.sky.com, itmedia.co.jp, axios.com, aap.com.au, veja.abril.com.br, economictimes.com, indianexpress.com, standard.co.uk, elespanol.com, clarin.com, tokyo-np.co.jp, ouest-france.fr, welt.de, voanews.com, washingtonexaminer.com, tribune.com.pk, sueddeutsche.de, faz.net, thetimes.co.uk, dawn.com, tagesspiegel.de, news.com.au, pagina12.com.ar, elperiodico.com, israelhayom.co.il, tf1info.fr, globalnews.ca, radiofrance.fr, timesnownews.com, tvn24.pl, lasexta.com, msnbc.com, usnews.com}\newline

These domains were cited in 220k unique notes (after pre-processing), representing 23.2\% of the notes.\newline

To test the exhaustiveness of this list of 139 media outlets we obtained the list of media outlets active around the world, curated by the \textit{BBC Media Guides}. This list includes 2061 web domains of news outlets, covering 154 countries. Combined with our manually curated list, we establish a set of 2134 unique news domains, of which 979 are used at least once in Community Notes as sources. The addition of this second list of medias did not significantly altered the percentage of notes that used news media articles as sources, when considered the union of both lists.
Overall, domains from this extended list of news media outlets are used in 246k Community Notes (after pre-processing), constituting the most common reference type (25.9\%) ahead of user-content on X (17.4\%) and Wikipedia articles (8.9\%), demonstrating the prevalence of editorial journalism in X's crowd-sourced fact-checking system.

\subsubsection{Expert fact-checking}
\label{subsubsec:fact_checks}

To contrast Community Notes with expert fact-checking, we identified notes citing expert fact-checking sources. To this end, we compiled a list of fact-checking sources by combining Duke University's Reporters' Lab directory \cite{ReportersLabList}, which catalogs over 400 geographically diverse fact-checking websites, with news agencies and signatories of the International Fact-Checking Network \cite{IFCNSignatories}.\newline

To avoid including URLs of general news coverage sources, we manually curated the most frequently used domains from this compiled list. Specifically, we segmented websites into those that published exclusively fact-checks or maintained dedicated fact-checking sections (e.g., reuters.com/fact-check), and those publishing both general news coverage and fact-checking articles. For news organizations that publish fact-checks alongside general coverage without dedicated fact-checking sections in their URL structure, we required the presence of the regular expression ``fact.*check'' in URLs; including translations. This filtering approach ensures that, e.g., \url{cnn.com} articles are not considered fact-checks except for those matching this pattern, such as, \url{edition.cnn.com/2025/04/23/politics/price-of-eggs-gas-trump-fact-check}. 

From this curated list of over 500 geographically diverse fact-checking organizations, we identified 46,869 Community Notes referencing them as sources within the entire set of Community Notes published prior to March 1, 2025 (including ``Note Not Needed'' notes). The most frequently cited fact-checks are published by \texttt{snopes.com}, \texttt{reuters.com}, \texttt{politifact.com}, \texttt{apnews.com}, and \texttt{usatoday.com}.

After following X's pre-processing guidelines \cite{birdwatch, TwitterGithubMain} and excluding ``Note Not Needed" notes, Community Notes citing fact-checking websites constitute 3.5\% (33,329) of notes considered by the Community Notes algorithm.


\subsection{Topics of posts and notes}
\label{subsec:post_topic}

To characterize the textual content of the proposed Community Notes and their associated posts, we first categorized posts into 19 distinct topics using Antypas et al.'s model \cite{dimosthenis}, after translating all posts to English using Facebook's \textit{M2M100} translation model \cite{m2m100}. These models were selected for their open-weight architectures and top-tier performance; we refer to \cite{dimosthenis} and \cite{m2m100} respectively for details on model training.\newline

To validate topic interpretation, we present below, for the ten most frequent topics, the 15 words the most over-represented, per the chi-square statistic, in the posts of each topic, compared to the overall corpus.

\begin{itemize}
    \item \textbf{News And Social Concern:} \textit{trump, israel, president, government, biden, hamas, ukraine, election, israeli, minister, party, police, russia, donald, kamala};
    \item \textbf{Sports:} \textit{football, league, messi, player, olympics, ronaldo, players, olympic, goals, season, match, cristiano, champions, madrid, sports};
    \item \textbf{Diaries And Daily Life:} \textit{going, thing, sorry, happy, think, little, woman, looking, thank, friend, birthday, girlfriend, morning, father, child};
    \item \textbf{Other Hobbies:} \textit{issue, previous, paypay, featured, prophet, trusted, allah, going, outer, sunshine, follow, reply, gujarat, spirit, likes};
    \item \textbf{Science And Technology:} \textit{iphone, apple, earth, space, solar, technology, android, google, spacex, science, battery, phone, microsoft, tesla, software};
    \item \textbf{Fitness And Health:} \textit{cancer, vaccine, autism, vaccines, disease, health, vaccination, covid, doctor, medical, vaccinated, study, doctors, brain, effects};
    \item \textbf{Celebrity And Pop Culture:} \textit{taylor, nicki, swift, drake, diddy, minaj, megan, selena, bieber, meghan, kendrick, ariana, cardi, kanye, rihanna};
    \item \textbf{Business And Entrepreneurs:} \textit{money, bitcoin, paypay, tesla, company, economy, business, companies, crypto, stock, select, trading, market, million, transfer};
    \item \textbf{Film, Tv And Video:} \textit{movie, anime, disney, movies, netflix, marvel, video, spider, episode, simpsons, animation, films, superman, series, cinema};
    \item \textbf{Gaming:} \textit{playstation, games, fortnite, nintendo, gaming, rockstar, console, trailer, roblox, steam, minecraft, gamers, switch, gtavi, starfield}.
\end{itemize}

For brevity, we refer to posts classified as ``News and Social Concern" ---defined by \cite{dimosthenis} as content related to awareness, activism, dialogue, and discussion of social issues and newsworthy events--- as ``political" posts.\newline

\begin{figure}[h]
\centering
\includegraphics[width=\textwidth]{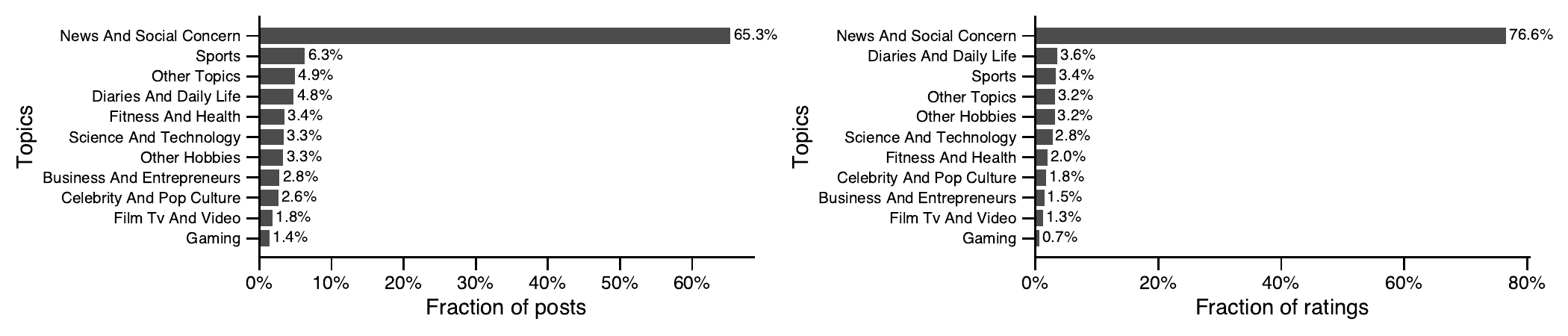}
\caption{Distribution of topics discussed in annotated posts according to \textit{dimosthenis}: (left) fraction of posts and (right) fraction of community note ratings associated with the posts. The 10 most frequent topics are displayed; less frequent topics are aggregated as ``Other Topics''.}
\label{fig:distrib_topic}
\end{figure}

Figure~\ref{fig:distrib_topic} shows the topic distribution of annotated posts (regardless of note status). Political content dominates with $65.2\%$ of posts and 76.6\% of ratings. Sports ranks second at $6.3\%$ of posts, followed by ``Diaries \& Daily Life" (slice of life, everyday content that illustrates personal opinions, feelings, occasions, and lifestyles).\newline

\begin{figure}[h]
\centering
\includegraphics[width=\textwidth]{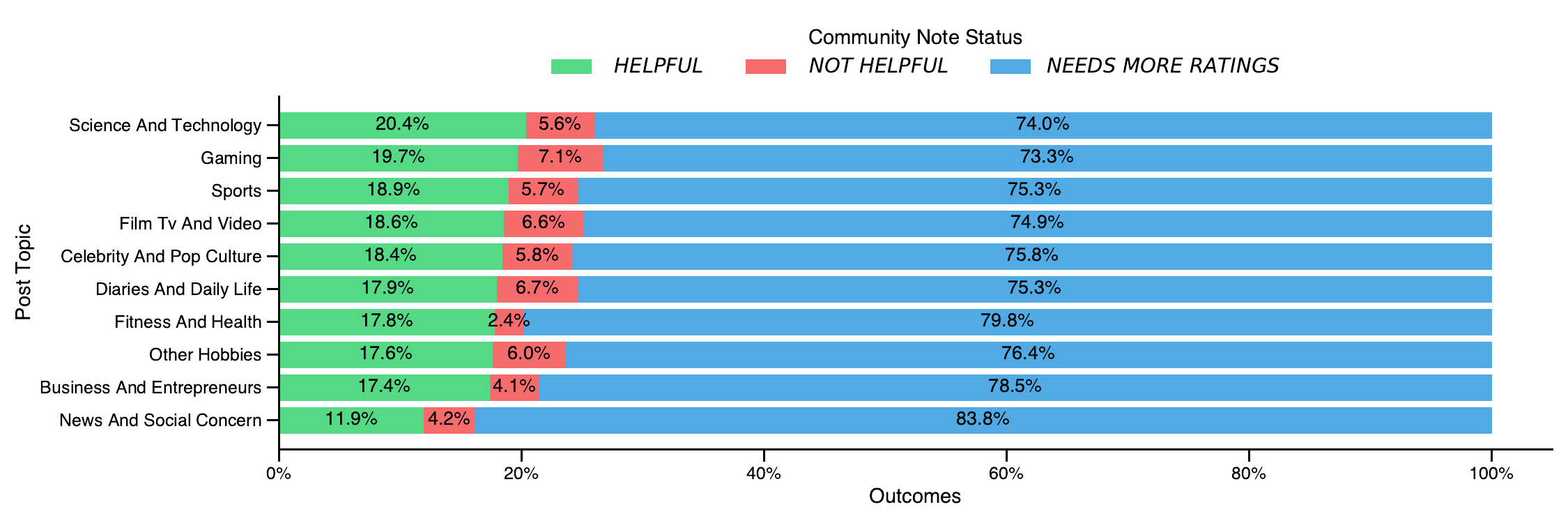}
\caption{Distribution of note statuses across post topics.}
\label{fig:distrib_topic_status}
\end{figure}

Next, we investigate the relationship between post topics and Community Notes rating outcomes. Figure~\ref{fig:distrib_topic_status} displays the distribution of note status across topics. Since only one note can be displayed under a post, we prioritize these statuses in the following order: \textit{Helpful Status}, \textit{Not Helpful Status}, and \textit{Needs More Ratings Status}. For instance, if a post has two associated notes one with \textit{Helpful Status} and the other with \textit{Needs More Ratings Status} the post is classified as having a note with \textit{Helpful Status}.\newline

While only 11.9\% of annotated political posts, having been retrieved, receive a note with \textit{Helpful Status}, this rate increases to 18.5\% for non-political posts; statistical significance confirmed through one-vs-rest chi-squared tests (p-value $< 10^{-4}$).\newline

This observation may support different hypotheses. It may reflect less consensus on political posts compared to other topics. Alternatively, it may reveal a higher propensity of users to delete their posts upon receiving a note with \textit{Helpful Status} when discussing political issues rather than sports or sciences. Finally, this result may be a direct consequence of X's Community Notes algorithm. Indeed, as the vast majority of Community Notes are associated with political posts, rater latent ideology $\theta_r$ learned by X's systems is likely to capture their political stance, rather than their views on other topics, such as their favorite soccer team. Consequently, when rating notes on non-political topics, users with diverse ``political viewpoints" ---but potentially similar views on the specific topic being discussed--- may rate the note as \textit{Helpful}, leading to its selection by X's Community Notes system. These non-exclusive factors that may explain the lower rate of notes with \textit{Helpful Status} notes on political posts cannot be definitively tested due to lack of access to deleted posts.\newline

\subsection{Classifying political notes}
\label{subsec:post_manifesto}

To achieve a finer granularity in our analysis of political posts, we leveraged the Manifesto Project coding scheme \cite{MANIFESTO} and a language model trained on this taxonomy. Specifically, we employed \textit{Manifestoberta} \cite{Burst:2023}, a language model based on the \textit{XLM-RoBERTa} architecture \cite{xlm_roberta}, trained on 1.6 million annotated statements from the Manifesto Corpus \cite{ManifestoProject}, which provides robust performance in classifying political texts \cite{Burst:2023}. By translating all political posts to English, through Facebook's \textit{M2M100} translation model \cite{m2m100}, we ensure equivalent classification performance across post original languages.\newline

\textit{Manifestoberta} provides, for each analyzed document (notes in our case), a probability score of mentioning a topic category among those existing in its topic ontology. We implemented a classification threshold at 0.5, assigning 230k political posts to a Manifesto Project category. To validate the inferred categories, we extracted, for the ten most frequent categories, the 15 most over-represented words, per the chi-square statistic, in the post of each category, compared to the overall corpus. The \textit{Manifestoberta} topic scheme differentiates topics that address issues with positive or negative attitudes towards them.

\begin{itemize}
    \item \textbf{305 - Political Authority.} \textit{trump, president, biden, donald, party, harris, kamala, minister, milei, trudeau};
    \item \textbf{605 - Law And Order: Positive.} \textit{police, crime, arrested, violence, murder, prison, shooting, criminal, sexual, officers};
    \item \textbf{503 - Equality: Positive.} \textit{trans, black, women, racist, racism, white, gender, transgender, lgbtq, woman};
    \item \textbf{501 - Environmental Protection: Positive.} \textit{climate, water, warming, emissions, change, polluted, ocean, animals, pollution, weather};
    \item \textbf{201 - Freedom And Human Rights.} \textit{freedom, speech, palestinian, rights, censorship, obtain, israeli, journalists, expression, journalist};
    \item \textbf{304 - Political Corruption.} \textit{corruption, corrupt, fraud, money, million, hunter, epstein, scandal, biden, election};
    \item \textbf{504 - Welfare State Expansion.} \textit{vaccine, covid, vaccines, vaccinated, deaths, vaccination, cancer, health, colona, pension};
    \item \textbf{603 - Traditional Morality: Positive.} \textit{abortion, jesus, church, christian, christians, catholic, christ, christianity, religious, abortions};
    \item \textbf{202 - Democracy.} \textit{democracy, elections, parliamentary, voting, votes, election, democratic, electoral, ballots, voters};
    \item \textbf{104 - Military: Positive.} \textit{missile, russian, missiles, military, aircraft, hezbollah, ukrainian, fighter, forces, defense}.
\end{itemize}

\begin{figure}[h]
\centering
\includegraphics[width=\textwidth]{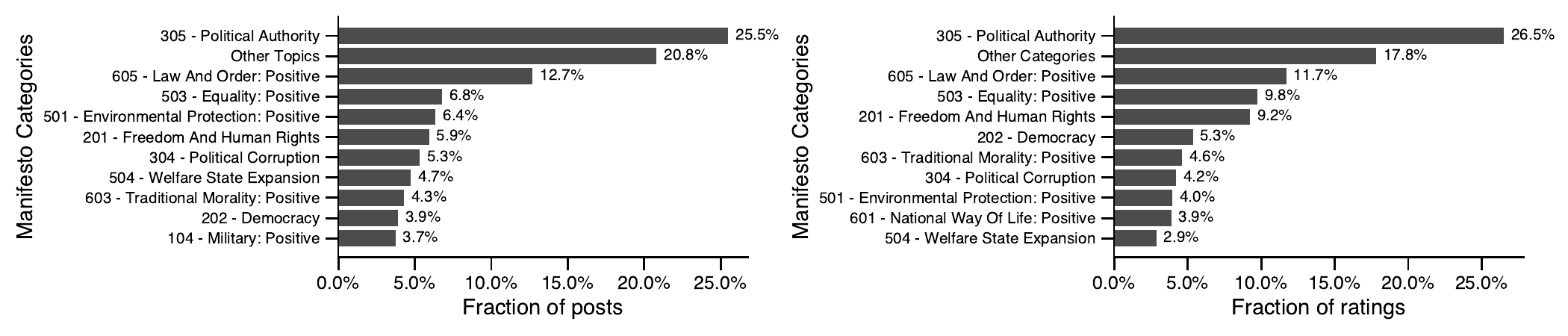}
\caption{Distribution of topics discussed in political posts according to \textit{Manifestoberta}: (left) fraction of posts and (right) fraction of community note ratings associated with the posts. The 10 most frequent topics are displayed; less frequent topics are aggregated as ``Other Categories''.}
\label{fig:distrib_label_posts_ratings_top10}
\end{figure}

Figure~\ref{fig:distrib_label_posts_ratings_top10} displays the distribution of \textit{Manifestoberta} categories in retrieved political posts. Within our corpus, ``Political authority", relating to governance competence and political power (with posts featuring political figures' names) emerges as the most frequent category with 25.5\% of posts and 26.5\% of ratings. ``Law and order" ranks second, defined as ``favorable mentions of strict law enforcement, and tougher actions against domestic crime", as corroborated by the extracted keywords. ``Freedom and human rights" ranks third, which in our corpus of annotated posts primarily addresses free speech censorship, and the Israel-Gaza conflict, as evidenced by the extracted keywords.\newline

\begin{figure}[h]
\centering
\includegraphics[width=\textwidth]{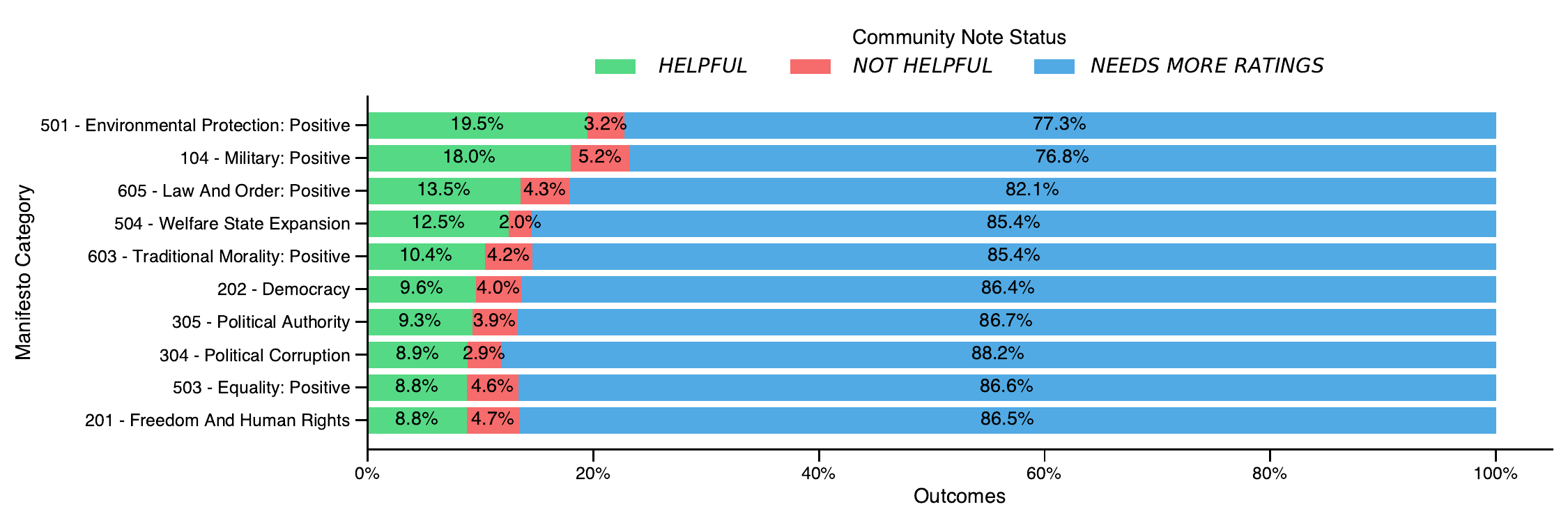}
\caption{Distribution of note statuses across category of political posts.}
\label{fig:note_status_per_label}
\end{figure}

Figure~\ref{fig:note_status_per_label} presents the distribution of notes status associated to political posts across categories. Posts discussing ``Freedom and Human Rights", ``Equality", and ``Political Corruption" exhibit the lowest proportion of note with \textit{Helpful Status}. Conversely, posts about ``Environmental Protection", ``Military" and ``Law and Order" demonstrate higher consensus, resulting in a greater proportion of notes achieving \textit{Helpful Status} in X's Community Notes system.\newline

\noindent{} Interpretations of these results are subject to the same caveats outlined in subsection \ref{subsec:post_topic}.


\subsection{Country segmentation}
\label{subsec:country_split}

\subsubsection{Attributing notes and raters to countries}
\label{subsubsec:country_split_methods}

To enable an analysis disaggregated by national setting, we segment Community Notes by country. In the absence of explicit metadata from X, we develop a multi-stage approach relying on note language, source websites, and contributor rating patterns.\newline

\begin{figure}[h]
\centering
\includegraphics[width=\textwidth]{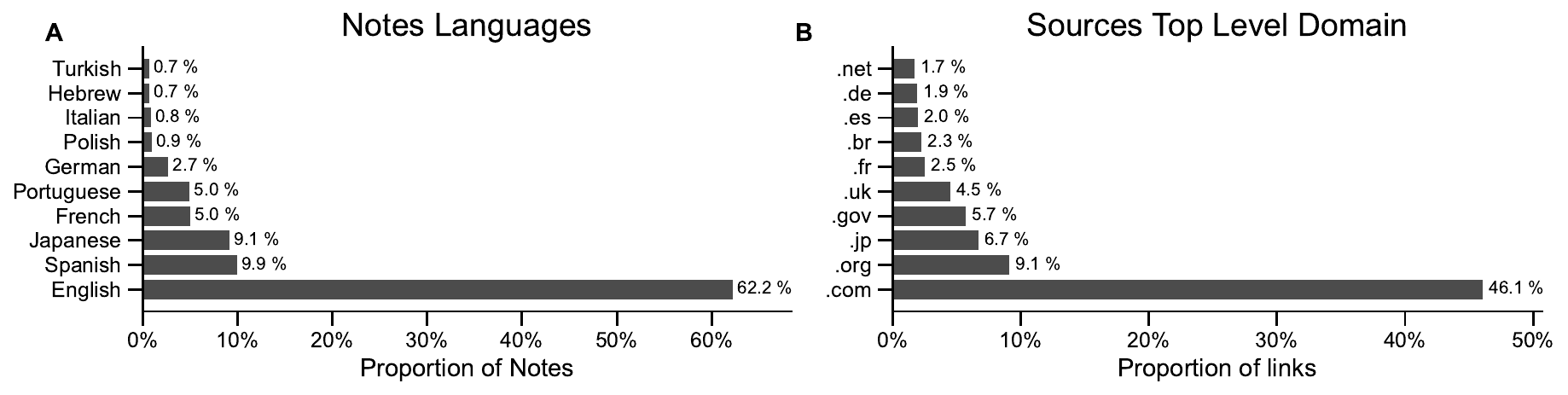}
\caption{Distribution of Community Notes language \textbf{(A)} and top-level domains of websites used as sources \textbf{(B)}. Statistics compiled from 1.2 million Community Notes submitted prior to March 1, 2025, considering the original post as ``misinformed or potentially misleading".}
\label{fig:country_accounts_notes}
\end{figure}

First, we combined notes language and the top-level domain of websites cited in notes (when present); we display in Figure \ref{fig:country_accounts_notes} the distribution of these features. A note is associated with a specific country when it meets two criteria: (1) it is written in one of the country's national languages, and (2) it cites either a national news outlet or a website under the national top-level domain. For instance, a note written in Japanese and containing a link to \texttt{sankei.com} (a Japanese newspaper) or to any website with the \texttt{.jp} domain extension (e.g., \texttt{nhk.or.jp}) will be assigned to Japan. From the complete set of Community Notes published prior to March 2025, this first conservative heuristic leave 71.2\% of notes unassigned ---those with missing or multiple inconsistent national identifiers, e.g., notes solely using \texttt{x.com} as source.\newline

Second, we map raters to countries based on their rating behavior. Each rater is assigned to the country whose notes (as identified in the first pass) they rate most frequently, restricting our analysis to rater who have rated at least 5 notes assigned to a country in the first pass.\newline

Finally, we extend this labeling to all notes and raters through an iterative classification. Unassigned notes receive country labels based on the majority country of their raters, while raters are assigned to countries based on which country's notes they most frequently evaluate.\newline

\begin{figure}[h]
\centering
\includegraphics[width=\textwidth]{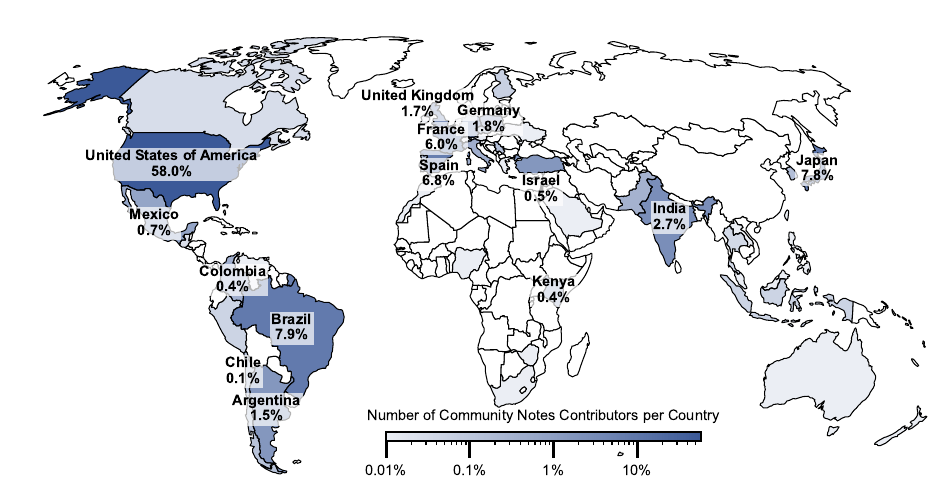}
\caption{Distribution of estimated number of Community Notes contributors per country. See \ref{subsubsec:country_split_methods} for methodology details.}
\label{fig:contributors_country}
\end{figure}

This multi-stage approach is necessary to expand the initial conservative but robust country assignments (based on explicit website and language indicators) to the complete dataset using rating patterns. Ultimately, a note is associated with a country when it is rated predominantly by contributors who typically rate notes with that country's language and national websites. Figures \ref{fig:fig_1_description} and \ref{fig:contributors_country} display the number of Community Notes written and the number of contributors across countries.\newline

There are inherent limitations to this approach, relying on the assumption that raters predominantly evaluate content from their own countries and fail to capture cross-country engagement of some raters. Yet, in the European Union, we observe a strong Pearson correlation ($\rho=0.823, ~ p<10^{-4}$) between the number of Community Note contributors per country and the number of active logged-in users, as disclosed under European transparency obligations.

\subsubsection{Accounts receiving most notes per country}

To corroborate the country segmentation, we report in ten countries, the X accounts having the highest number of submitted Community Notes (irrespective of their ``helpfulness" status):

\begin{itemize}

\item \textbf{United States}: \texttt{@elonmusk}, \texttt{@BGatesIsaPyscho}, \texttt{@jacksonhinklle}, \texttt{@iluminatibot}, \texttt{@EndWokeness}, \texttt{@KamalaHQ}, \texttt{@JoeBiden}, \texttt{@VigilantFox}, \texttt{@libsoftiktok}, \texttt{@POTUS}

\item \textbf{Japan}: \texttt{@kharaguchi}, \texttt{@kirinjisinken}, \texttt{@tweetsoku1}, \texttt{@jhmdrei}, \texttt{@himuro398}, \texttt{@JINKOUZOUKA\_jp}, \texttt{@mattariver1}, \texttt{@livedoornews}, \texttt{@newssharing1}, \texttt{@CS60osaka1}

\item \textbf{United Kingdom}: \texttt{@Conservatives}, \texttt{@RishiSunak}, \texttt{@BladeoftheS}, \texttt{@PolitlcsUK}, \texttt{@Nigel\_Farage}, \texttt{@BGatesIsaPyscho}, \texttt{@darrengrimes\_}, \texttt{@Keir\_Starmer}, \texttt{@TRobinsonNewEra}, \texttt{@elonmusk}

\item \textbf{Spain}: \texttt{@vitoquiles}, \texttt{@el\_pais}, \texttt{@CapitanBitcoin}, \texttt{@eldiarioes}, \texttt{@elmundoes}, \texttt{@FonsiLoaiza}, \texttt{@sanchezcastejon}, \texttt{@wallstwolverine}, \texttt{@abc\_es}, \texttt{@AlertaNews24}

\item \textbf{Brazil}: \texttt{@choquei}, \texttt{@folha}, \texttt{@LulaOficial}, \texttt{@Metropoles}, \texttt{@GloboNews}, \texttt{@felipeneto}, \texttt{@siteptbr}, \texttt{@gleisi}, \texttt{@CR7Brasil}, \texttt{@NewsLiberdade}

\item \textbf{France}: \texttt{@CerfiaFR}, \texttt{@silvano\_trotta}, \texttt{@BFMTV}, \texttt{@JLMelenchon}, \texttt{@AlertesInfos}, \texttt{@RimaHas}, \texttt{@ALeaument}, \texttt{@f\_philippot}, \texttt{@sandrousseau}, \texttt{@franceinfo}

\item \textbf{Germany}: \texttt{@jreichelt}, \texttt{@tagesschau}, \texttt{@Markus\_Soeder}, \texttt{@niusde\_}, \texttt{@ZDFheute}, \texttt{@Karl\_Lauterbach}, \texttt{@Alice\_Weidel}, \texttt{@polenz\_r}, \texttt{@SHomburg}, \texttt{@Bundeskanzler}

\item \textbf{India}: \texttt{@MrSinha\_}, \texttt{@zoo\_bear}, \texttt{@INCIndia}, \texttt{@IndianTechGuide}, \texttt{@dhruv\_rathee}, \texttt{@ANI}, \texttt{@rishibagree}, \texttt{@INCKerala}, \texttt{@RahulGandhi}, \texttt{@TimesAlgebraIND}

\item \textbf{Argentina}: \texttt{@JMilei}, \texttt{@porqueTTarg}, \texttt{@therealbuni}, \texttt{@ElTrumpista}, \texttt{@laderechadiario}, \texttt{@infobae}, \texttt{@clarincom}, \texttt{@TugoNews},  \texttt{@GordoDan\_}, \texttt{@TraductorTeAma}

\item \textbf{Israel}: \texttt{@YinonMagal}, \texttt{@netanyahu}, \texttt{@bezalelsm}, \texttt{@itamarbengvir}, \texttt{@ShaykhSulaiman}, \texttt{@jacksonhinklle}, \texttt{@Now14Israel}, \texttt{@TallyGotliv}, \texttt{@rothmar}, \texttt{@amit\_segal}, 

\end{itemize}

\begin{figure}[h]
\centering
\includegraphics[width=\textwidth]{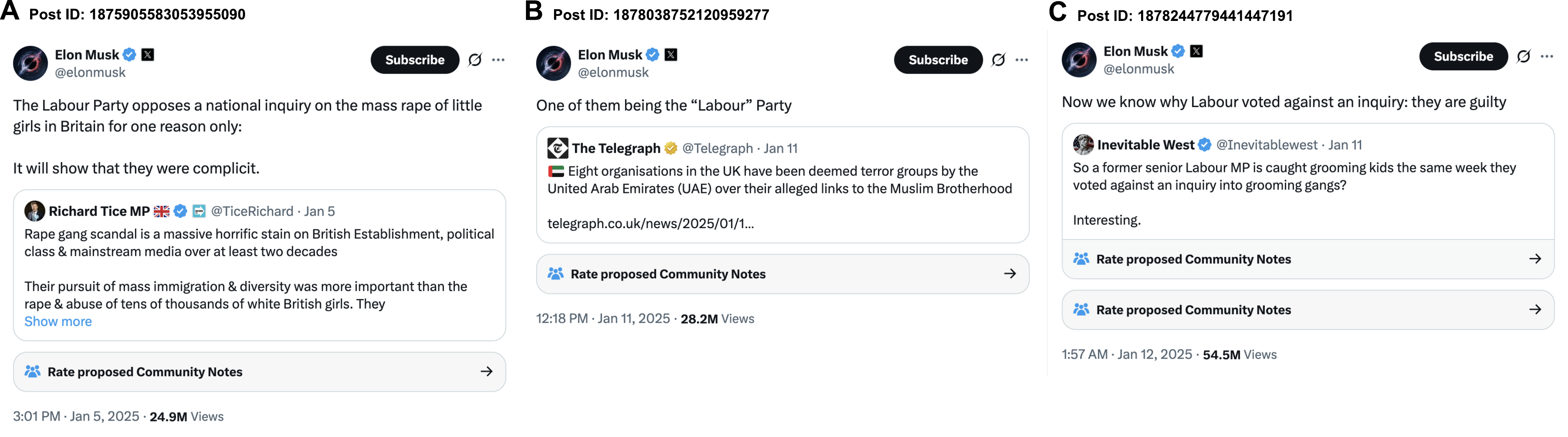}
\caption{Example of \texttt{@elonmusk} post classified as United Kingdom per the methodology outlined in \ref{subsubsec:country_split_methods}.}
\label{fig:uk_elon}
\end{figure}

Manual inspection confirms that these accounts align with their inferred countries of origin. Notably, Elon Musk's account \texttt{@elonmusk} ranks among the most commented in both the United States and the United Kingdom. Notes associated with Elon Musk's posts classified as United Kingdom are attached to content discussing UK politics and contain references to UK news outlets, while notes associated with US-related content are classified accordingly. Figure \ref{fig:uk_elon} illustrates examples of posts whose associated notes were classified as United Kingdom. Similar patterns emerge with \texttt{@BGatesIsaPyscho}, an account that publishes both US and UK related content, and \texttt{@jacksonhinklle}, whose content addresses Israel-Palestine issues and appears among the most commented accounts in both the US and Israel splits.

\section{Outcomes of the algorithmic resolution}
\label{sec:outcomes_si}

In this section we turn our attention to the analysis of notes according to the status they are attributed by X's Community Notes algorithm, and publicly disclosed by X.


\subsection{Outcomes by country}
\label{subsec:outcomes_country}

\begin{figure}[h]
\centering
\includegraphics[width=\textwidth]{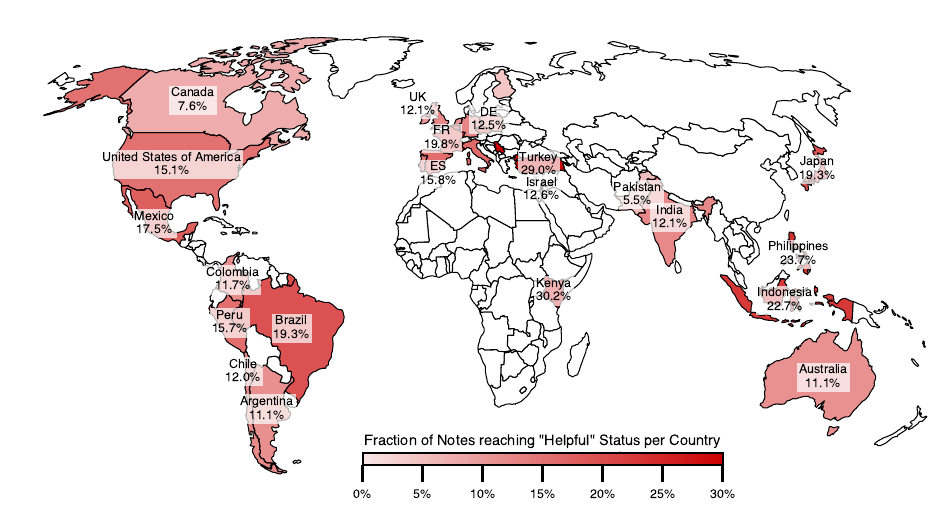}
\caption{Fraction of annotated posts with a note with \textit{Helpful Status} per country; limited to countries with at least a thousand annotated posts, as per \ref{subsubsec:country_split_methods}, after pre-processing and excluding ``Note Not Needed" notes.}
\label{fig:distrib_category_status}
\end{figure}

Leveraging the country-mapping of Community Notes described in Section~\ref{subsubsec:country_split_methods}, we present in Figure~\ref{fig:distrib_category_status} the fraction of posts ending up with a note with \textit{Helpful Status} across the 30 countries with at least one thousand annotated posts; computation restricted to the notes considered by X's Community Notes algorithm i.e. after preprocessing \cite{TwitterGithubMain}.\newline

We observe significant variation between countries, from 29.0\% of posts receiving a note with \textit{Helpful Status} in Turkey and 19.8\% in France to 12.6\% in Israel and 11.1\% in Argentina.


\subsection{Outcomes by topic}
\label{subsec:outcomes_topic}

To achieve finer granularity than the general topic characterization in Sections \ref{subsec:post_topic} \& \ref{subsec:post_manifesto} and circumvent post-deletion observation bias, we examine word-level differences between Community Notes that achieved \textit{Helpful Status} and those that did not. We compute chi-square statistics for words of four letters or longer in note content, focusing on the ten countries with the most Community Notes. All keywords were translated into English.

\textbf{\emph{United States}}
\begin{itemize}
\item \textbf{Over-Represented words}: account, scam, Tesla, fake, steal, video, assets, scams, impersonating, pretext, publicity, scammers, elonmusk, promotion, promotions
\item \textbf{Under-Represented words}: Gaza, political, Donald, Israeli, opinion, quot, context, president, vaccines, evidence, Biden, COVID, Israel, Hamas, Trump
\end{itemize}

\textbf{\emph{Japan}}
\begin{itemize}
\item \textbf{Over-Represented words}: earthquake, prediction, forecasting, method, fraud, deception, prize, winning, occurrence, weather, happen, sell, atmosphere, message, time
\item \textbf{Under-Represented words}: deliberation, deputy, reaction, problem, authority, election, evaluation, mild, causality, nature, certification, meeting, relief, party, system
\end{itemize}

\textbf{\emph{United Kingdom}}
\begin{itemize}
\item \textbf{Over-Represented words}: shipped, products, scam, unfit, stores, store, generated, purchasing, selling, priced, image, drop, video, caution, exercise
\item \textbf{Under-Represented words}: opinion, violence, political, Israel, gender, male, change, quot, trans, government, people, women, Hamas, party, Labour
\end{itemize}

\textbf{\emph{Spain}}
\begin{itemize}
\item \textbf{Over-Represented words}: video, image, corresponds, original, scam, photo, TikTok, digitally, cryptocurrencies, famous, Instagram, Biden, altered, artificial, real
\item \textbf{Under-Represented words}: Montero, parties, Hamas, right, Popular, Quiles, convicted, Pedro, Vito, hate, extreme, crime, Ayuso, PSOE, government
\end{itemize}

\textbf{\emph{Brazil}}
\begin{itemize}
\item \textbf{Over-Represented words}: bets, guidelines, house, stake, hidden, site, adult, game, tiger, prohibited, games, pirate, gang, violating, report
\item \textbf{Under-Represented words}: tax, Jair, facts, note, freedom, Hamas, left, Israel, minister, federal, deputy, president, Lula, Bolsonaro, government
\end{itemize}

\textbf{\emph{France}}
\begin{itemize}
\item \textbf{Over-Represented words}: acts, account, validate, site, thread, steering, photo, user, usurps, scam, date, original, image, blue, redone
\item \textbf{Under-Represented words}: victims, racism, opinion, quot, hostages, right, racist, Hassan, Palestinians, terrorist, Gaza, political, party, left, Hamas
\end{itemize}

\textbf{\emph{Germany}}
\begin{itemize}
\item \textbf{Over-Represented words}: celebrities, scam, advertisement, crypto, replicated, advertise, portals, paid, real, fraud, video, page, acts, abused, world
\item \textbf{Under-Represented words}: study, insult, electricity, climate, government, percent, politics, influence, context, mister, Germany, power, Correctiv, party, Habeck
\end{itemize}

\textbf{\emph{India}}
\begin{itemize}
\item \textbf{Over-Represented words}: video, account, websites, tokens, original, pump, dump, impersonating, garbage, forex, sell, digitally, schemes, heavy, advance
\item \textbf{Under-Represented words}: party, constitution, spreading, propaganda, opposition, Muslim, political, Modi, Hindus, Rahul, people, government, Hindu, Gandhi, Congress
\end{itemize}

\textbf{\emph{Argentina}}
\begin{itemize}
\item \textbf{Over-Represented words}: Instagram, image, photo, video, original, corresponds, Infobae, Franco, digitally, window, Colapinto, patio, altered, false, Tesla
\item \textbf{Under-Represented words}: Javier, measures, system, president, Massa, party, change, quot, senators, debt, candidate, advances, freedom, government, Milei
\end{itemize}

\textbf{\emph{Israel}}
\begin{itemize}
\item \textbf{Over-Represented words}: refused, Gvir, return, convicted, police, examples, millions, Avishi, trust, border, Rothman, drowning, protester, execution, raised
\item \textbf{Under-Represented words}: discussed, Yinon, video, writer, spokesman, Jews, mistaken, probably, brothers, quot, actually, Gaza, original, that, instead
\end{itemize}

We observe that across countries, scam and doctored media-related keywords are over-represented in notes with \textit{Helpful Status} compared to those that do not, while politically divisive issues, such as the Israel-Palestine conflict, are under-represented in notes with \textit{Helpful Status}.\newline

\noindent{}To quantify this observation we curated the following sets of keywords (some are intentionally shortened for expanded matching):

\begin{itemize}
    \item \textbf{US Politics}: ``Biden",``Kamala",``Trump", ``Democrat, ``Republican"
    \item \textbf{Israel-Palestine Conflict}: ``Israel",``Gaza",``Netanyahu", ``Hamas, ``IDF", ``Hezbollah", ``Tel Aviv", ``Palestin", ``UNRWA", ``Nakba"
    \item \textbf{Russia-Ukraine Conflict}: ``Russi",``Ukrain", ``Putin, ``Zelensky", ``Moscow", ``Kyiv",``Kremlin", ``NATO", 
    \item \textbf{Scam}: ``scam",``fraud",``impersonating"
\end{itemize}

\noindent{}These sets of manually curated keywords are not comprehensive and prioritized precision over recall. We then extracted, in the United States, Community Notes containing one of these keywords, and determine for each topic the fraction of notes reaching \textit{Helpful Status}:

\begin{itemize}
\item \textbf{US Politics}: 6.0\% of notes reached \textit{Helpful Status}.
\item \textbf{Israel-Palestine Conflict}: 5.5\% of notes reached \textit{Helpful Status}.
\item \textbf{Russia-Ukraine Conflict}: 11.7\% of notes reached \textit{Helpful Status}.
\item \textbf{Scam}: 39.0\% of notes reached \textit{Helpful Status}.
\end{itemize}

The rate at which notes reach \textit{Helpful Status} is statistically lower for those related to US Politics, Israel-Palestine Conflict, and Russia-Ukraine Conflict compared to those associated with Scam, as assessed by pairwise permutation test, $p<0.005$.

Additionally, we consider the subset of community notes related to elections held in the United States (2024 presidential election), United Kingdom (2024 general election), France (2024 legislative election), and Germany (2025 federal elections). For each election, we curated the following lists of keywords (translated into English) related to the elections and political parties: 

\begin{itemize}

    \item \textbf{United States} \textit{election, vote, voting, voter, ballot, campaign, Trump, Harris, Kamala, Donald Trump, Kamala Harris, presidential, president, POTUS, swing state, battleground, electoral, electoral college, Republican, Democrat, Democratic, GOP, midterm, primary, caucus, poll, polling, Biden, Joe Biden, debate, rally, MAGA, election day, early voting, mail-in ballot, absentee ballot, Pennsylvania, Michigan, Wisconsin, Arizona, Georgia, Nevada, North Carolina, electoral votes, red wave, blue wave, turnout, registered voter, Vance, JD Vance, Walz, Tim Walz, running mate, VP, vice president, inauguration, transition, concession, recount, projection, race called, swing district, Senate race, House race, congressional, gubernatorial, ballot measure, referendum, voter suppression, voter fraud, election integrity, certification, electoral count, November 5}

    \item \textbf{United Kingdom} \textit{general election, UK election, Britain election, British election, Parliament, parliamentary, House of Commons, Westminster, MP, member of parliament, constituency, seat, Labour, Labour Party, Conservative, Tory, Lib Dem, Liberal Democrats, Reform UK, Reform, Green Party, Greens, SNP, Scottish National Party, Plaid Cymru, DUP, Democratic Unionist Party, Sinn Féin, Alliance Party, UUP, Ulster Unionist Party, Keir Starmer, Rishi Sunak, Ed Davey, Nigel Farage, John Swinney, Rhun ap Iorwerth, Carla Denyer, Adrian Ramsay, Kemi Badenoch, Liz Truss, Boris Johnson, Jeremy Corbyn, Theresa May, Rachel Reeves, Angela Rayner, Prime Minister, PM, first-past-the-post, FPTP, proportional representation, PR, vote, voting, voter, ballot, polling, poll, turnout, majority, landslide, hung parliament, coalition, minority government, government formation, opposition, manifesto, campaign, election campaign, snap election, dissolution, prorogation, July 4, May 22, King Charles, monarch, 10 Downing Street, Downing Street, Number 10, swing state, swing seat, marginal seat, safe seat, red wall, blue wall, tactical voting, exit poll, by-election, Scotland, Wales, England, Northern Ireland, Scottish, Welsh, NHS, National Health Service, immigration, Brexit, cost of living, economy, tax, taxation, Rwanda, Rwanda scheme, D-Day, debate, televised debate, voter ID, photo ID, postal vote, early voting, constituency boundary, boundary review, redistricting, declaration, count, recount, incumbent, challenger, cabinet, shadow cabinet, minister, secretary of state, backbencher, frontbencher, whip, Speaker, Chancellor, Home Secretary, Foreign Secretary, election 2024}

    \item \textbf{France}: \textit{legislative election, French election, election, National Assembly, deputies, RN, National Rally, Marine Le Pen, Jordan Bardella, Emmanuel Macron, NFP, New Popular Front, Popular Front, LFI, France Insoumise, Jean-Luc Mélenchon, PS, Socialist Party, Olivier Faure, Ensemble, Renaissance, Gabriel Attal, Michel Barnier, François Bayrou, Élisabeth Borne, LR, Republicans, PCF, Communist Party, EELV, Ecologists, Greens, Marine Tondelier, Horizons, MoDem, snap election, dissolution, cohabitation, hung parliament, first round, second round, runoff, three-way race, four-way race, June 30, July 7, turnout, participation, constituency, vote, voting, far-right, radical right, left-wing, coalition, alliance, government formation, Prime Minister, President, parliament, majority, absolute majority, minority government, opposition, projections, polls, 2024 election}

    \item \textbf{Germany}: \textit{German federal election, Bundestag, German election, CDU, CSU, CDU/CSU, SPD, AfD, Greens, FDP, The Left, BSW, Sahra Wagenknecht, Friedrich Merz, Olaf Scholz, Alice Weidel, Christian Lindner, traffic light coalition, grand coalition, coalition, chancellor, vote, voting, first vote, second vote, constituency, Federal Returning Officer, confidence vote, snap election, early election, February 23, voter turnout, government, opposition, Federal President, Frank-Walter Steinmeier, Bavaria, Baden-Württemberg, North Rhine-Westphalia, Saxony, Thuringia, Brandenburg, Saxony-Anhalt, Mecklenburg-Western Pomerania, Berlin, Hamburg, Hesse, Lower Saxony, Rhineland-Palatinate, Saarland, Schleswig-Holstein, Bremen, East Germany, West Germany, right-wing populism, right-wing, far-right, firewall, proportional representation, electoral threshold, five percent threshold, 5\% threshold, overhang mandate, balance seat, direct mandate, state list, party list, manifesto, electoral program, coalition negotiations, coalition talks, government formation, minority government, election 2025}

\end{itemize}

For robustness, we bootstrap over keywords by selecting community notes containing a random sample of 50\% of keywords, measuring the fraction reaching helpful status, and reporting the average fraction and 95\% confidence interval over 250 repetitions.


\subsection{Outcomes by account}
\label{subsec:outcomes_account}

For each X account, we contrast the observed number of posts receiving note reaching \textit{Helpful Status} with the expected frequency derived from the population average. Using the chi-square statistic, we identify the 10 most significantly over-moderated and under-moderated accounts across five countries: the United States, Japan, Spain, France, and Brazil. \newline

Importantly, these results are conditioned on posts being available for collection. As such, under-moderated accounts may reflect either lower consensus among raters, a higher tendency of users to delete posts deemed misleading (i.e., those associated with notes with \textit{Helpful Status}), or some combination of both factors.\newline

\paragraph{United States}
\begin{itemize}
    \item \textbf{Under-Moderated}: \texttt{@elonmusk} (2.6\%), \texttt{@KamalaHQ} (0.1\%), \texttt{@VigilantFox} (1.6\%), \texttt{@libsoftiktok} (3.3\%), \texttt{@EndWokeness} (4.0\%), \texttt{@RadioGenoa} (2.1\%), \texttt{@KamalaHarris} (1.2\%), \texttt{@JoeBiden} (1.6\%), \texttt{@POTUS} (2.4\%), \texttt{@krassenstein} (1.8\%)
    \item \textbf{Over-Moderated}: \texttt{@BGatesIsaPyscho} (33.1\%), \texttt{@iluminatibot} (26.1\%), \texttt{@thehealthb0t} (28.8\%), \texttt{@DailyLoud} (41.4\%), \texttt{@Evony\_TKR} (78.1\%), \texttt{@DramaAlert} (33.1\%), \texttt{@QTHESTORMM} (52.0\%), \texttt{@realstewpeters} (29.3\%), \texttt{@dom\_lucre} (23.0\%), \texttt{@tassagency\_en} (30.1\%)
\end{itemize}

\paragraph{Japan}
\begin{itemize}
    \item \textbf{Under-Moderated}: \texttt{@takavet1} (0.0\%), \texttt{@RyuichiYoneyama} (0.0\%), \texttt{@tamakiyuichiro} (1.9\%), \texttt{@mas\_\_yamazaki} (0.0\%), \texttt{@umekichkun} (2.0\%), \texttt{@siroiwannko1} (0.0\%), \texttt{@himasoraakane} (0.0\%), \texttt{@nekoruck} (0.0\%), \texttt{@bad\_texter} (2.2\%), \texttt{@noruzo\_tetsudo} (3.4\%) 
    \item \textbf{Over-Moderated}: \texttt{@kirinjisinken} (81.6\%), \texttt{@tweetsoku1} (40.0\%), \texttt{@geoscience16} (47.5\%), \texttt{@kkkfff1234k} (45.1\%), \texttt{@hibakuyada} (41.3\%), \texttt{@takui\_aka} (36.2\%), \texttt{@jhmdrei} (30.8\%), \texttt{@giantearthquake} (39.3\%), \texttt{@foodloss\_sos} (83.3\%), \texttt{@himuro398} (26.4\%)
\end{itemize}

\paragraph{Spain}
\begin{itemize}
    \item \textbf{Under-Moderated}: \texttt{@wallstwolverine} (2.4\%), \texttt{@eldiarioes} (3.4\%), \texttt{@PSOE} (3.3\%), \texttt{@IdiazAyuso} (0.0\%), \texttt{@sanchezcastejon} (2.0\%), \texttt{@iescolar} (2.6\%), \texttt{@Santi\_ABASCAL} (0.0\%), \texttt{@oscar\_puente\_} (1.3\%), \texttt{@vox\_es} (2.9\%), \texttt{@AlanBarrosoA} (1.4\%)
    \item \textbf{Over-Moderated}: \texttt{@ceciarmy} (55.0\%), \texttt{@A3Noticias} (29.2\%), \texttt{@Tebasjavier} (53.3\%), \texttt{@larazon\_es} (29.8\%), \texttt{@isaacfouto} (43.8\%), \texttt{@MediterraneoDGT} (26.8\%), \texttt{@abc\_es} (18.5\%), \texttt{@EFEnoticias} (19.0\%), \texttt{@20m} (27.3\%), \texttt{@\_anapastor\_} (27.6\%)
\end{itemize}

\paragraph{France}
\begin{itemize}
    \item \textbf{Under-Moderated}: \texttt{@CerfiaFR} (5.4\%), \texttt{@RimaHas} (3.9\%), \texttt{@AlertesInfos} (5.9\%), \texttt{@julienbahloul} (0.0\%), \texttt{@CaronAymericoff} (3.7\%), \texttt{@EmmanuelMacron} (1.5\%), \texttt{@CordierAlice2} (0.0\%), \texttt{@Pediavenir} (0.0\%), \texttt{@ArnaultRaphael} (4.9\%), \texttt{@emma\_ducros} (0.0\%)
    \item \textbf{Over-Moderated}: \texttt{@silvano\_trotta} (26.6\%), \texttt{@55Bellechasse} (60.9\%), \texttt{@VictorSinclair3} (34.7\%), \texttt{@\_IDVL} (54.5\%), \texttt{@CestTerrifiant} (55.6\%), \texttt{@greenpeacefr} (66.7\%), \texttt{@BPartisans} (32.4\%), \texttt{@Reporterre} (57.9\%), \texttt{@75secondes} (48.3\%), \texttt{@pierre\_jacquel2} (50.0\%)
\end{itemize}

\paragraph{Brazil}
\begin{itemize}
    \item \textbf{Under-Moderated}: \texttt{@LulaOficial} (0.8\%), \texttt{@GloboNews} (3.9\%), \texttt{@vinicios\_betiol} (1.7\%), \texttt{@pesquisas\_elige} (1.4\%), \texttt{@felipeneto} (2.7\%), \texttt{@revistaforum} (0.0\%), \texttt{@jairbolsonaro} (0.9\%), \texttt{@PedroRonchi2} (2.9\%), \texttt{@Estadao} (3.7\%), \texttt{@gleisi} (4.2\%) 
    \item \textbf{Over-Moderated}: \texttt{@CR7Brasil} (76.7\%), \texttt{@7gamesbet} (88.9\%), \texttt{@quero\_escolher} (86.7\%), \texttt{@choquei} (21.4\%), \texttt{@AndreGA\_Pe} (40.7\%), \texttt{@samengo\_ofc} (72.7\%), \texttt{@siteptbr} (26.3\%), \texttt{@updatecharts} (40.5\%), \texttt{@joaquimrib1801} (44.8\%), \texttt{@mspbra} (27.1\%)
\end{itemize}

Across different countries, a consistent observation emerges: politically divisive accounts ---whether associated with \texttt{@KamalaHarris}, \texttt{@tamakiyuichiro}, \texttt{@sanchezcastejon}, \texttt{@EmmanuelMacron}, or \texttt{@LulaOficial}--- display smaller fraction of post with note having reached \textit{Helpful Status} compared to the general population. Conversely, entertainment accounts (\texttt{@DramaAlert}, \texttt{@CR7Brasil}) and conspiracy theory promoters (e.g., \texttt{@BGatesIsaPsycho}, \texttt{@silvano\_trotta}) tend to generate high agreement and exhibit elevated moderation rates.

\section{Ideology scaling of X users}
\label{sec:scaling_si}

To explore how the latent ideology dimension computed by the Community Notes model relates to Left-Right and Anti-Elite positions, we employ the ideological scaling of X users developed by Ramaciotti et al. \cite{ramaciotti2022inferring}. This approach extends Barberá's \cite{barbera2015birds} methodology to multidimensional settings, enabling identification and calibration with political survey data across countries.

\subsection{Multidimensional ideology scaling}
\label{subsec:epo_dataset}

We collected, through the API of X, the set of users following Members of Parliament (MPs) on X across 23 countries: \textit{France, Germany, United States, United Kingdom, Spain, Brazil, Australia, Italy, Israel, Argentina, Denmark, Sweden, Canada, Poland, the Netherlands, Switzerland, South Africa, Nigeria, Finland, Mexico, New Zealand, Japan, and Belgium.} This sampling strategy enables the selection of a large number of users engaged with national politics; we report in Table \ref{tab:epo_x_cn} the number of such users per country. In the following, we consider only users following at least three MPs and that were followed by at least 25 accounts, following Barberá's protocol in his 2015 studies \cite{barbera2015birds, Barber2015Posting, barbera2015understanding}.\newline

We then estimate the ideal positions of users and MPs by performing, as Barberá \cite{barbera2015birds}, a Correspondence Analysis \cite{Greenacre2007} on the follower-MP adjacency matrix. As demonstrated by Carroll et al. \cite{Carroll1997}, this procedure yields approximations to the ideal points estimated through the following homophilic data generation process:

\begin{equation}
    P\left(\text{U}_i\rightarrow \text{MP}_j|\alpha_i,\beta_j,\gamma,\varphi_i,\varphi_j\right) = \text{logistic}\left(\alpha_i + \beta_j - \gamma \|\varphi_i - \varphi_j \|^2 \right),
    \label{eq:homophily_model}
\end{equation}\newline

\noindent{}where, $U_i\rightarrow MP_j$ stands for the event in which user $\text{U}_i$ follows $\text{MP}_j$, $\alpha_i$ is the level of \textit{activity} of user $\text{U}_i$ measured in number of followed accounts, $\beta_j$ is the level of \textit{popularity} of $\text{MP}_j$, measured in number of followers, $\gamma$ is a scale variable, and $\varphi_i$ and $\varphi_j$ are the multidimensional positions of $\text{U}_i$ and $\text{MP}_j$ in some unobservable multidimensional space. This data generation process is homophilic in the sense that $P\left(\text{U}_i\rightarrow \text{MP}_j\right)$ increases as the distance between unobservable positions $\|\varphi_i - \varphi_j\|$ in some latent space decreases.\newline

Most applications of this scaling method, especially those using social media data, rely on a single-dimensional model for unobservable positions $\varphi$, relying the assumption that the the recovered dimension will match the leading axis of polarization in the national setting under study.
While this assumption is usually validated in studies set in the United States \cite{Ramaciotti2024_us3d}, empirical spatial analyses have shown that more general settings, in particular those in European countries \cite{Bakker2012}, cannot be reduced to a single ideological dimension (e.g., Left-Right).\newline


\subsection{Calibration and validation}

Ramaciotti et al. \cite{ramaciotti2022inferring} extended this scaling procedure to capture multiple political dimensions by considering the subsequent principal directions $\delta_2, \delta_3,...$ identified through the Correspondence Analysis.
We then calibrate these latent dimensions with political survey data. This step is required because, in Eq.~\eqref{eq:homophily_model}, the probability $P\left(\text{U}_i\rightarrow \text{MP}_j\right)$ suffers from an identification issue. Indeed, the model is invariant to isometries, such as flips, translations, and rotations on positions $\varphi_i$ and $\varphi_j$, meaning that any such transformation would yield identical probability values. Consequently, the raw positions $\varphi_i$ lack intrinsic political meaning and cannot serve as reliable indicators for specific political stances or ideologies (ideology being defined as an organized structure of attitudes \cite{campbell1960american,converse1964nature}). For instance, the relation $\varphi_i > \varphi_j$ could equally indicate that user $i$ is more Left-leaning than user $j$, or precisely the opposite, rendering the model's output ambiguous without external anchoring. This identification issue is particularly limiting in cross-country comparative analyses where a dimension of interest for comparison (e.g., a Left-Right) dimension, might present itself as different spatial directions in the latent homophily space of positions $\varphi$ in equation Eq.~\eqref{eq:homophily_model}. This limitation is made all the more salient when the phenomenon of interest, such as factuality of claims, is expected to depend on a leading dimension of polarization, and dimensions linked to trust in elites and institutions \cite{ramaciotti2023geometry}.
\newline

To resolve this identification problem, we use the Global Party Survey \cite{norris2020measuring}---an expert survey positioning parties along several issue and ideology dimensions---to map positions $\varphi_j$ from the latent homophily spaces to positions along the dimensions of the expert survey. Specifically, we map the average position in the latent homophily space of MPs from a party to the position of the party along the dimensions assessed in the Survey.\newline

We consider the Global Party Survey for our study because of its international coverage and available coherence metrics with other studies on party positions. We retain for our study the Left-Right dimension (variable \textit{V4\_Scale}) and the dimension measuring populist rhetoric, capturing attitudes towards elites and institutions (among other populist traits; variable \textit{V8\_Scale}).\newline

\begin{figure}[h!]
\centering
\includegraphics[width=\textwidth]{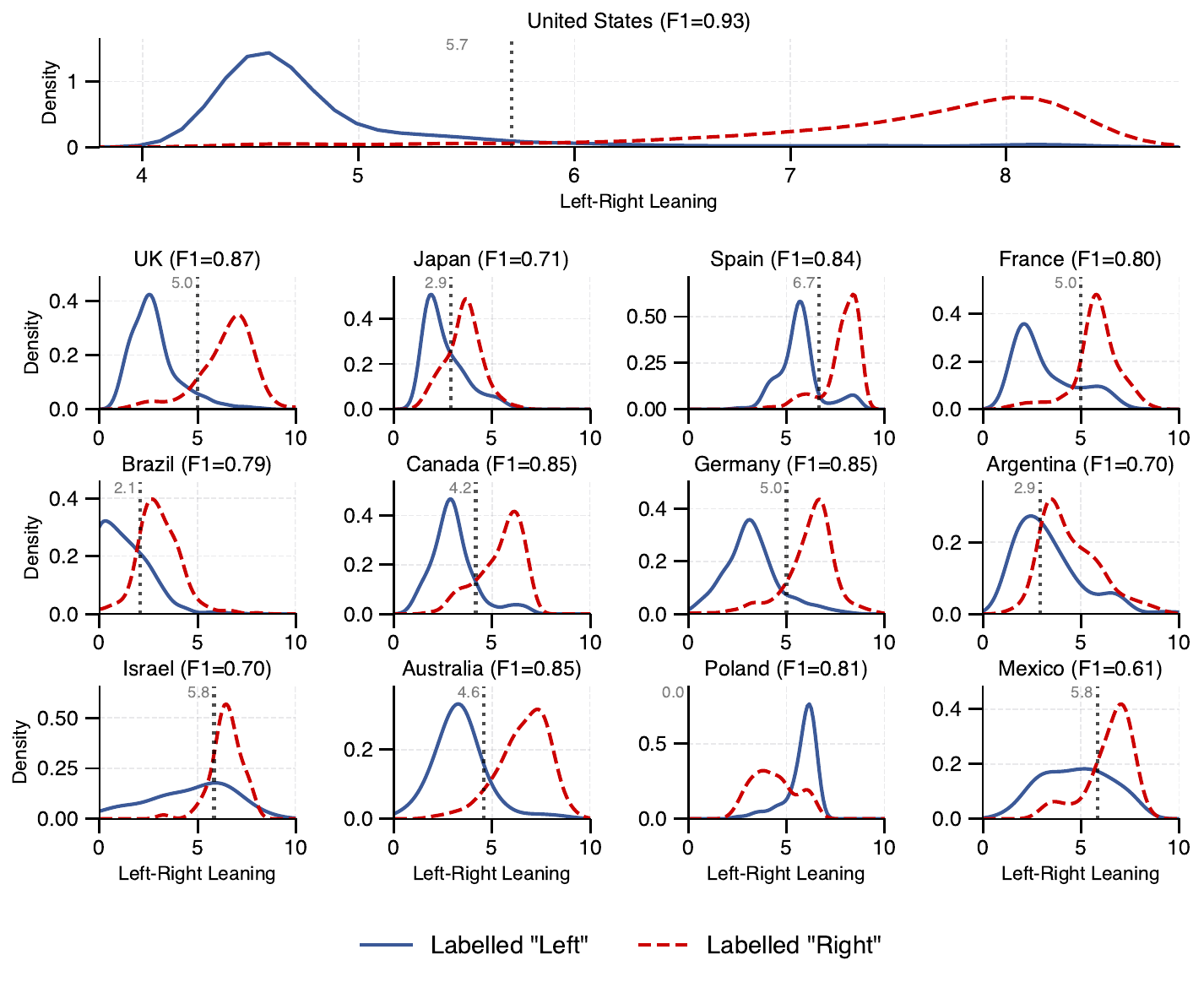}
\caption{Distribution of the inferred Left-Right leaning segmenting users labeled by LLM Zephyr 7B $\beta$ \cite{Zephyr} as ``Right-leaning'' or ``Left-leaning''. We report the optimal threshold best segmenting these populations as well as the associated F1-score.}
\label{fig:validation_epo_left_right}
\end{figure}

To validate the positions of users in our study along the Left-Right position, we leverage an independent signal not used to infer users' leaning, namely, the textual content of users' profile bios. Specifically, we prompted a Large Language Model, in our case Zephyr 7B $\beta$ \cite{Zephyr}, to identify a subset of users that could be attributed with the labels ``Left'' or ``Right''. We display in Figure \ref{fig:validation_epo_left_right} the distribution of inferred Left-Right leaning for users labeled by the LLM as ``Left'' or ``Right''.
Our validation method shows that this Left-Right dimension separates users that can be labeled as binary being either ``Left'' or ``Right'' leaning on the basis of what they write in their profile bios. We quantify this spatial separation with the F1 performance of an logistic regression classifier.

\begin{figure}[h]
\centering
\includegraphics[width=\textwidth]{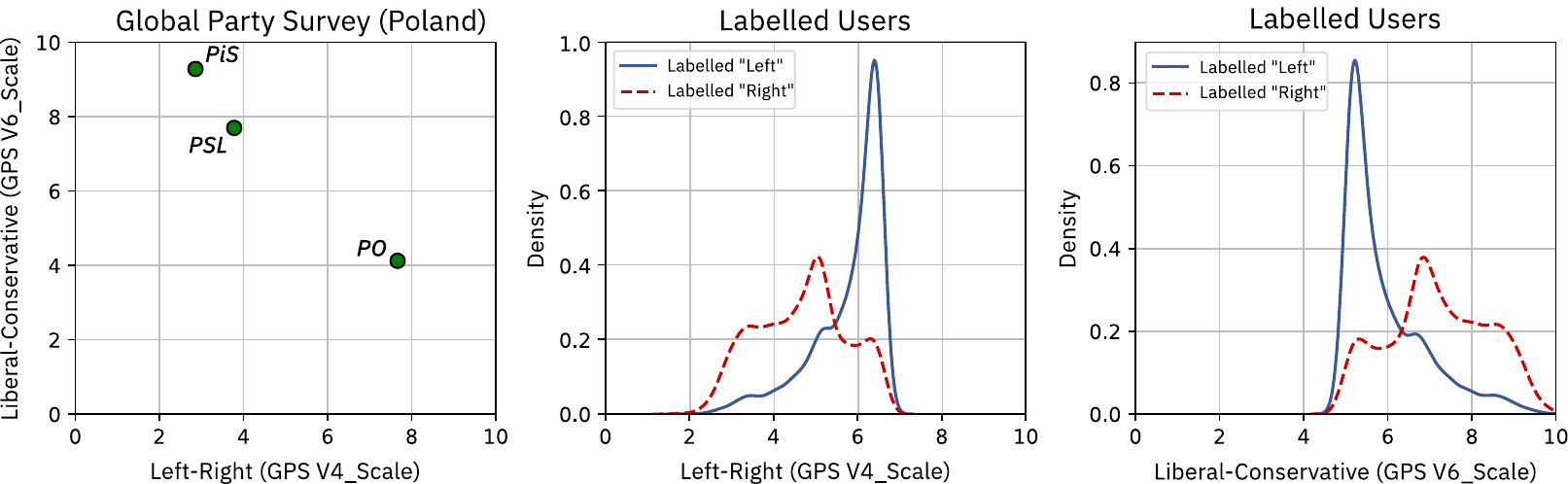}
\caption{Additional details for the validation of ideology scaling data for Poland in Figure~\ref{fig:poland_gps_valid}. Left panel shows that, for parties available in the survey used for space identification (Global Party Survey), Left-Right and Liberal-Conservative positions are inverted. Center and right panels show cues of Left and Right identification provided by users in the social media profiles are coherent with the Liberal-Conservative dimension, and inverted for the Left-Right dimension.}
\label{fig:poland_gps_valid}
\end{figure}

It is noteworthy that, while the Left-Right dimension correctly separates users including Left or Right identification cues in the social media profiles, the order is inverted for Poland in Figure~\ref{fig:validation_epo_left_right}.
This inversion is anticipated by the literature in political sciences, and observed empirically in different studies. 
In most Eastern-European post-communist countries, the most structuring political dimension can be identified as spanning from Left-Authoritarian to Right-Libertarian positions, in contrast with Western-European political space structured between Left-Libertarian to Right-Authoritarian positions \cite{marks2006party,Bakker2012,malka2019cultural}.
For instance, the Law and Justice Party (\textit{Prawo i Sprawiedliwość}, or PiS), is recognized to be a Right-wing party that supports nonetheless several economically Left-leaning policies \cite{turner2017sage}.
Figure~\ref{fig:poland_gps_valid} provides further explanation into this inversion in our data. In the survey used for the identification of the latent space of the ideology scaling model (GPS), positions of available parties (three) along Left-Right and Liberal-Conservative are negatively correlated.
This means that, in our ideology scaling data of X users, Left-Right and Liberal-Conservative dimensions are related by a flip and translation operation.
Among the dimensions available in the GPS data, the Liberal-Conservative dimensions provides the second viable dimension structuring Left or Right identification.
An examination of substance of the questions presented to respondents of the survey surface a predominance of economic issues in Left-Right positioning (including, e.g., issues such as taxation, privatization and welfare state), and a predominance of cultural issues in Liberal Conservative positioning (including, e.g., mentions to order, tradition, abortion, and same-sex marriage).
Figure~\ref{fig:poland_gps_valid} further shows that order in self-identification into Left and Right groups is inverted (with respect to other countries) along the Left-Right dimension, while being coherently ordered in the Liberal-Conservative dimension.
Because of the sign invariance of the latent ideology dimension in the Community Notes modes (i.e., $\theta_n$ and $\theta_r$), this inversion has no consequences in our conclusions, as they hinge on the alignment and not the orientation of dimensions.

\subsection{Community Notes population with inferred ideology}
\label{subsubsec:epo_x_cn}

Community Notes contributors are not publicly linked to their X handles, preventing direct comparison between rater latent ideology $\theta_r$ and their political positions inferred from followed MPs. Instead, we compare the latent ideology of each note $\theta_n$ with the ideology of the X account that authored the original post to which the note is attached.\newline

To enable an analysis disaggregated by national setting, we segment posts by country. Specifically, we define a post and its associated Community Note as involved in a country's political debate when the author of the post follows at least three Members of Parliament from that country and the post is written in one of the country's national languages.\newline

This country segmentation methodology differs from the approach outlined in Section \ref{subsubsec:country_split_methods}, serving distinct purposes and presenting different limitations. This procedure assigns country classifications at the post level, which then extends to all associated notes, necessarily requiring post metadata. This creates a practical constraint: as of March 2025, 18.2\% of unique posts with Community Notes were unavailable for collection ---due to their deletion--- preventing author identification and country assignment. As detailed previously in Section \ref{subsec:data_collection_post}, posts associated with note with \textit{Helpful Status} are more likely to be deleted. This limitation explains our preference for note-specific country mapping whenever possible, particularly when characterizing global Community Notes usage patterns. Nevertheless, by necessity we employ post-specific country mapping to enable comparisons between X's learned ideological representations and established ideological scaling datasets.\newline

Among the 23 geographically diverse countries initially considered in this study, listed in Section \ref{subsec:epo_dataset}, we focus on the 13 countries that have at least one thousand unique X accounts receiving Community Notes with inferred ideology: \textit{United States, United Kingdom, Japan, Spain, France, Brazil, Canada, Germany, Argentina, Israel, Australia, Poland}, and  \textit{Mexico}. We report in Table~\ref{tab:epo_x_cn} the number of notes, posts, and X accounts with inferred ideological positions across all 13 countries.\newline

\begin{table}[h]
\centering
\begin{tabular}{|l|r|r|r|r|}
\hline
Country & Number of & Number of & Number of Posts & Total Number of Accounts \\
& Notes & Posts & Authors & with Inferred Ideology \\
\hline
\hline
Argentina & 12578 & 9430 & 1819 & 1410839 \\
Australia & 11927 & 9064 & 1345 & 554224 \\
Brazil & 28690 & 21188 & 2925 & 6142800 \\
Canada & 25797 & 18991 & 2603 & 786589 \\
France & 32152 & 25184 & 3763 & 979089 \\
Germany & 25712 & 19663 & 3298 & 1169464 \\
Israel & 14920 & 11194 & 1686 & 407378 \\
Japan & 27397 & 21960 & 4259 & 1489166 \\
Mexico & 7500 & 5643 & 1275 & 2120278 \\
Poland & 10928 & 8704 & 1536 & 732669 \\
Spain & 38289 & 28377 & 4095 & 1723836 \\
Uk & 59352 & 43655 & 6813 & 2007915 \\
Us & 166951 & 117235 & 14582 & 10927393 \\
\hline
\end{tabular}
\caption{Number of notes, posts, post authors, and the total number of X accounts with inferred ideological positions (including those without Community Notes) by country.}
\label{tab:epo_x_cn}
\end{table}

\begin{figure}[h]
\centering
\includegraphics[width=\textwidth]{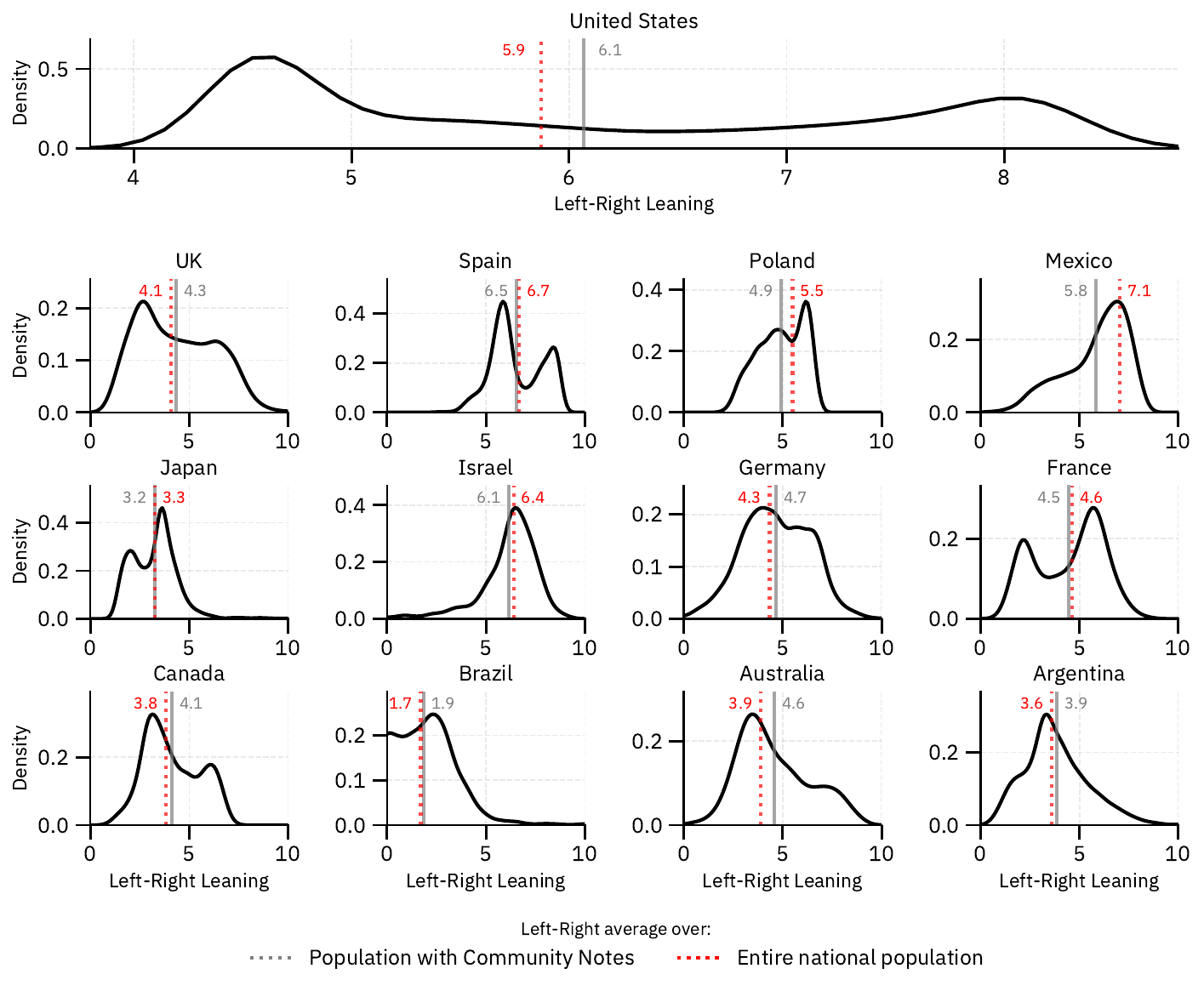}
\caption{Distribution of the inferred Left-Right leaning of X users who received at least one proposed Community Note. We report the average Left-Right leaning of the annotated population (plain grey line) and of the global population following 3 MPs (dotted red line).}
\label{fig:compare_left_right_pop_epo_cn}
\end{figure}

\begin{figure}[h]
\centering
\includegraphics[width=\textwidth]{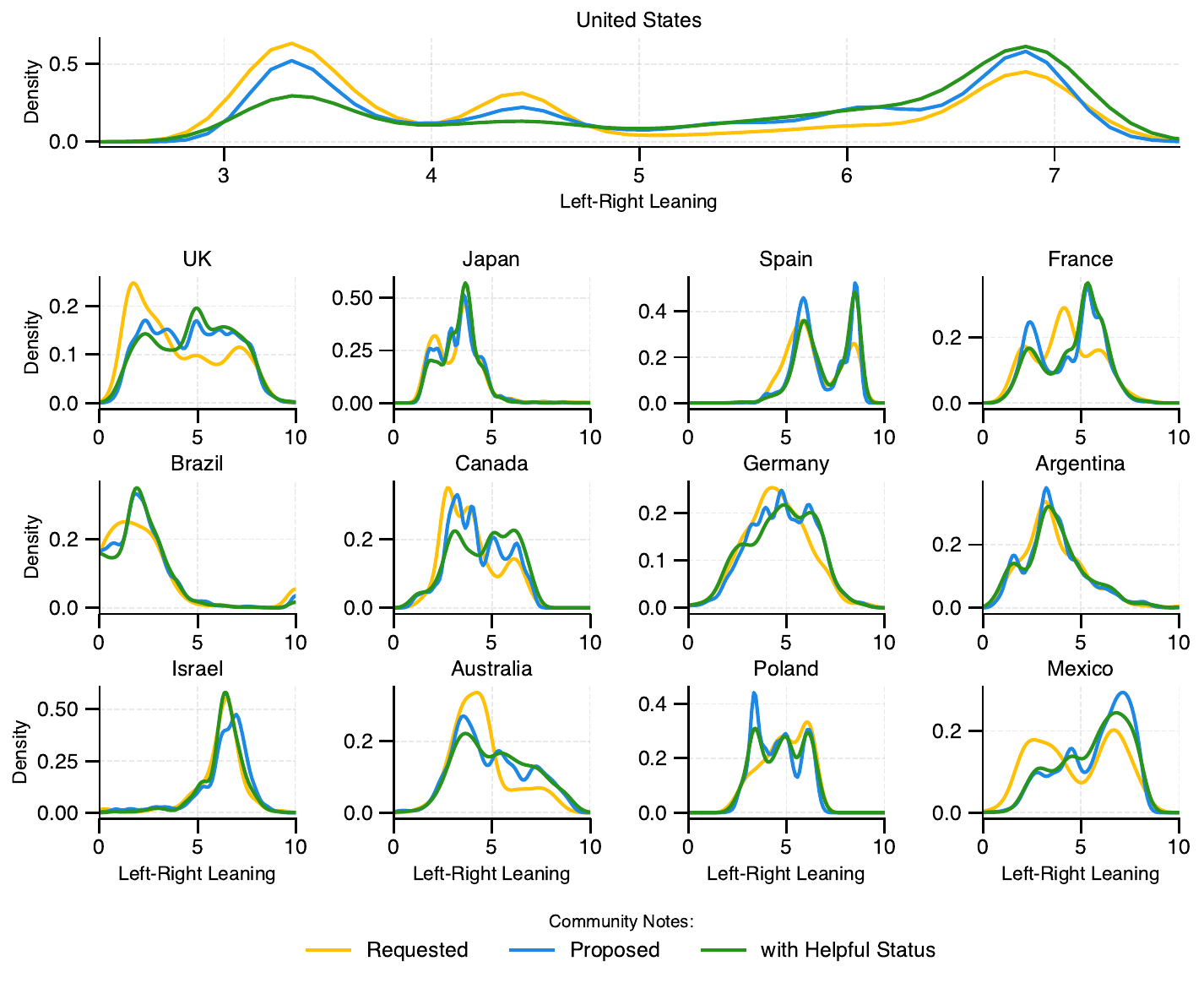}
\caption{Distribution of the inferred Left-Right ideology of X users having authored posts for which Community Notes i) were requested, ii) were proposed, and iii) reached \textit{Helpful} status.}
\label{fig:left_right_requested_proposed_helpful}
\end{figure}

Figure \ref{fig:compare_left_right_pop_epo_cn} displays the distribution of Left-Right ideology among X users associated with at least one Community Note (irrespective of ``helpfulness" status), overlaid with the average Left-Right leaning of the general population following MPs. This comparison shows general alignment between users associated with proposed Community Notes and the broader population of users following at least three national MPs. Additionally, 
across all 13 countries in our dataset, we report in Figure \ref{fig:left_right_requested_proposed_helpful} the distribution of Left-Right ideology among X accounts having authored posts for which Community Notes i) were requested, ii) were proposed, and iii) reached \textit{Helpful} status.

\section{Inferring the Community Notes model}
\label{sec:training_cn}

We now turn to the analysis of the parameters inferred by X's Community Notes algorithm from contributor ratings.


\subsection{Matrix factorization}

X's Community Notes system aims to surface notes deemed \textit{Helpful} by users across ``diverse viewpoints" \cite{CommunityNotesX}, implementing a bridging-based ranking approach \cite{ovadya2023}. This objective is operationalized through matrix factorization, which learns an opinion space, positioning raters and notes in an unsupervised manner, to explain observed ratings. The core matrix factorization underlying X's Community Notes system is formalized in Equation~$\ref{eq:matrix_facto}$ \cite{birdwatch}:

\begin{equation}\label{eq:matrix_facto}
\underbrace{\hat{\eta}_{rn}}_{\text{Predicted Rating}} = 
\underbrace{\beta_0}_{\text{Global Bias}} +
\underbrace{\beta_r}_{\substack{\text{Rater Bias} \\ \text{``leniency"}}}+
\underbrace{\beta_n}_{\substack{\text{Note Bias} \\ \text{``helpfulness"}}} +
\underbrace{\theta_n \cdot \theta_r}_{\substack{\text{Note-Rater} \\ \text{Ideological Alignment}}} \end{equation}

With the actual ratings $\eta_{un}$ numerically encoded as 1 for \textit{Helpful}, 0.5 for \textit{Somewhat Helpful} and 0 for \textit{Not Helpful}. The baseline ratings are captured by the bias terms, $\beta_0, ~\beta_r,~ \beta_n$. A high value of $\beta_r$ indicates an high propensity for the rater $r$ to rate notes as \textit{Helpful}, while a high $\beta_n$ reflects a note's tendency to receive \textit{Helpful} ratings across the rater base. $\theta_n$ and $\theta_r$ encode notes and raters latent ideology.\newline

The product $\theta_r \cdot \theta_n$ quantifies ideology alignment, with the rational that when raters share the note's ideology, they are more likely to rate it as \textit{Helpful}. When $|\theta_n|$ is large, voting behaviors on that note varies a lot in function of raters' latent ideology $\theta_r$. To maximize this effect, X applies, during training, stronger regularization on bias terms $\beta_0, ~\beta_r,~ \beta_n$ compared to latent ideology terms $\theta_r, ~\theta_n$ \cite{birdwatch, TwitterGithubMain}.\newline

Through this design, when ratings on a note depend on raters' latent ideology (e.g., raters with $\theta_r > 0$ rate the note \textit{Helpful} while those with $\theta_r < 0$ rate it \textit{Not Helpful}), the note latent ideology $\theta_n$ will align with raters' (e.g. $\theta_n>0$). Correspondingly, the bias term $\beta_n$ will be small due to regularization.\newline
Conversely, when consensus emerges (i.e. raters with different $\theta_r$ signs agree), $|\theta_n|$ will shrink to minimize rater leaning impact on predicted ratings $\hat{\eta}_{rn}$, while $|\beta_n|$ will be high: if raters agree the note is \textit{Helpful} then $\beta_n > 0$, if they agree it is \textit{Not Helpful} then $\beta_n < 0$.\newline

Finally, X's Community Notes system assign \textit{Helpful Status}/\textit{Not Helpful Status} to notes when their bias term $\beta_n$ exceeds specified thresholds. For intermediate cases where rater leaning $\theta_r$ effectively predicts rating patterns, X's system withholds classification pending additional diverse ratings, so-called \textit{Needs More Ratings Status}.


\subsection{Computational implementation} 
\label{subsec:training_output}

While X's Community Notes algorithmic implementation is open-source \cite{TwitterGithubMain}, it incorporates multiple heuristics and experimental features that increase model complexity. For simplicity, we solely implement the core matrix factorization from Equation~$\ref{eq:matrix_facto}$. We will validate inferred values of the Community Notes models by comparing the outcomes produced by the model we train, with those observed in the data disclosed by X.\newline

Following X's approach \cite{birdwatch, TwitterGithubMain}, we learn biases $\beta_0,~\beta_r,~\beta_n$ and latent ideologies $\theta_r,~\theta_n$ by minimizing the loss function in Equation~\ref{eq:loss}, with regularization five times higher on biases than on latent ideologies:\newline

\begin{equation}\label{eq:loss}
\min_{\{\beta,\theta\}} \sum_{\eta_{rn}} \underbrace{(\eta_{rn} - \hat{\eta}_{rn})^2}_{\text{Reconstruction Error}} + \quad  \underbrace{5\lambda(\beta_0^2 + \beta_r^2 + \beta_n^2) +  \lambda(\theta_r^2 + \theta_n^2)}_{\text{Regularization}}.
\end{equation}

In their publication, X's researchers justify higher regularization on biases compared to latent ideologies to minimize ``risks to [the] community and reputation from increasing visibility of low quality notes''  \cite{birdwatch}. Indeed, by increasing regularization on note and rater biases ($\beta_n$, $\beta_r$), this encourages latent ideologies ($\theta_n$, $\theta_r$) to explain more of the rating patterns. As such, the bias terms, particularly $\beta_n$, can be interpreted as the idiosyncratic ``quality'' of the note—-i.e., the propensity of raters to rate it ``helpful'' beyond their ideological preferences. However, we should emphasize that the numerical factor of 5 in Equation~\eqref{eq:loss} is arbitrarily set, and tweaking it can directly impact the fraction of notes perceived to ``foster consensus''.

Contrary to X's production system that processes ratings sequentially, we aggregate all ratings until March 1, 2025. This larger training dataset requires adjusting the regularization magnitude $\lambda$ compared to X's production system.\newline

\begin{figure}[h]
\centering
\includegraphics[width=\textwidth]{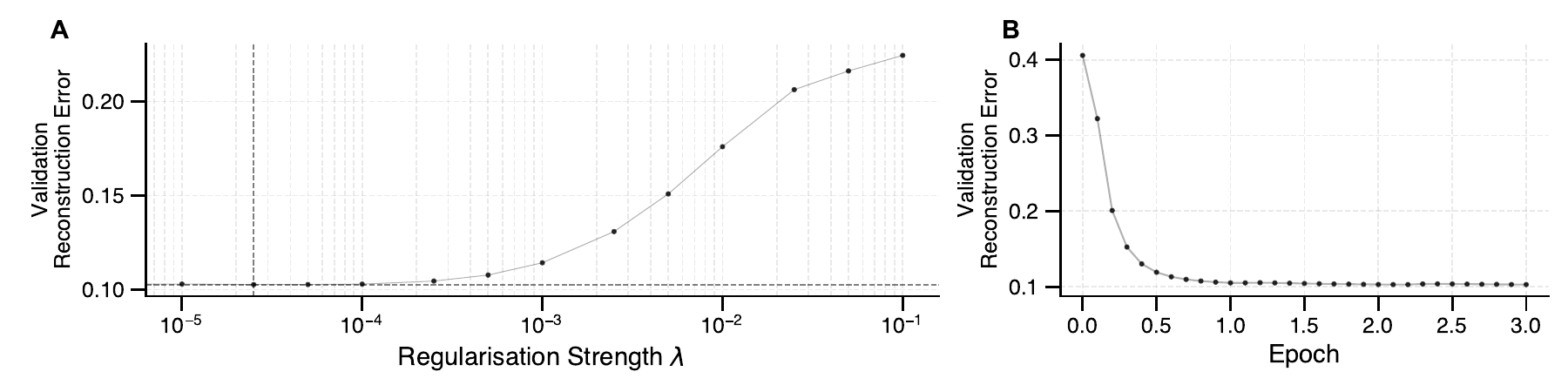}
\caption{Hyperparameter tuning of the regularization strength $\lambda$ \textbf{(A)} and number of training epoch \textbf{(B)} evaluated using reconstruction error on a held-out dataset.}
\label{fig:hypertuning}
\end{figure}

The validation loss plateaus quickly during the training, to ensure convergence and model stability, we performed three epochs. We tuned the hyperparameter $\lambda$, while preserving X's 5:1 ratio between biases and latent ideologies regularization, by measuring the reconstruction error $|\eta_{rn} - \hat{\eta}_{rn}|$ on 90/10 train/test splits, after three epochs. In Figure~\ref{fig:hypertuning}, the optimal value of $\lambda = 2.5 \times 10^{-5}$ is identified.\newline

\begin{figure}[h]
\centering
\includegraphics[width=\textwidth]{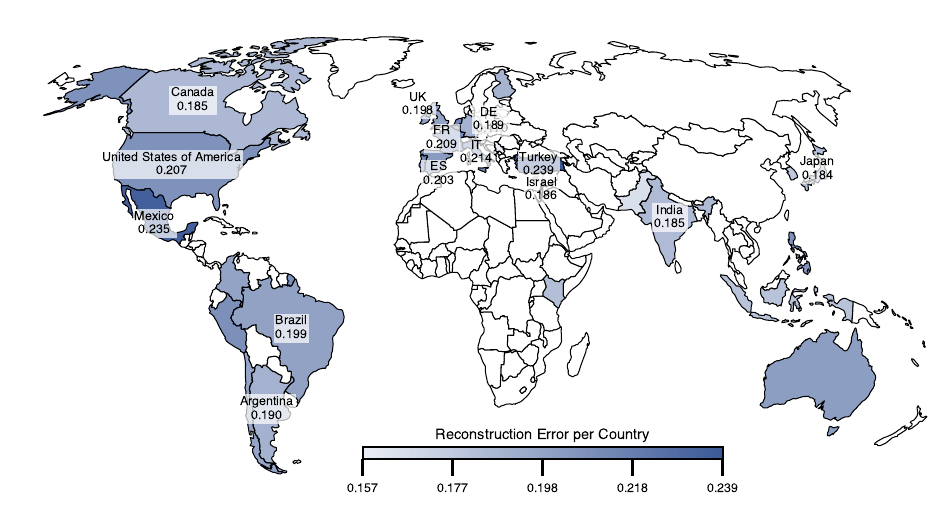}
\caption{Average reconstruction error $|\eta_{un} - \hat{\eta}_{un}|$, between the predicted rating and raters' effective ratings, per country (see  S.\ref{subsubsec:country_split_methods}).}
\label{fig:reconstruction_error}
\end{figure}

Informed by the hyperparameter tuning, we then trained the model for three epochs with a regularization strength of $\lambda = 2.5 \times 10^{-5}$ at a learning rate of $2.5\times10^{-3}$. After training, the average reconstruction error $|\eta_{un} - \hat{\eta}_{un}|$ between the predicted rating $\hat{\eta}_{rn} = \beta_0 + \beta_n + \beta_r + \theta_n \cdot \theta_r$ and raters' effective ratings $\eta_{un}$ averages at $0.204$. Figure \ref{fig:reconstruction_error} displays the average reconstruction error per country (segmenting according to the methodology in \ref{subsubsec:country_split_methods}), showcasing comparable performance across countries.

\subsection{Validating the parameters of the model} 
\label{subsec:training_results}

\begin{figure}[h!]
\centering
\includegraphics[width=\textwidth]{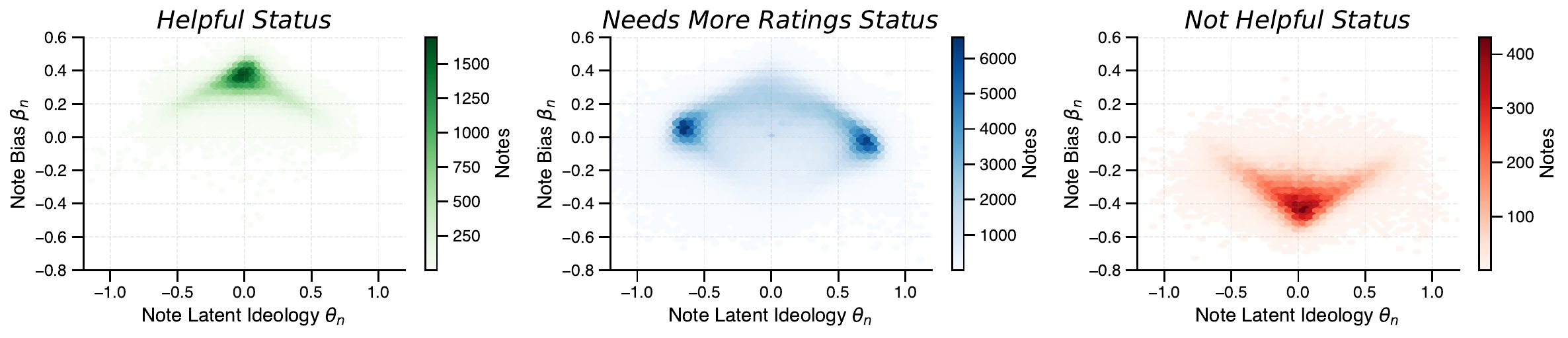}
\caption{Distribution of notes bias $\beta_n$ and latent ideology $\theta_n$, segmented by status assigned by X: \textit{Helpful Status}, \textit{Needs More Ratings Status}, or \textit{Not Helpful Status}.}

\label{fig:comparaison_w_X}
\end{figure}

The distribution of notes and rater biases and latent ideology inferred from Community Notes ratings, displayed in Figure~\ref{fig:comparaison_w_X}, mirrors those reported in X's article \cite{birdwatch} and aligns with expectations from X's Community Notes source code \cite{TwitterGithubMain}.\newline

\begin{figure}[h!]
\centering
\includegraphics[width=\textwidth]{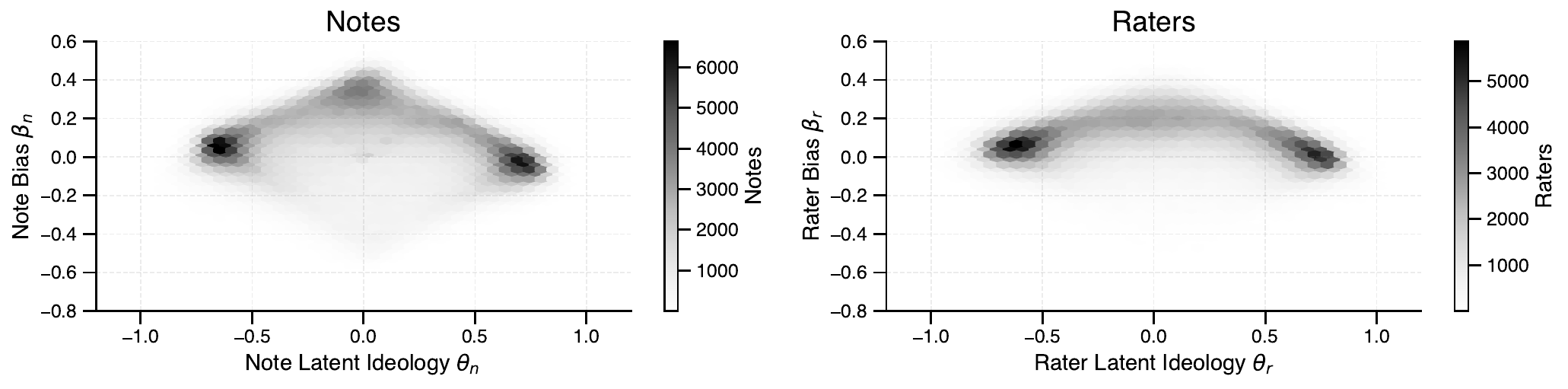}
\caption{Distribution of notes (raters) bias $\beta_n$ ($\beta_r$) and latent ideology $\theta_n$ ($\theta_r$).}
\label{fig:distrib_embeddings}
\end{figure}

We segment in Figure~\ref{fig:comparaison_w_X} notes by X's Community Notes outcomes: \textit{Helpful Status}, \textit{Needs More Ratings Status}, or \textit{Not Helpful Status}. Conform to expectation notes with \textit{Helpful Status} have high positive bias $\beta_n$, those with \textit{Not Helpful Status} have large negative bias, and notes with \textit{Needs More Ratings Status} have high latent ideology $|\theta_n|$. Quantitatively, 90\% of notes with \textit{Helpful Status} have bias terms $\beta_n>0.180$, 90\% of notes with \textit{Not Helpful Status} have $\theta_n<-0.159$, and 69\% of notes with \textit{Needs More Ratings Status} fall between these thresholds. The inferred note bias, $\beta_n$ allows prediction of whether X assigned a note \textit{Helpful Status} with One-vs-the-rest AUC of 0.92 and \textit{Not Helpful Status} with One-vs-the-rest AUC of 0.97. \newline

The differences between the numerical thresholds and those in \cite{birdwatch} stem from the difference in dataset size. While we learn note and rater representations across all available ratings, X's Community Notes production system aims to surface, in a timely fashion, \textit{Helpful} notes to users by considering recent ratings. Additionally, X implements heuristics constraining note status changes, which we did not implement, focusing solely on the core matrix factorization of Equation~\ref{eq:matrix_facto}.


\section{Analysis of the latent ideology dimension learned by Community Notes}
\label{sec:learned_viewpoints}

The matrix factorization used by X's Community Notes system aims to learn a continuous opinion space from rating patterns \cite{birdwatch}. Given that Community Notes predominantly address political content (see Section~\ref{subsec:post_topic}), this learned opinion space is expected to capture raters' and notes' political leanings and the surfaced notes to be rated as \textit{Helpful} by users across the political spectrum. Through surveys and self-reported political affiliations, X's researchers validated this objective during the pilot project \cite{birdwatch}. Allen et al. \cite{Allen2022} further substantiated ---during the pilot phase with $\sim5000$ notes--- these findings by showing that users political partisanship strongly predicts their ratings.\newline

Despite growing literature on crowd-sourced fact-checking, the opinion space learned by X's Community Notes system ---which drives note selection--- remains unexamined. This section characterizes this learned space and evaluates its alignment with Twitter users' political leanings. To this end, we first present in Section \ref{subsec:epo_dataset} the ideological scaling data leveraged in this analysis, and then compare it with the learned representations in Sections \ref{subsec:op_1d} \& \ref{subsec:op_multi}.\newline


\subsection{Left-Right dimension}
\label{subsec:op_1d}

To foster participation in Community Notes, contributors' profiles are not publicly linked to their X accounts, preventing direct comparison between rater latent ideology $\theta_r$ and their political positions inferred from followed MPs. Consequently, we are left to compare the latent ideology of Community Notes $\theta_n$ with the political leaning of X accounts that authored the posts to which notes are attached, restricting our analysis to non-deleted posts.\newline

As a first qualitative analysis, we report the position of popular accounts in five countries over this latent ideology scale. Specifically, for each country, we compute the average latent ideology $\theta_n$ of community notes associated with each X account's posts. We then divide the range of average $\theta_n$ values into 10 bins and manually inspect accounts within each bin. To protect privacy while providing illustrative examples, we focus on accounts that have received at least 10 community notes (irrespective of status), have the highest follower counts within their respective bins, and whose political leanings are manifestly public. To validate the reliability of the average latent ideology $\theta_n$, one can observe that for accounts with at least 10 notes, 83.6\% of those notes have the same latent ideological sign ($\text{sign}(\theta_n)$).

\begin{figure}[h!]
\centering
\includegraphics[width=\textwidth]{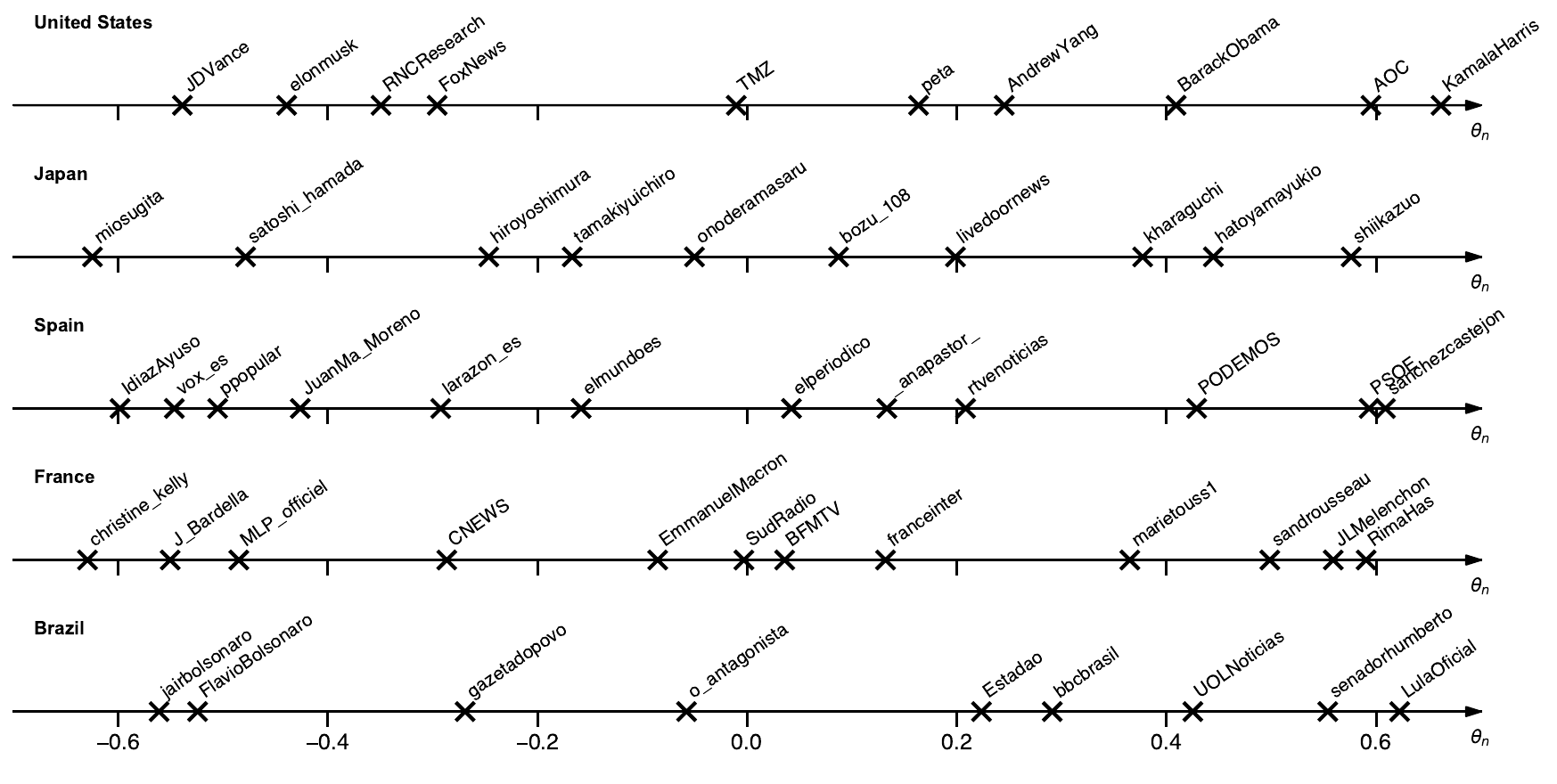}
\caption{Average positions of notes added to posts of political figures and news outlets per country. The position of accounts depend on the average latent ideology of notes $\theta_n$ in the Community Notes models in the United States, Brazil, Japan, France and Spain, given to them.}
\label{fig:scaling_usrs}
\end{figure}

We observe in Figure~\ref{fig:scaling_usrs} that Community Notes associated with posts published by Right-leaning accounts tend to have negative latent ideology ($\theta_n<0$), while those associated with Left-leaning authored posts have positive latent ideology ($\theta_n>0$). For instance, notes added to posts by U.S. Vice President J. D. Vance, Brazil's former President Jair Bolsonaro, France's far-Right leader Jordan Bardella, and Spain's far-Right party Vox have negative latent ideology $\theta_n$.
Driven by homophily, this indicates that Left-leaning Community Notes contributors have negative latent ideology $\theta_r<0$, so that they rate \textit{Helpful} notes under Right-leaning posts but \textit{Not Helpful} notes under Left-leaning posts, and conversely Right-leaning contributors have positive latent ideology $\theta_r>0$.\newline

The second observation is that, while different countries could in principle exhibit different sign orientations in how $\theta_r$ and $\theta_n$ map onto Left-Right, all countries in our study share the same orientation: $\theta_n<0$ under post of Right-leaning users and $\theta_n>0$ under post of Left-leaning users. Importantly, this alignment emerges naturally from the non-supervised inference procedure without any explicit constraints. This observation can be explained by cross-country rating patterns. Since the Community Notes algorithm considers ratings as a whole rather than segmenting them by country, raters' latent ideological positions $\theta_r$ must align across national settings to best explain their rating behavior.

\begin{figure}[h]
\centering
\includegraphics[width=\textwidth]{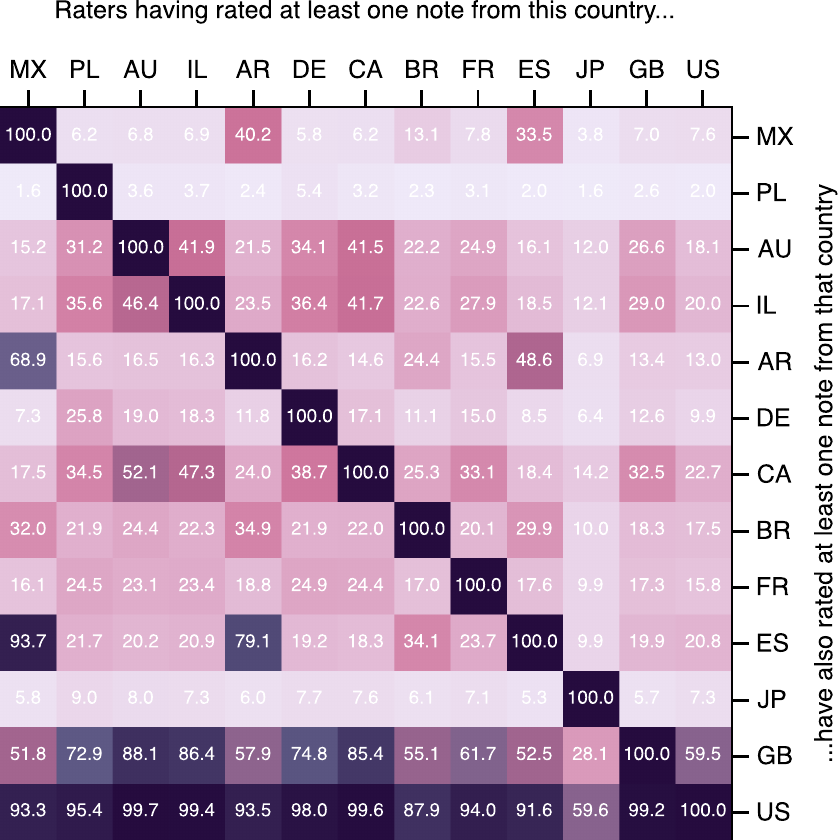}
\caption{Cross-country rating patterns among Community Notes contributors. The figure shows the percentage of users who rated at least one note from a given country (columns) and also rated at least one note from another country (rows). For instance, 59.6\% of users who rated at least one Japanese Community Note also rated at least one US note, while only 7.3\% of users who rated at least one US note also rated at least one Japanese note.}
\label{fig:cross_country_ratings}
\end{figure}

To corroborate this interpretation, we display cross-country rating patterns in Figure \ref{fig:cross_country_ratings}. Specifically, we show the fraction of users who rated at least one note from a given country and also rated at least one note from another country. We observe that an overwhelming proportion of contributors rated US notes in addition to notes from their own countries.

It should be emphasized that the predicted ratings $\hat{\eta}_{rn}$ in Equation \ref{eq:matrix_facto} are invariant to a joint sign inversion of latent ideologies: $(\theta_n, \theta_r) \mapsto (-\theta_n, -\theta_r)$. As such, while we report Left-leaning contributors as having negative latent ideology ($\theta_r<0$) and Right-leaning contributors as having positive latent ideology ($\theta_r>0$), the exact opposite assignment could hold without any impact on the model's operation. All subsequent discussion holds without loss of generality under either sign convention.

\subsubsection{United States}

\paragraph{Alignment with the Left-Right dimension in political survey data.}
\label{paragraph:alignment}

In the United States, 95.0\% of Community Notes added to posts authored by Republican members of 118th United States Congress have negative latent ideology $\theta_n<0$, while 94.6\% of those associated with Democratic members of Congress have positive latent ideology $\theta_n>0$ (excluding ``Note Not Needed" notes).\newline

Similarly, considering accounts following at least 10 US MPs, 82.2\% of notes given to posts published by accounts following a majority of Democratic MPs have a positive latent ideology ($\theta_n>0$); while 85.1\% of notes given to posts published by accounts following a majority of Republican MPs have a negative latent ideology ($\theta_n<0$).
\newline

Relying on the continuous Left-Right dimension inferred in Section~\ref{sec:scaling_si} we can further quantify this alignment. When framed as a prediction task, the position of a user on the Left-Right political spectrum in the United States predicts the sign of $\theta_n$ (the latent ideology of the Community Notes associated with their post) with an AUC-ROC of $0.822 \pm 0.009$ (standard deviations estimated through 10-fold cross-validation).\newline

\begin{figure}[h]
\centering
\includegraphics[width=\textwidth]{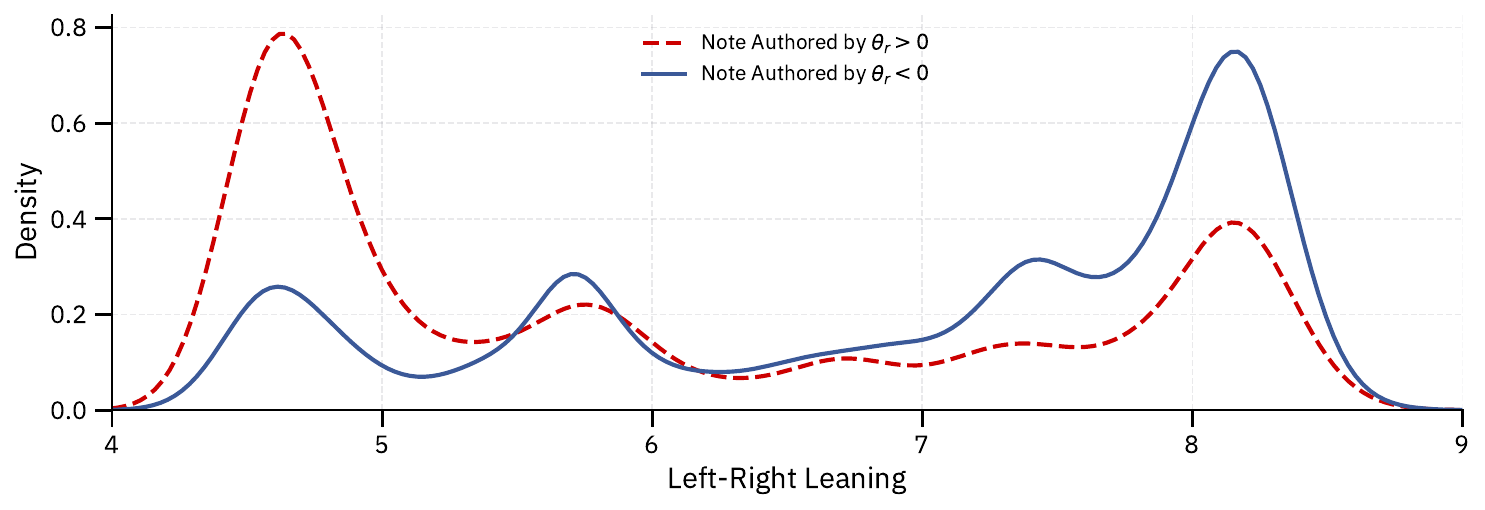}
\caption{Distribution of Left-Right leaning of X accounts having authored posts to which Community Notes were written by contributors with $\theta_r<0$ (plain blue) and $\theta_r>0$ (dashed red).}
\label{fig:us_note_authors}
\end{figure}

To corroborate the Left-Right alignment of note latent ideology, we examine a signal not leveraged in the inference of such ideology: note authorship. Indeed, X Community Notes system only considers ratings, not which contributor wrote the community notes being rated. Therefore, examining authorship pattern of contributors with $\theta_r>0$ or $\theta_r<0$ allows us to further characterize the underlying dynamics. We observe in Figure~\ref{fig:us_note_authors} that users with $\theta_r<0$, who rate \textit{Helpful} notes under Right-leaning accounts, are also the ones writing notes under Right-leaning accounts; aligning with the observations made by Allen et al. during the pilot program \cite{Allen2021}. Conversely, users with $\theta_r>0$, who rate \textit{Helpful} notes under Left-leaning accounts, are also the ones writing notes under Left-leaning accounts.

\paragraph{Media Bias.}
\label{paragraph:media_bias}

The alignment between the ideological space learned by X's Community Notes system and established political dimensions can be further substantiated relying on a signal independent from poster following patterns, namely, the use of news media outlet as sources in Community Notes.\newline

\begin{figure}[h]
\centering
\includegraphics[width=\textwidth]{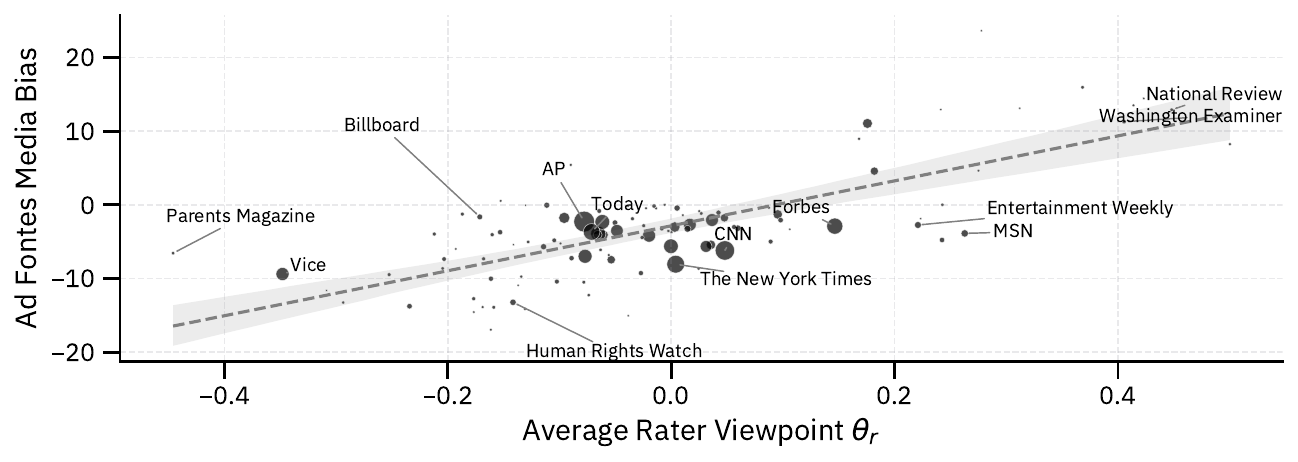}
\caption{Ad Fontes Media \cite{Otero2025} Left-Right bias for news media outlets as function of the average latent ideology $\theta_r$ of Community Note contributors who authored notes using those domains as sources. Negative values indicate Left-leaning bias, positive values indicate Right-leaning bias. The size of each node is proportional to the number of Community Notes referencing each domain.}
\label{fig:media_bias}
\end{figure}

Specifically, we consider the 109 US news media outlets, used in at least 100 Community Notes, whose ideological skew was also evaluated by Ad Fontes Media \cite{Otero2025}. We subsequently compare the Left-Right bias of the news outlets, as assessed by Ad Fontes Media, and the average latent ideology $\theta_r$ of the Community Notes contributors who authored notes using these media as sources. As displayed in Figure~\ref{fig:media_bias}, we observe a strong Pearson correlation between these quantities: $\rho=0.744$ $p<10^{-4}$, with a $[0.65, ~0.82]$ Confidence Interval at 95\% for correlation, estimated via Fisher transformation.\newline

Importantly, this analysis relies on sources used in Community Notes ---data disclosed by X--- without requiring additional data collection compared to post-based metrics. This approach avoids biases from users removing posts after the note reached \textit{Helpful Status} notes, providing complementary robustness to the previously documented alignment.

\paragraph{Fact-Checks.}
\label{paragraph:fact_checks}

\begin{figure}[h!]
\centering
\includegraphics[width=\textwidth]{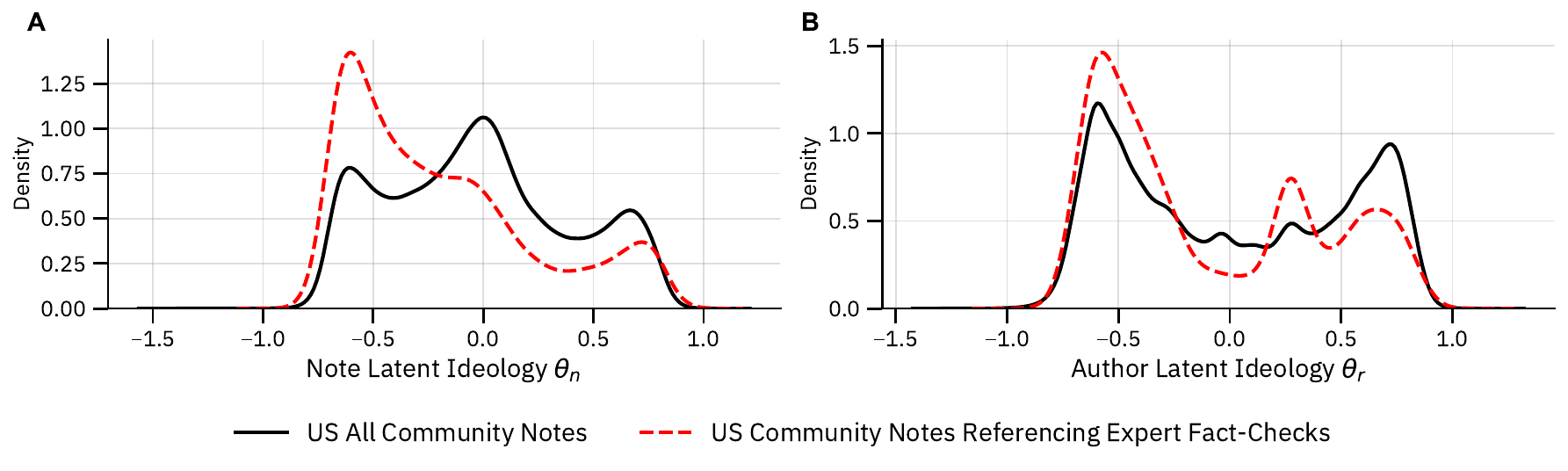}
\caption{Distribution of note latent ideology $\theta_n$ (\textbf{A}) and note author latent ideology $\theta_r$ (\textbf{B}) for notes containing expert-authored fact-checks.}
\label{fig:fact_check}
\end{figure}

We inspect the latent ideology distribution of Community Notes referencing expert fact-checking authored by 160 organizations; see curation in Section \ref{subsubsec:fact_checks}. Figure \ref{fig:fact_check} displays the distribution of note latent ideology $\theta_n$ and of contributors who authored the Community Notes $\theta_r$. We observe that Community Notes containing expert fact-checks span the full ideological axis. Yet, a majority of notes are authored by Left-leaning contributors, specifically $60.3\%~ [58.0, ~62.8]$ of notes are authored by contributors with $\theta_r<0$; 95\% confidence interval over 100 bootstraps with replacement over the notes and the set of fact-checking organizations considered in the analysis.\newline

Notes incorporating expert fact-checking sources reach \textit{Helpful Status} at significantly higher rates ($21.3\%~[18.3\%,~24.1\%]$) than average, with this effect more pronounced for notes authored by Right-leaning contributors ($\theta_r>0$): $26.9\%~[23.3\%,~30.8\%]$ versus $17.6\%~[14.0\%,~20.7\%]$  for Left-leaning contributors ($\theta_r<0$); 95\% confidence interval over 100 bootstraps with replacement over the notes and the set of fact-checking organizations considered in the analysis.

\paragraph{Deletion Rate.}
\label{paragraph:deletion_rate}

In the main text, we reported that notes with \textit{Helpful Status} appeared more frequently on posts published by Right-leaning X accounts compared to Left-leaning accounts. Specifically, in the United States, 8.3\% of annotated posts authored by Left-leaning users receive notes that achieve \textit{Helpful Status}, compared to 12.9\% for Right-leaning users, similar to what reported Renault et al. \cite{Renault2025}. However, this observation requires careful interpretation due to a critical methodological limitation: we can only infer the political leaning of an account if the X post was available at the time of data collection. When posts are deleted, the identity of the account that authored the post discussed in the note is not disclosed.

The observed asymmetry may be explained, at least partially, by differences in post deletion behavior. If Left-leaning users delete their posts after being flagged as ``misinforming or potentially misleading" more frequently than Right-leaning users, an observer would perceive that Left-leaning users are less frequently flagged for misinformation.

We illustrate this bias by constructing a fictitious example that matches our own observations. Consider a scenario where the true rate of posts receiving notes with \textit{Helpful Status} ($f_{\text{helpful}}$) is 13\% for both Left and Right-leaning users, and both delete their non-misleading posts at the same rate ($d_{\text{not helpful}}=15\%$). However, suppose Left-leaning users delete $d^{\text{Left}}_{\text{helpful}}=48.5\%$ of their posts associated with notes having \textit{Helpful Status}, while Right-leaning users delete only $d^{\text{Right}}_{\text{helpful}}=16\%$ of such posts.

Under these conditions, the observed proportion of notes having \textit{Helpful Status} associated with Left-leaning accounts would be:
\begin{align*}
\text{Observed}^{\text{Left}}_{\text{helpful}}  & = \frac{f_{\text{helpful}} \times (1- d^{\text{Left}}_{\text{helpful}})}{f_{\text{helpful}} \times (1- d^{\text{Left}}_{\text{helpful}}) + (1 - f_{\text{helpful}}) \times (1-d_{\text{not helpful}})} \\
                                 & = \frac{0.13 \times (1- 0.485)}{0.13 \times (1- 0.485) + (1-0.13) \times (1-0.15)} \\
                                 & = 8.3\%
\end{align*}
    
\noindent{}Similarly, for Right-leaning accounts:
\begin{align*}
\text{Observed}^{\text{Right}}_{\text{helpful}} & = \frac{0.13 \times (1- 0.16)}{0.13 \times (1- 0.16) + (1-0.13) \times (1-0.15)} \\
                         & = 12.9\%
\end{align*}

These hypothetical values align with our empirical observations, suggesting that differential deletion rates \textit{could} explain the apparent asymmetry.\newline

To investigate this possibility, we leveraged the ideological positioning inferred by X's Community Notes algorithm. All notes, including those associated with (now) deleted posts, are positioned on the same ideological axis, and we previously showed that notes with $\theta_n<0$ are generally associated with posts authored by Right-leaning accounts, while notes with $\theta_n>0$ are associated with posts authored by Left-leaning accounts.

Segmenting notes based on their inferred ideology $\theta_n$ as a proxy for the post authors' political leaning, we observe distinct deletion patterns:

\noindent{}Posts (presumably written by Right-leaning accounts) associated with Community Notes with $\theta_n<0$ :
\begin{itemize}
    \item having reached \textit{Helpful Status}, were deleted in 35.9\% of cases
    \item having not reached \textit{Helpful Status}, were deleted in 16.9\% of cases
\end{itemize}

\noindent{}Posts (presumably written by Left-leaning accounts) associated with Community Notes with $\theta_n>0$ :
\begin{itemize}
    \item having reached \textit{Helpful Status}, were deleted in 45.2\% of cases
    \item having not reached \textit{Helpful Status}, were deleted in 19.3\% of cases
\end{itemize}

A permutation test over note latent ideology signs ($sign(\theta_n)$) confirmed the statistical significance of these differences.

Under these deletion rates, Left-leaning accounts would appear to have notes with \textit{Helpful Status} on 8.3\% of their retrievable posts if the true rate is 11.8\%, while Right-leaning accounts would appear to have such notes on 12.9\% of their retrievable posts if the true rate is 15.6\%.

Importantly, we cannot extrapolate these deletion rates to the general population. As described by Wojcieszak et al. \cite{Wojcieszak2022}, ``most users do not follow political elites on X; those who do show overwhelming preferences for ideological congruity.'' Therefore, Left-leaning accounts heavily invested in political discussions and following multiple MPs (and thus positioned on our Left-Right scale) \textit{may} delete posts flagged as misleading at rates even higher than the estimate of 45.2\%.

Finally, it should be noted that X's Community Notes algorithm is blind to the identity of both the author of the post and that of the contributor having proposed the note, inferring an ideological space for notes and raters solely based on rating patterns. Because of the mathematical symmetry between ($\theta_n$, $\theta_r$) and (-$\theta_n$, -$\theta_r$) in Equation \ref{eq:matrix_facto}, there is no \textit{a priori} reason to assume different ``helpfulness'' rates based on latent ideology. In fact, the fraction of Community Notes with $\theta_n<0$ reaching \textit{Helpful Status} does not statistically differ from those with $\theta_n>0$, and the same holds for notes authored by contributors with $\theta_r<0$ and $\theta_r>0$.

Considering the evidence at hand, rather than interpreting the observed differences as definitive evidence that Right-leaning users are flagged more often than Left-leaning users for sharing misinformation, as did Renault et al. \cite{Renault2025}, we should carefully understand these differences as composite measures reflecting both the true rates of agreement leading to \textit{Helpful Status} of notes and differential deletion behaviors across the political spectrum. Based on collected public information, it is not possible to quantify with certainty any potential partisan asymmetry in misinformation sharing on X as flagged by Community Notes. Further experiments collecting posts as they receive community notes, prior to reaching \textit{Helpful Status}, would allow settling this alternative explanation.


\subsection{Anti-Elite dimension}
\label{subsec:op_multi}

We extend the previous analysis to all 13 countries for which we  possess ideological scaling data with significant overlap with submitted Community Notes. 
We include in our analysis an additional dimension identified in the literature as relevant for the spread of misinformation \cite{ramaciotti2023geometry}.
Through the AUC-ROC metric, we quantify the predictive power of various political dimensions in determining the sign of a Community Notes latent ideology $\theta_n$ based on the position of the post author along three dimensions:

\begin{itemize}
    \item $\delta_1$, the most structuring ideological dimension and which best explains MPs following patterns;
    \item Left-Right dimension of the Global Party Survey;
    \item Anti-Elite dimension of the Global Party Survey.
\end{itemize}

As shown in Table~\ref{tab:alignement_metrics}, in each country, the latent ideology $\theta_n$ learned by X's Community Notes system closely aligns with the national structuring political dimension $\delta_1$, and, to varying extents, with established political dimensions, i.e., Left-Right, and Anti-Elite.

For instance, in France, the ideological dimension learned by X's Community Notes system aligns strongly with users' Left-Right views, with an AUC-ROC of $0.789 \pm 0.023$, but weakly with users' stance on Anti-Elite (AUC-ROC of $0.590 \pm 0.045$). Conversely, in Argentina, Left-Right leaning poorly predict note latent ideology sign, while an X account's position on the Anti-Elite axis predicts the latent ideology sign of associated Community Notes with a high AUC-ROC of $0.859 \pm 0.016$.

\begin{table}[h]
\centering
\resizebox{\textwidth}{!}{
\begin{tabular}{|l|c|c|c|c|}
\hline
Country & AUC $\delta_1$ & AUC Left-Right & AUC AntiElite & AUC Left-Right+AntiElite \\
\hline
\hline
Argentina & 0.863 ± 0.025 & 0.682 ± 0.031 & 0.860 ± 0.024 & 0.863 ± 0.026 \\
Australia & 0.825 ± 0.036 & 0.809 ± 0.039 & 0.806 ± 0.033 & 0.824 ± 0.034 \\
Brazil & 0.831 ± 0.027 & 0.726 ± 0.032 & 0.552 ± 0.029 & 0.767 ± 0.035 \\
Canada & 0.836 ± 0.033 & 0.839 ± 0.027 & 0.832 ± 0.025 & 0.842 ± 0.025 \\
France & 0.775 ± 0.020 & 0.787 ± 0.016 & 0.589 ± 0.031 & 0.816 ± 0.019 \\
Germany & 0.840 ± 0.019 & 0.817 ± 0.020 & 0.772 ± 0.027 & 0.852 ± 0.021 \\
Israel & 0.729 ± 0.037 & 0.624 ± 0.037 & 0.719 ± 0.047 & 0.718 ± 0.041 \\
Japan & 0.731 ± 0.024 & 0.760 ± 0.020 & 0.715 ± 0.023 & 0.757 ± 0.020 \\
Mexico & 0.818 ± 0.031 & 0.758 ± 0.035 & 0.814 ± 0.030 & 0.822 ± 0.028 \\
Poland & 0.850 ± 0.041 & 0.838 ± 0.039 & 0.846 ± 0.042 & 0.850 ± 0.040 \\
Spain & 0.842 ± 0.023 & 0.850 ± 0.018 & 0.787 ± 0.021 & 0.855 ± 0.019 \\
UK & 0.780 ± 0.015 & 0.793 ± 0.015 & 0.735 ± 0.017 & 0.793 ± 0.015 \\
US & 0.822 ± 0.009 & 0.822 ± 0.008 & 0.822 ± 0.008 & 0.822 ± 0.008 \\
\hline
\end{tabular}
}
\caption{AUC-ROC in predicting the majority latent ideology sign ($\text{sign}(\theta_n)$) of notes associated with X accounts based on their position along the $\delta_1$ axis, Left-Right axis, Anti-Elite axis, and both Left-Right and Anti-Elite axes combined. Standard deviations estimated through 10-fold cross-validation.}
\label{tab:alignement_metrics}
\end{table}

\begin{figure}[h]
\centering
\includegraphics[width=\textwidth]{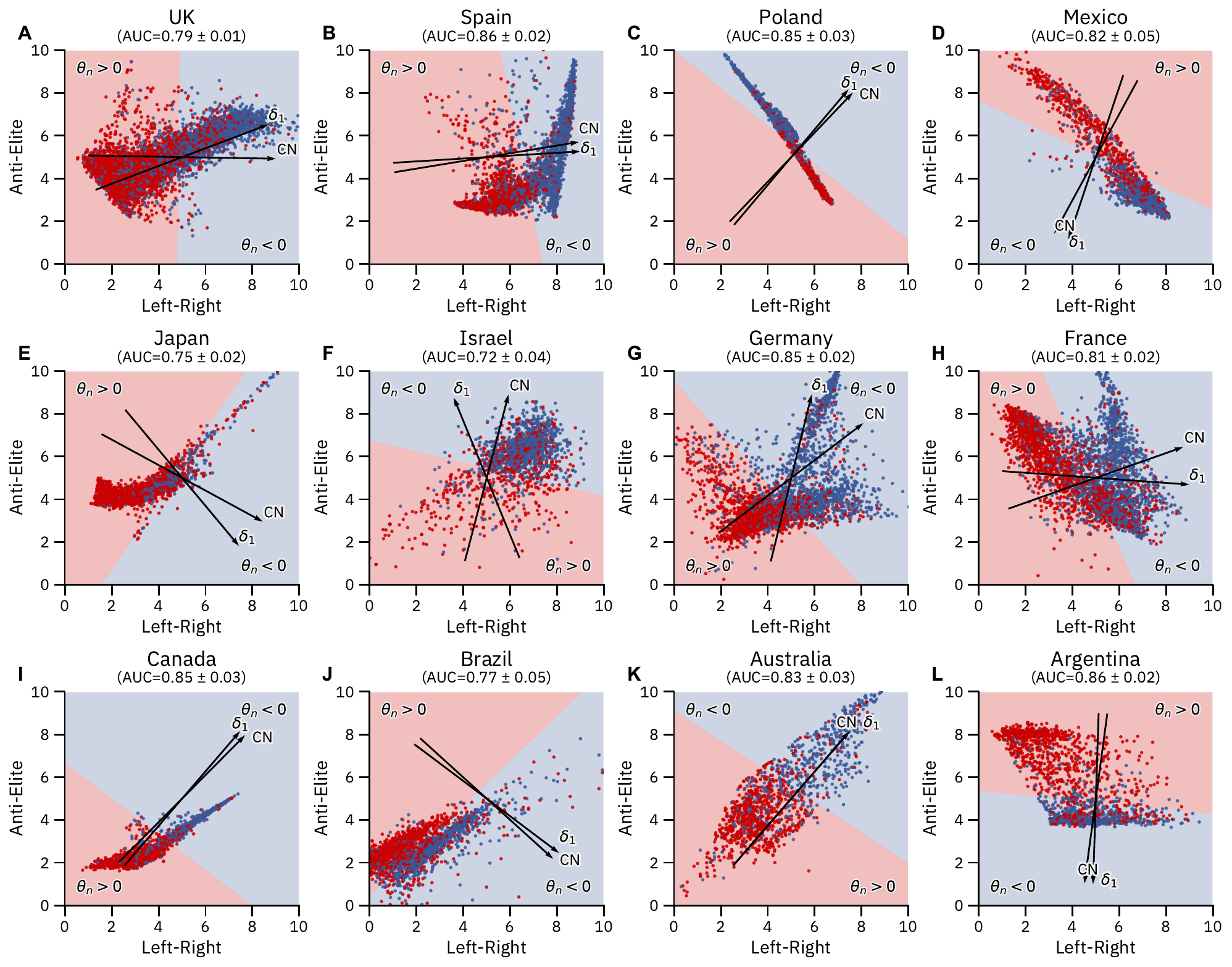}
\caption{X accounts positioned in the (Left-Right; Anti-Elite) 2D plane, segmented by the majority sign of the latent ideology $\theta_n$ of their associated Community Notes (red for $\theta_n<0$ and blue for $\theta_n>0$). The direction that best separates notes by majority outcome, learned through logistic regression, is displayed as $CN$ along with the corresponding AUC score for sign prediction based on account positions in the 2D plane (standard deviation from 10-fold cross-validation). The direction structuring each country's political landscape $\delta_1$ is projected onto the plane.}
\label{fig:twelve_countries_plot}
\end{figure}

For greater interpretability, we project in Figure~\ref{fig:twelve_countries_plot} the X accounts onto the 2D plane defined by the left-right and Anti-Elite axes, then trained a logistic regression model to predict the sign of the latent ideology of Community Notes associated with X accounts using their coordinates on this plane as input. To bolster the robustness of the analysis, we performed 10-fold cross validation and report the average AUC-ROC of this prediction task with standard deviation across the 10 folds. Overall, with an average AUC-ROC of $0.814 \pm 0.034$, we observe that across national political settings, X's Community Notes system successfully captures users' ideological leanings.\newline

\begin{figure}[h]
\centering
\includegraphics[width=\textwidth]{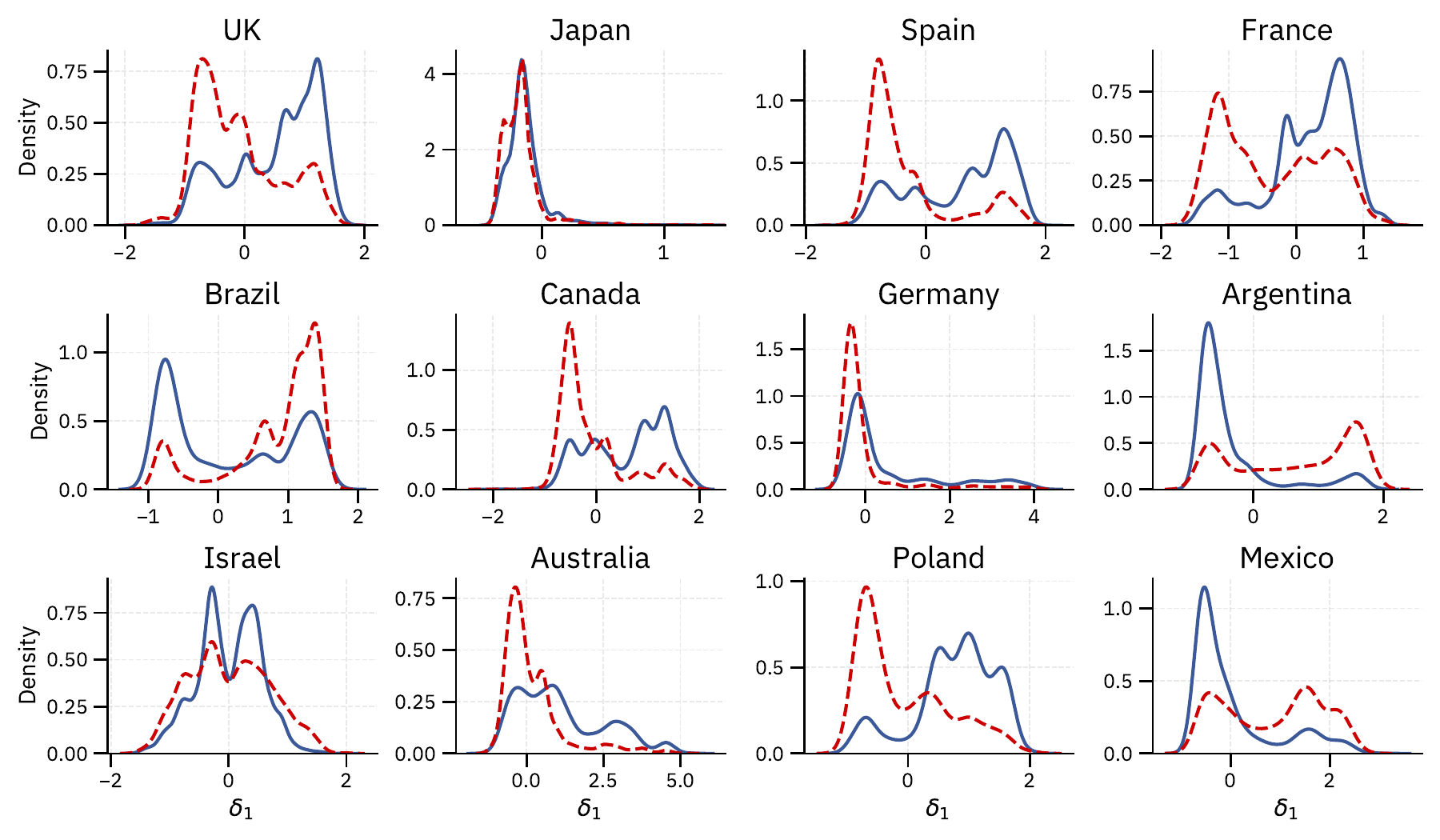}
\caption{Distribution of the position of X accounts along $\delta_1$, segmented by the Community Notes author latent ideology sign ($\theta_r<0$ plain blue and $\theta_r>0$ dashed red).}
\label{fig:delta1_note_authors}
\end{figure}

Finally, as coherent validation, we rely on a signal not considered in X Community Notes systems: note authorship. We display in Figure \ref{fig:delta1_note_authors} the position along the structuring dimension $\delta_1$ of X accounts associated with Community Notes authored by contributors with $\theta_r<0$ and $\theta_r>0$. Overall, we observe a stark distinction between these two sets of contributors in the UK, Spain, France, Brazil, Canada, Argentina, Poland, and Mexico, while the distinction is less pronounced in Japan, Germany, Israel, and Australia. This indicates, in relation to Table~\ref{tab:alignement_metrics}, that in the latter countries, the authoring of Community Notes is not as aligned with the rating of Community Notes, or alternatively that ratings of Community Notes are not structured along the same dimension as MP-following patterns.

